# Decidability and Undecidability Results
# for Propositional Schemata

**Vincent Aravantinos**                         Vincent.Aravantinos@imag.fr
**Ricardo Caferra**                               Ricardo.Caferra@imag.fr
**Nicolas Peltier**                               Nicolas.Peltier@imag.fr
*Université de Grenoble (LIG/CNRS)*
*Bât. IMAG C, 220, rue de la Chimie*
*38400 Saint Martin d'Hères, France*

## Abstract

We define a logic of *propositional formula schemata* adding to the syntax of propositional logic indexed propositions (e.g., $p_i$) and iterated connectives $\bigvee$ or $\bigwedge$ ranging over intervals parameterized by arithmetic variables (e.g., $\bigwedge_{i=1}^{n} p_i$, where $n$ is a *parameter*). The satisfiability problem is shown to be undecidable for this new logic, but we introduce a very general class of schemata, called *bound-linear*, for which this problem becomes decidable. This result is obtained by reduction to a particular class of schemata called *regular*, for which we provide a sound and complete terminating proof procedure. This schemata calculus (called STAB) allows one to capture *proof patterns* corresponding to a large class of problems specified in propositional logic. We also show that the satisfiability problem becomes again undecidable for slight extensions of this class, thus demonstrating that bound-linear schemata represent a good compromise between expressivity and decidability.

## 1. Introduction

Being able to solve classes of problems – possibly efficiently and elegantly – strongly depends on the language in which they are specified. This is decisive in a lot of applications of Artificial Intelligence. One language long used by humans is that of *schemata*. As very general characterizations of the notion of schema would be useless, we have focused on a particular class of schemata arising naturally in practice, quite expressive and (as will be shown) with "good" computational properties. These schemata are those generated by unbounded repetitions of patterns, we call them 'iterated schemata'.

We motivate our approach via an example, frequently used and well-known by the AI community: circuit verification. Circuit verification problems are often modeled as sequences of propositional problems parameterized by a natural number $n$ that encodes the size of the data (e.g., the number of bits, number of layers in the circuit, etc.). We call these sequences *iterated schemata*, or simply *schemata*. A typical example is an $n$-bit sequential adder circuit i.e. a circuit which computes the sum of two bit-vectors of length $n$. Such a circuit is built by composing $n$ 1-bit adders. The $i^{th}$ bits of each operand are written $p_i$ and $q_i$. $r_i$ is the $i^{th}$ bit of the result and $c_{i+1}$ is carried over to the next bit (thus $c_1 = 0$). We set the notations ($\oplus$ denotes exclusive or):

$$Sum_i(p, q, c, r) \stackrel{\text{def}}{=} r_i \Leftrightarrow (p_i \oplus q_i) \oplus c_i$$

and

$$Carry_i(p,q,c) \stackrel{\text{def}}{=} c_{i+1} \Leftrightarrow (p_i \wedge q_i) \vee (c_i \wedge p_i) \vee (c_i \wedge q_i).$$

Then the formula:

$$Adder(p,q,c,r) \stackrel{\text{def}}{=} \bigwedge_{i=1}^{n} Sum_i(p,q,c,r) \wedge \bigwedge_{i=1}^{n} Carry_i(p,q,c) \wedge \neg c_1$$

with the constraint $n \geq 1$, schematises the adder circuit (it states that $r$ encodes the sum of $p$ and $q$). *Adder* contains iterations ranging on intervals depending on $n$. If $n$ is instantiated by a natural number then the expression reduces to a propositional formula. Therefore each *instance* of this schema can be solved in propositional logic. However, proving that the schema is unsatisfiable (or satisfiable) *for every instance of $n$* is much harder. This problem cannot be specified in propositional logic and, as we shall see, this is even out of the scope of first-order logic. It can be expressed in higher order logics but it is well-known that such languages are less suitable for automation (see Section 3 for details).

Such iteration schemata are ubiquitous in formalized reasoning. Problems over finite domains can be specified as generic propositional formulae fitting the same pattern, the parameter being the (finite but *unbounded*) size of the domain. Among these patterns, those corresponding to the pigeonhole principle, Ramsey theory, coloring graphs problems or constraint programming specifications such as the $n$-queens problem (Marriott, Nethercote, Rafeh, Stuckey, García de la Banda, & Wallace, 2008) should be mentioned. Iterated schemata are also extremely useful for the formalization of mathematical proofs, because they allow one to express infinite proof sequences, which can avoid, for instance, explicit use of the induction principle. This idea has been used, e.g., in the work of Hetzl, Leitsch, Weller, and Woltzenlogel Paleo (2008).

In this paper we present the first (to the best of our knowledge) thorough analysis of propositional iterated schemata. We define a logic handling arithmetic variables, indexed propositions and iterated connectives. The satisfiability problem is obviously semi-decidable in the sense that a (straightforward) algorithm exists to enumerate all satisfiable schemata (i.e. all schemata with a satisfiable instance). However the set of (unrestricted) unsatisfiable schemata is not recursively enumerable. Thus we restrict ourselves to a particular class of schemata, called *bound-linear* and we provide a decision procedure for this class. This procedure is based on a reduction to a very simple class of schemata, called *regular*, for which a tableaux-based proof procedure is presented. Then we provide some undecidability results for (rather natural) extensions of this class.

The rest of the paper is structured as follows.

- In Section 2 we introduce a logic (syntax and semantics) for handling propositional schemata and we establish some of its basic properties. The propositional symbols are indexed by arithmetic expressions (e.g., $p_{n+1}$) containing arithmetic variables. These variables can be either parameters (i.e. free variables), or bound variables introduced by generalized connectives of the form $\bigvee_{i=a}^{b}$ or $\bigwedge_{i=a}^{b}$. These connectives can be read as $\exists i \in [a,b]$ or $\forall i \in [a,b]$, where $a,b$ are arithmetic expressions possibly containing (free or bound) variables. We restrict ourselves to monadic schemata (i.e. the propositions are indexed by at most one expression) and to linear arithmetic expressions[1]. We then introduce a particular subclass of schemata, called *bound-linear*. Intuitively, a schema is bound-linear if every arithmetic expression occurring in it contains at most one *bound* variable. Furthermore, the coefficient of this variable in the expression should be $\pm 1$ (or 0). Thus expressions such as $1, n, 2n - i$ or $i + 2$ are allowed (where $n$ is the parameter and $i$ a bound variable), but $2i$ or $i + j$ (where $i, j$ are both bound) are not. The coefficient of the parameter $n$ is not constrained.

- Section 3 contains a brief survey of existing work on propositional schemata as well as (informal) comparisons with related logics.

- In Section 4 we introduce a simpler class of schemata, called regular, and we provide an algorithm to transform every bound-linear schema into a (sat-)equivalent regular schema.

- In Section 5 a tableaux-based proof procedure, called STAB (standing for **s**chemata **tab**leaux), is introduced for reasoning with propositional schemata. This proof procedure is sound and complete (w.r.t. *satisfiability*) and terminates on every regular schema. Together with the results in Section 4 this implies that the class of bound-linear schemata is decidable.

- Section 6 shows that relaxing very slightly the conditions on bound-linear schemata makes the satisfiability problem undecidable. Thus this class can be seen as "canonical", with a good trade-off between expressivity, simplicity of the definition and decidability.

- Finally, Section 7 summarizes the results and provides some lines of future work.

## 2. Schemata of Propositional Formulae

In this section, we introduce the syntax and semantics of propositional schemata.

### 2.1 Syntax

The set of *linear arithmetic expressions* (denoted by $\mathcal{N}$) is built as usual on the signature $0, s, +, -$ and on a fixed and countably infinite set of arithmetic variables $\mathcal{V}$, quotiented by the usual properties of the arithmetic symbols (e.g., $n + s(0) + n + s(s(s(0)))$ and $n + n + s(s(s(s(0))))$ are assumed to be equivalent). As usual, $s^\kappa(0)$ is denoted by $\kappa$ and $i + \ldots + i$ ($\kappa$ times) is $\kappa.i$. If $n$ is an arithmetic variable we denote by $\mathcal{N}_{\times n}$ the set of arithmetic expressions of the form $\alpha.n + \beta$ where $\alpha, \beta \in \mathbb{Z}$ (with possibly $\alpha = 0$) and by $\mathcal{N}_n$ the set of expressions of the form $n + \beta$ where $\beta \in \mathbb{Z}$. Obviously $\mathcal{N}_n \subset \mathcal{N}_{\times n} \subset \mathcal{N}$. If $n + \alpha, n + \beta \in \mathcal{N}_n$ we write $n + \alpha \leq n + \beta$ iff $\alpha \leq \beta$.

---

1. If one of these two conditions does not hold then the satisfiability problem is trivially undecidable. For instance, the Post correspondence problem can be easily encoded into schemata with non monadic variables (Aravantinos, Caferra, & Peltier, 2009b). Similarly, if non linear arithmetic expressions are considered then the 10th Hilbert's problem can be encoded.

For the sake of readability, we adopt the following conventions. Integers are denoted by Greek letters $\alpha, \beta, \gamma, \delta^2$, natural numbers by $\kappa$ or $\iota$, arithmetic variables by $i, j, k, n$, propositional variables by $p, q, r$ (with indices). Arithmetic expressions are denoted by $a, b, c, d$. Schemata are denoted by $\phi, \psi$. $\Pi$ and $\Gamma$ denote generic iteration connectives $\bigvee$ or $\bigwedge$.

### Definition 2.1 (Indexed propositions)

Let $\mathcal{P}$ be a fixed and countably infinite set of *propositional symbols*. An *indexed proposition* is an expression of the form $p_a$ where $p \in \mathcal{P}$ and $a$ is a linear arithmetic expression (the *index*). An indexed proposition $p_a$ s.t. $a \in \mathbb{Z}$ is called a *propositional variable*. A *literal* is an indexed proposition or its negation.

In contrast to our previous work (Aravantinos et al., 2009b) we only consider *monadic* propositions, i.e. every proposition has only one index.

### Definition 2.2 (Schemata)

The set of *formula schemata* is the smallest set satisfying the following properties.

- $\top$, $\bot$ are formula schemata.

- If $a, b$ are integer expressions then $a < b$ is a formula schema.

- Each indexed proposition is a formula schema.

- If $\phi_1$, $\phi_2$ are schemata then $\phi_1 \vee \phi_2$, $\phi_1 \wedge \phi_2$ and $\neg\phi_1$ are formula schemata.

- If $\phi$ is a formula schema not containing $<$, and if $a, b \in \mathcal{N}$, and $i$ is an arithmetic variable, then $\bigwedge_{i=a}^{b} \phi$ and $\bigvee_{i=a}^{b} \phi$ are formula schemata.

Notice that, by definition, every schema must be finite. Schemata of the form $a < b$, $p_a$ or $\top, \bot$ are called *atoms*. Schemata of the form $\bigwedge_{i=a}^{b} \phi$ and $\bigvee_{i=a}^{b} \phi$ are called *iterations*, $a$ and $b$ are the *bounds* of the iteration and $b - a$ is its *length* (notice that $b - a$ may contain variables). A schema is an *arithmetic formula* iff it contains no iteration and if every atom occurring in it is of the form $\top, \bot$ or $a < b$. In particular, every boolean combination of arithmetic atoms is a schema. $a \leq b$ (or $b \geq a$) and $a = b$ are used as abbreviations for $\neg(b < a)$ and $\neg(b < a) \wedge \neg(a < b)$ respectively. As for arithmetic expressions, arithmetic formulae are taken up to arithmetic equivalence, e.g., $n = 1$ and $n < 2 \wedge n > 0$ are considered identical. The usual priority rules apply to disambiguate the reading of formula schemata. Analogously to first-order logic quantifiers, the iteration operators have the highest priority (e.g., $\bigwedge_{i=1}^{n} p_i \vee p_n \wedge \neg p_1$ should be read as $(\bigwedge_{i=1}^{n} p_i) \vee (p_n \wedge \neg p_1)$).

### Example 2.3

$$\phi = q_1 \wedge \bigwedge_{i=1}^{n} \left( p_{i+2n} \wedge \bigvee_{j=n}^{2n+1} (\neg q_{n-j} \vee q_{j+1}) \right) \wedge n \geq 0 \quad \text{is a formula schema.}$$

$q_1$, $p_i$, $q_j$ and $q_{j+1}$ are indexed propositions. $\bigwedge_{i=1}^{n} \left( p_{i+2n} \wedge \bigvee_{j=n}^{2n+1} (\neg q_{n-j} \vee q_{j+1}) \right)$ and $\bigvee_{j=n}^{2n+1} (\neg q_{n-j} \vee q_{j+1})$ are the only iterations occurring in $S$.

---

2. This slightly unusual convention is used to avoid confusion between arithmetic variables and integers.

**Remark 2.4**
Notice that the arithmetic atoms of the form $a < b$ can only occur outside the iterations, i.e. $n \geq 1 \Rightarrow \bigvee_{i=1}^{n} p_i$ is allowed, but neither $\bigvee_{i=1}^{n}(i \leq 3 \vee p_i)$ nor $\bigvee_{i=1}^{n}(n \geq 1 \Rightarrow p_i)$. This restriction is only used to simplify technicalities. As we shall see in Definition 2.5 (semantics of schemata), an arithmetic atom of the form $a < b$ is equivalent to the schema $\bigvee_{i=a+1}^{b} \top$.

A variable $i$ is *bound* in $\phi$ if $\phi$ contains an iteration of the form $\Pi_{i=a}^{b}\psi$ ($\Pi \in \{\bigvee, \bigwedge\}$), it is *free* (or is a *parameter* of $\phi$) if it has an occurrence in $\phi$ which is not in the scope of an iteration $\Pi_{i=a}^{b}\psi$. From now on, we assume that no variable is simultaneously free and bound in a schema $\phi$ (thus schemata such as $p_n \wedge \bigvee_{n=1}^{10} \neg p_n$ are not well-formed) and that if $\Pi_{i=a}^{b}\psi$ and $\Gamma_{j=c}^{d}\psi'$ (where $\Pi, \Gamma \in \{\bigvee, \bigwedge\}$) are two distinct iterations occurring in $\phi$ then $i$ and $j$ are distinct.

A *substitution* is a function mapping every arithmetic variable to a linear arithmetic expression. We write $[a_1/i_1, \ldots, a_\kappa/i_\kappa]$ for the substitution mapping respectively $i_1, \ldots, i_\kappa$ to $a_1, \ldots, a_\kappa$. The application of a substitution $\sigma$ to a schema (or arithmetic expression) $\phi$ is defined as usual and denoted by $\phi\sigma$. Notice that if $a$ is an arithmetic expression and $\sigma$ a substitution mapping every variable in $a$ to a ground term (i.e. a term with no variable) then $a\sigma$ is an integer (since we identify, e.g., $2 - 1$ and $1$).

The previous notation is also used to denote the replacement of subexpressions: If $\phi$ is a schema, $\psi$ is an expression (schema or arithmetic expression) occurring in $\phi$ and $\psi'$ is an expression of the same type as $\psi$, then $\phi[\psi'/\psi]$ denotes the formula obtained by replacing all the occurrences of $\psi$ in $\phi$ by $\psi'$.

## 2.2 Semantics

An *interpretation of the schemata language* is a function mapping every integer variable to an integer and every propositional variable to a truth value $\mathbf{T}$ or $\mathbf{F}$. If $\mathcal{I}$ is an interpretation and $\sigma$ a substitution, we denote by $\mathcal{I}\sigma$ the interpretation defined as follows: $\mathcal{I}\sigma$ and $\mathcal{I}$ coincide on every propositional variable and for every variable $n$, $\mathcal{I}\sigma(n) \stackrel{\text{def}}{=} \mathcal{I}(n\sigma)$. Consider for instance the following interpretation $\mathcal{I}$:

$$n \mapsto 5$$
$$m \mapsto 2$$
$$p_1 \mapsto \mathbf{T}$$
$$p_2 \mapsto \mathbf{F}$$
$$p_3 \mapsto \mathbf{F}$$
$$p_4 \mapsto \mathbf{F}$$

and whose definition is unsignificant for other (integer or propositional) variables. Let also be $\sigma$ the substitution $\{n \mapsto n - 1, m \mapsto m - 2\}$. Then $\mathcal{I}\sigma$ is:

$$n \mapsto 4$$
$$m \mapsto 0$$
$$p_1 \mapsto \mathbf{T}$$
$$p_2 \mapsto \mathbf{F}$$
$$p_3 \mapsto \mathbf{F}$$
$$p_4 \mapsto \mathbf{F}$$

If $\mathcal{I}$ is an interpretation, we denote by $\sigma_\mathcal{I}$ the restriction of $\mathcal{I}$ to $\mathcal{V}$, i.e. the substitution mapping every variable $n$ to $\mathcal{I}(n)$. If $a$ is an arithmetic expression, we denote by $[\![a]\!]_\mathcal{I}$ the expression $a\sigma_\mathcal{I}$. Since $a\sigma_\mathcal{I}$ is ground, it is (equivalent to) an integer.

### Definition 2.5 (Semantics)

The *truth value* $[\![\phi]\!]_\mathcal{I}$ of a propositional schema in an interpretation $\mathcal{I}$ is inductively defined as:

- $[\![\top]\!]_\mathcal{I} = \mathbf{T}$, $[\![\bot]\!]_\mathcal{I} = \mathbf{F}$

- $[\![a < b]\!]_\mathcal{I} = \mathbf{T}$ iff $[\![a]\!]_\mathcal{I} < [\![b]\!]_\mathcal{I}$.

- $[\![p_a]\!]_\mathcal{I} = \mathcal{I}(p_{[\![a]\!]_\mathcal{I}})$ for $p \in \mathcal{P}$.

- $[\![\neg\Phi]\!]_\mathcal{I} = \mathbf{T}$ iff $[\![\Phi]\!]_\mathcal{I} = \mathbf{F}$.

- $[\![\Phi \vee \Phi']\!]_\mathcal{I} = \mathbf{T}$ iff $[\![\Phi]\!]_\mathcal{I} = \mathbf{T}$ or $[\![\Phi']\!]_\mathcal{I} = \mathbf{T}$.

- $[\![\Phi \wedge \Phi']\!]_\mathcal{I} = \mathbf{T}$ iff $[\![\Phi]\!]_\mathcal{I} = \mathbf{T}$ and $[\![\Phi']\!]_\mathcal{I} = \mathbf{T}$.

- $[\![\bigvee_{i=a}^{b} \phi]\!]_\mathcal{I} = \mathbf{T}$ iff there is an integer $\alpha$ s.t. $[\![a]\!]_\mathcal{I} \leq \alpha \leq [\![b]\!]_\mathcal{I}$ and $[\![\phi]\!]_{\mathcal{I}[\alpha/i]} = \mathbf{T}$.

- $[\![\bigwedge_{i=a}^{b} \phi]\!]_\mathcal{I} = \mathbf{T}$ iff for every integer $\alpha$ s.t. $[\![a]\!]_\mathcal{I} \leq \alpha \leq [\![b]\!]_\mathcal{I}$: $[\![\phi]\!]_{\mathcal{I}[\alpha/i]} = \mathbf{T}$.

A schema $\phi$ is *satisfiable* iff there is an interpretation $\mathcal{I}$ s.t. $[\![\phi]\!]_\mathcal{I} = \mathbf{T}$. $\mathcal{I}$ is called a *model* of $\phi$ (written $\mathcal{I} \models \phi$). Two schemata $\phi, \psi$ are *equivalent* (written $\phi \equiv \psi$) iff $\mathcal{I} \models \phi \Leftrightarrow \mathcal{I} \models \psi$. $\phi$ and $\psi$ are *sat-equivalent* (written $\phi \equiv_S \psi$) iff $\phi$ and $\psi$ are both satisfiable or both unsatisfiable.

In the following, we assume that for every free variable $n$ in $\phi$ and for every model $\mathcal{I}$ of $\phi$, $\mathcal{I}(n) \in \mathbb{N}$. This can be ensured by explicitly adding the arithmetic atom $n \geq 0$ to $\phi$[3].

Let $\mathcal{S}$ be the following system of rewrite rules:

---

3. Thus we assume that parameters are mapped to natural numbers. This convention is convenient because it allows one to use mathematical induction on the parameters (see Section 5.2). It is not restrictive since a schema $\phi$ where $n \in \mathbb{Z}$ could be replaced by the (equivalent) disjunction of the schemata $\phi \wedge n \geq 0$ and $\phi[-m/n] \wedge m \geq 0$ (i.e. in the case in which $n$ is negative, every occurrence of $n$ is simply replaced by $-m$).

$$\mathcal{S} = \begin{cases} \bigvee_{i=\alpha}^{\beta} \phi & \to & \bot & \text{if } \alpha, \beta \in \mathbb{Z}, \beta < \alpha \\ \bigwedge_{i=\alpha}^{\beta} \phi & \to & \top & \text{if } \alpha, \beta \in \mathbb{Z}, \beta < \alpha \\ \bigvee_{i=\alpha}^{\beta} \phi & \to & (\bigvee_{i=\alpha}^{\beta-1} \phi) \vee \phi[\beta/i] & \text{if } \alpha, \beta \in \mathbb{Z}, \beta \geq \alpha \\ \bigwedge_{i=\alpha}^{\beta} \phi & \to & (\bigwedge_{i=\alpha}^{\beta-1} \phi) \wedge \phi[\beta/i] & \text{if } \alpha, \beta \in \mathbb{Z}, \beta \geq \alpha \end{cases}$$

For instance the following formula:

$$\neg p_1 \wedge \bigwedge_{i=1}^{3} (p_i \Rightarrow p_{i+1})$$

is rewritten into:

$$\neg p_1 \wedge (p_1 \Rightarrow p_2) \wedge (p_2 \Rightarrow p_3) \wedge (p_3 \Rightarrow p_4)$$

Notice that no rule of $\mathcal{S}$ applies on $\neg p_1 \wedge \bigwedge_{i=1}^{n} (p_i \Rightarrow p_{i+1})$ as the upper bound of the iteration contains a parameter. $\mathcal{S}$ is actually designed to be used only on schemas whose parameters have been instantiated by a number.

## Proposition 2.6
$\mathcal{S}$ is convergent and preserves equivalence.

### Proof
Termination is immediate since the length of an iteration strictly decreases at each step. Confluence is obvious since the critical pairs are trivially joinable. The fact that the obtained schema is equivalent to the original one is a straightforward consequence of Definition 2.5. $\square$

We denote by $\phi{\downarrow}_{\mathcal{S}}$ the (unique) normal form of $\phi$. If $\sigma$ is a substitution mapping every free variable in $\phi$ to a natural number, $\phi\sigma{\downarrow}_{\mathcal{S}}$ is called a *propositional realization* of $\phi$.

It is trivially semi-decidable to know if a schema is satisfiable:

## Proposition 2.7
The set of satisfiable schemata is recursively enumerable.

### Proof
By Definition 2.5, for every interpretation $\mathcal{I}$ and for every schema $\phi$, we have $(\mathcal{I} \models \phi) \Leftrightarrow (\mathcal{I} \models \phi\sigma)$, where $\sigma = \sigma_{\mathcal{I}}$. Thus $\phi$ is satisfiable iff there exists a substitution $\sigma$ such that $\phi\sigma$ is satisfiable. We now prove that there exists an algorithm for checking the satisfiability of $\phi\sigma$. By Proposition 2.6, we have $\phi\sigma \equiv \phi\sigma{\downarrow}_{\mathcal{S}}$. By definition of $\sigma$, $\phi\sigma$ contains no free variable. Let $\Pi_{i=a}^{b}\phi$ be an outermost iteration in $\phi$. By definition $a$ and $b$ must be ground, thus one of the rules in $\mathcal{S}$ applies which is impossible. Thus $\phi\sigma{\downarrow}_{\mathcal{S}}$ contains no iteration hence $\phi\sigma{\downarrow}_{\mathcal{S}}$ is a propositional formula (in the usual sense) built on the set of propositional variables. Consequently, there exists an algorithm to check whether the formula $\phi\sigma{\downarrow}_{\mathcal{S}} \equiv \phi\sigma$ is satisfiable or not. Since the set of ground substitutions is recursively enumerable, and since $\phi$ is satisfiable iff $\phi\sigma$ is satisfiable for at least one substitution $\sigma$, this implies that it is semi-decidable to check whether $\phi$ is satisfiable or not. $\square$

For every schema $\phi$ and for every substitution $\sigma$ we denote by $[\phi]_\sigma$ the formula $\phi\sigma{\downarrow}_{\mathcal{S}}$.

For every arithmetic expression $a$ (possibly containing bound variables) in a schema $\phi$, we compute an interval $[min_\phi(a), max_\phi(a)]$ where $\min_\phi(a), \max_\phi(a)$ are arithmetic expressions only containing variables that are free in $\phi$. The intuition is that $a$ always "belongs" to this interval. Lemma 2.8 formalizes this property.

- If $a$ is an integer or a variable that is free in $\phi$ then $\min_\phi(a) \overset{\text{def}}{=} \max_\phi(a) \overset{\text{def}}{=} a$.

- If $a$ is of the form $b+c$ then $\min_\phi(a) \overset{\text{def}}{=} \min_\phi(b) + \min_\phi(c)$ and $\max_\phi(a) \overset{\text{def}}{=} \max_\phi(b) + \max_\phi(c)$.

- If $a$ is of the form $-b$ then $\max_\phi(a) \overset{\text{def}}{=} -\min_\phi(b)$ and $\min_\phi(a) \overset{\text{def}}{=} -\max_\phi(b)$.

- If $i$ is a bound variable, occurring in an iteration of the form $\Pi_{i=a}^{b}\phi$ then $\min_\phi(i) \overset{\text{def}}{=} \min_\phi(a)$ and $\max_\phi(i) \overset{\text{def}}{=} \max_\phi(b)$.

A ground substitution $\sigma'$ is a $\phi$-*expansion* of another ground substitution $\sigma$ for a subschema $\psi$ in $\phi$ iff for every variable $i$ that is bound in $\psi$, $\sigma'(i) \in [\sigma(\min_\phi(i)), \sigma(\max_\phi(i))]$ (since $\sigma, \sigma'$ are ground, the expressions $\sigma'(i), \sigma(\min_\phi(i)), \sigma(\max_\phi(i))$ are considered as integers). The intuition behind $\phi$-expansions is the following: A substitution $\sigma$ does not affect the bound variables of a schema; so the values given by $\sigma$ to such bound variables are unsignificant; on the contrary, the definition of a $\phi$-expansion $\sigma'$ imposes that:

1. the value given to a variable $i$ bound in $\phi$ indeed falls in the set of values that $i$ can take in the context of $\phi$ ;

2. the value given by $\sigma'$ to a variable free in $\phi$ is the same as the one given by $\sigma$.

W.r.t. substitution application, there is no difference between $\sigma$ and $\sigma'$. The next lemma shows the importance of $\phi$-expansions.

**Lemma 2.8**
Let $\phi$ be a schema and let $i$ be a variable (possibly bound) occurring in $\phi$. The expressions $\min_\phi(i)$ and $\max_\phi(i)$ are well-defined. Moreover, for every ground substitution $\sigma$ and for all atoms $p_\alpha$ occurring in $[\phi]_\sigma$ there exist an atom $p_a$ occurring in $\phi$ and a $\phi$-expansion $\sigma'$ of $\sigma$ for $p_a$ s.t. $\sigma'(a) = \alpha$.

PROOF
This is an immediate consequence of Definition 2.5 (by a straightforward induction on the depth of the schema). □

We write $IC(\phi)$ (standing for "Interval Constraints") for the conjunction of arithmetic constraints of the form $\min_\phi(i) \leq i \wedge i \leq \max_\phi(i)$ where $i$ is a variable that is bound in $\phi$. $IC(\phi)$ can be extended to sets of schemata by handling them as conjunctions.

Consider, e.g., $\phi = p_0 \wedge \bigwedge_{i=1}^{n-1}(p_{i+1} \wedge \neg q_i)$. We have: $\min_\phi(i) = 1$ and $\max_\phi(i) = n-1$. Consider furthermore $\sigma = \{n \mapsto 4\}$ and $p_\alpha = p_3$. Then we can take $p_a = p_{i+1}$ (which indeed occurs in $\phi$) and $\sigma' = \{n \mapsto 4, i \mapsto 2\}$.

We see informally the use of $\phi$-expansions: they allow, in some sense, to make the connection between a propositional variable occurring in the instance of a schema and the indexed proposition where it "comes from".

## 2.3 The Class of Bound-Linear Schemata

As we shall see (in, e.g., Theorem 6.2) the satisfiability problem is undecidable for schemata. In order to characterize a decidable subclass, we introduce the following definition:

**Definition 2.9**

A schema $\phi$ is *bound-linear* iff the following conditions hold:

1. $\phi$ contains at most one free arithmetic variable $n$ (called the *parameter* of $\phi$).

2. Every non arithmetic atom in $\phi$ is of the form $p_{\alpha.n+\beta.i+\gamma}$ where $p \in \mathcal{P}$ and $i$ is a bound variable, $\alpha, \gamma \in \mathbb{Z}$ and $\beta \in \{-1, 0, 1\}$.

3. If $\Pi_{i=a}^{b}\psi$ is an iteration in $\phi$ (where $\Pi \in \{\bigvee, \bigwedge\}$) then $a, b$ are respectively of the form $\alpha.n+\beta$ and $\gamma.n+\delta+\epsilon.j$ where $\alpha, \beta, \gamma, \delta \in \mathbb{Z}$, $\epsilon \in \{-1, 0, 1\}$ and $j$ is a bound variable.

This class is comprehensive enough with respect to decidable satisfiability. The key point is that all the indices and iteration bounds contain *at most* one bound variable. Furthermore, the coefficient of this variable must be 1 (or 0).

## 2.4 Expressiveness of Bound-Linear Schemata

In order to show evidence that the class of bound-linear schemata is not an artificial or too narrow one, we provide in this section some examples of problems that can be naturally encoded into bound-linear schemata.

It is easy to check that the schema $Adder(p, q, c, r)$ defined in the Introduction (formalizing a sequential adder) is bound-linear. Various properties of this circuit can be encoded. For instance, the following schema checks that 0 is a (left) neutral element:

$$(Adder(p, q, c, r) \wedge \bigwedge_{i=1}^{n} \neg p_i) \Rightarrow \bigwedge_{i=1}^{n} (r_i \Leftrightarrow q_i)$$

The schema below checks that the adder is a function i.e. that the sum of two operands is unique.

$$(Adder(p, q, c, r) \wedge Adder(p, q, c', r')) \Rightarrow \bigwedge_{i=1}^{n} (r_i \Leftrightarrow r'_i)$$

The next one checks that it is commutative:

$$(Adder(p, q, c, r) \wedge Adder(q, p, c', r')) \Rightarrow \bigwedge_{i=1}^{n} (r_i \Leftrightarrow r'_i)$$

Many similar circuits can be formalized in a similar way, such as a carry look-ahead adder (a faster version of the $n$-bit adder that reduces the amount of time required to compute carry bits):

$$CLA\text{-}Adder(p, q, c) \stackrel{\text{def}}{=} \bigwedge_{i=1}^{n} (r_i \Leftrightarrow ((p_i \oplus q_i) \oplus c_i)) \wedge \bigwedge_{i=1}^{n} (c_{i+1} \Leftrightarrow (p_i \wedge q_i) \vee (c_i \wedge (p_i \vee q_i)))$$

The equivalence of the two definitions is encoded as follows:

$$(Adder(p, q, c, r) \land CLA\text{-}Adder(p, q, c', r')) \Rightarrow \bigwedge_{i=1}^{n} (r_i \Leftrightarrow r'_i)$$

Comparison between two natural numbers can easily be formalized, e.g. $r_n$ holds iff $p \geq q$:

$$r_0 \land \bigwedge_{i=1}^{n} (r_i \Leftrightarrow (r_{i-1} \land (p_i \Leftrightarrow q_i) \lor p_i \land \neg q_i))$$

By composing the previous schemata, any (quantifier-free) formula of Presburger arithmetic can be encoded.

More generally, one can formalize every circuit composed by serially putting together $n$ layers of the same basic circuit. These circuits are usually defined inductively, which can be easily encoded into our formalism with a formula of the form:

$$(p_0 \Leftrightarrow \phi_{base}) \land \bigwedge_{i=0}^{n-1} (p_{i+1} \Leftrightarrow \phi_{ind}),$$

where $\phi_{base}$ and $\phi_{ind}$ are the formulae corresponding to the base case and inductive case, respectively. $\phi_{ind}$ contains some occurrences of $p_i$ and encodes the basic circuit to be composed in sequence. Of course, for most complex circuits, $p_i$ may be replaced by a *vector* of bits $p_i, q_i, r_i$ defined inductively from the $p_{i-1}, q_{i-1}, r_{i-1}, \dots$. Such inductively-defined circuits appear very frequently in practice (Gupta & Fisher, 1993).

If the index of the proposition denotes the time, then various finite state sequential systems can be encoded. The state of the system is described by a set of propositional variables, and $p_i$ encodes the value of $p$ at step $i$. The parameter $n$ denotes the number of steps in the transformation (which is assumed to be finite but unbounded). The transition function from state $i$ to $i + 1$ can easily be formalized by a bound-linear schema. For instance, the inclusion of two automata can be encoded (the parameter being the length of the run). We provide another example. Consider a register with three cells $p, q, r$ and assume that there are two possible actions rl and rr that rotate the values of the cells to the left and to the right respectively. The behavior of this system is modeled by the following schema (the propositions $rl_i$ and $rr_i$ indicate which action is applied at step $i$). First $L(i)$ expresses the state of the registers at time $i$ depending on their state at time $i - 1$, when $rl_i$ has been applied to it:

$$L(i) \equiv rl_i \Rightarrow ((p_i \Leftrightarrow q_{i-1}) \land (q_i \Leftrightarrow r_{i-1}) \land (r_i \Leftrightarrow p_{i-1}))$$

Then $R(i)$ is similar for rr:

$$R(i) \equiv rr_i \Rightarrow ((p_i \Leftrightarrow r_{i-1}) \land (q_i \Leftrightarrow p_{i-1}) \land (r_i \Leftrightarrow q_{i-1}))$$

Finally, we state that this holds at any time:

$$\phi_n \equiv \bigwedge_{i=1}^{n} L(i) \land \bigwedge_{i=1}^{n} R(i)$$

We can then express properties on such registers. For instance, the following formula states that $n$ rotations to the right followed by $n$ rotations to the left are equivalent to identity:

$$(\phi_{2n} \wedge \bigwedge_{i=1}^{n} \text{rr}_i \wedge \bigwedge_{i=n+1}^{2n} \text{rl}_i) \Rightarrow (p_0 \Leftrightarrow p_{2n}) \wedge (q_0 \Leftrightarrow q_{2n}) \wedge (r_0 \Leftrightarrow r_{2n})$$

## 3. Related Work

Different forms of schemata have been used by several authors, either in propositional logic (Baaz & Zach, 1994) or in first-order logic to obtain results in proof theory, in particular related to the number of proof lines (Parikh, 1973; Baaz, 1999; Krajicek & Pudlak, 1988; Orevkov, 1991). Parikh (1973) presents a notion of schematic systems, Baaz (1999) uses the concept of unification, Krajicek and Pudlak (1988) introduce the notion of 'proof skeleton', very similar to that of schema, Orevkov (1991) studies schemata in first-order Hilbert-type system. Pragmatically, schemata have been successfully used, e.g., in solving open questions in equivalential calculus (i.e. the field of formal logic concerned with the notion of equivalence) with the theorem-prover OTTER (Wos, Overbeek, Lush, & Boyle, 1992). However, to the best of our knowledge, the formal handling of such schemata at the object level has never been considered. Although the notion of 'schema' is recognized as an important one, it deserves more applied works in our opinion. Sometimes schemata are not sufficiently emphasized, e.g., in the work of Barendregt and Wiedijk (2005) a nice and deep analysis about the challenge of computer mathematics is given. The authors overview the state of the art (by describing and comparing most powerful existing systems in use) but structuring proofs is not explicitly mentioned (maybe this feature can be included in what they call "mathematical style" or "support reasoning with gaps"). In our approach to schemata it is clear that they are a way of structuring proofs and can also help to overcome one of the obstacles to the automation of reasoning pointed out by Wos (1988), i.e. *the size of deduction steps*.

There exist term languages expressive enough to denote iteration schemata as those introduced in Definition 2.2: In particular, **term schematisation languages** can be used to denote infinite sequences of structurally similar terms or formulae. For instance the primal grammar (Hermann & Galbavý, 1997) $\hat{f}(n) \rightarrow (p(n) \vee \hat{f}(n-1)), \hat{f}(0) \rightarrow \perp$ denotes the iteration $\bigvee_{i=1}^{n} p_i$. It is worth mentioning that this iteration cannot be denoted by other term schematisation languages (Chen, Hsiang, & Kong, 1990; Comon, 1995) because the inductive context is not constant. However, term schematisation languages do not allow to *reason* on such iterations (they are only useful to *represent* them).

Encoding schemata into **first-order logic** is a very natural idea, interpreting iterated connectives as bounded quantifiers. Additional axioms can be added to express arithmetic properties if needed. For instance the schema $(\bigvee_{i=1}^{n} p_i) \wedge (\bigwedge_{i=1}^{n} \neg p_i)$ can be encoded by $\exists i.(1 \le i \wedge i \le n \wedge p(i)) \wedge \forall i.(1 \le i \wedge i \le n \Rightarrow \neg p(i))$ which is obviously unsatisfiable. However, since inductive domains cannot be defined in first-order logic, such a translation necessarily introduces some unintended interpretations hence does not yield a complete procedure (satisfiability is not always preserved, although the unsatisfiability of the obtained formula necessarily entails the unsatisfiability of the original one). For instance, the schema $p_0 \wedge \bigwedge_{i=1}^{n}(p_{i-1} \Rightarrow p_i) \wedge \neg p_n$ is translated into $p(0) \wedge \forall i.(1 \le i \wedge i \le n \wedge p(i-1) \Rightarrow p(i) \wedge \neg p(n))$, which is actually satisfiable (we do not know that $n \in \mathbb{N}$ and there is no way to express this property). In order to obtain an unsatisfiable formula, some *inductive axioms* must be added to allow (necessarily restricted) applications of the induction principle. In this particular case, the proof can be obtained by a simple induction on $i$ using the inductive lemma $\forall i.(i \leq n \Rightarrow p(i))$, thus we could add the axiom: $[q(0) \wedge \forall i.(q(i) \Rightarrow q(i+1))] \Rightarrow \forall i.q(i)$ where $q(i) \equiv i \leq n \Rightarrow p(i)$. With this axiom, it is easy to check that the previous formula becomes unsatisfiable. However, in the general case it is hard to determine a priori the right axiom (if there is one). Actually the termination proof in Section 5 implicitly provides a way to determine candidate axioms (for the particular class of regular schemata): every looping node in the tableaux constructed by the proof procedure STAB (see Section 5) corresponds to an application of the induction principle, hence to an induction axiom. The termination proof precisely shows that the size of these inductive lemmata is bounded, thus the whole set of potential induction axioms could be in principle computed and added to the formula before the beginning of the search. But the practical interest of this transformation is obviously highly questionable.

Several procedures have been designed for proving **inductive theorems**, (Boyer & Moore, 1979; Bouhoula, Kounalis, & Rusinowitch, 1992; Comon, 2001; Bundy, van Harmelen, Horn, & Smaill, 1990; Bundy, 2001). Since schemata can be seen as an "explicit way" of handling mathematical induction, using such proof procedures for proving them is a very natural idea. In general, induction is used to define terms (e.g., recursive functions operating on inductive data structures), whereas in our case the formulae themselves are defined inductively. Obviously this problem could be solved by using an appropriate encoding of the formulae. However there are very few decidability results in inductive theorem proving and known classes (Giesl & Kapur, 2001) are not expressive enough to encode propositional schemata. Notice that most systems concentrate on universal quantifications, where we have to handle both iterated conjunctions (which can be interpreted as universal quantification on a finite domain) and iterated disjunctions (i.e. the analogous of existential quantifications). Adding existential quantification in inductive theorem proving is known to be a difficult problem. Most inductive theorem provers are designed to prove universal theorems of the form $\forall \vec{x}.\psi$ where $\psi$ is a quantifier-free formula (usually a clause) and the variables in $\vec{x}$ range over the set of (finite) terms. In our context, $\psi$ would contain finite quantification (over intervals constrained by $n$), corresponding to the iterated connectives. In particular, schemata may have several models, thus implicit induction (Comon, 2001) (which explicitly requires that the underlying Herbrand model is unique) cannot be (directly) used.

Of course, these problems can be overcome by encoding interpretations as terms (for instance by vectors or ordered lists of truth values) and schemata as functions mapping every interpretation to a truth value. Then inductive theorem provers may be used to prove inductive properties of these functions (showing for instance that their value is $\bot$ for every interpretation). However these provers are not complete (due to well-known theoretical limitations) thus the practical interest of this encoding is unclear. For instance, we have tried to use the theorem prover ACL2 to prove the validity of some of the benchmarks considered in Section 5, but it fails on all non trivial examples. We conjecture that this is not only due to efficiency problems, but that additional inductive lemmata are needed, which are very hard to determine in advance.

The above definitions should also remind the reader of **fixed point logics**. Indeed iterated schemata are obviously particular cases of fixed points, e.g., the schema $\bigwedge_{i=1}^{n} p_i$ might be represented as $(\mu X(i).i \leq 0 \vee (p(i) \wedge X(i-1)))(n)$. The "standard" fixed point logic is the (propositional) *modal $\mu$-calculus* (Bradfield & Stirling, 2007) in which many temporal logics can be encoded, e.g., LTL or CTL. However the involved logic is very different from ours and actually simpler from a theoretical point of view. Indeed modal $\mu$-calculus is decidable (and thus complete) whereas – as we shall see in Section 6 – iterated schemata are not (nor are they complete). Furthermore, our language allows one to use complex (though carefully restricted) arithmetic operations in the definition of the iterations, both in the indices and in the bounds. For instance we may relate the truth values of two propositions whose index are arbitrary far from each other (such as $p_i$ and $p_{n-i}$). As far as we are aware, these operations cannot be directly encoded into propositional $\mu$-calculus.

Actually iterated schemata share much more with **least fixpoint logic**, or LFP (Immerman, 1982), studied in finite model theory (Fagin, 1993; Ebbinghaus & Flum, 1999): LFP is a logic allowing to iterate first-order formulae maintaining constant the number of their variables. However we do no know of any *calculus* for deciding the satisfiability in LFP. We see two reasons for this: first, LFP is undecidable and not complete, second the purposes of this logic are mainly theoretical, hence the fact that research in this field has not focussed on decision procedures for some subclasses. In contrast with propositional $\mu$-calculus, **first-order $\mu$-calculus** (Park, 1976) clearly embeds iterated schemata (allowing for instance the above fixed-point expression of $\bigwedge_{i=1}^{n} p_i$), but no published research seems to be focused on the identification of complete subclasses. With a similar expressive power one also finds logics with inductive definitions (Aczel, 1977) which are quite widespread in proof assistants (Paulin-Mohring, 1993), but again out of the range of automated theorem provers. As far as we know the only study of a complete subclass in such fixed point logics is in the work of Baelde (2009), and iterated schemata definitely do not lie in this class nor can be reduced to it.

As we shall see in Section 5.2, completeness of bound-linear schemata (or more precisely regular schemata) lies in the detection of **cycles** during the proof search. This idea is not new, it is used, e.g., in tableaux methods dealing with modal logics in transitive frames (Goré, 1999), or $\mu$-calculi (Cleaveland, 1990; Bradfield & Stirling, 1992). However cycle detection in our work is quite different because we use it to prove *by induction*. Notice in particular that we cannot in general ensure termination (contrarily to the above methods). It is more relevant to consider our method as a particular instance of *cyclic proofs*, which are studied in proof theory precisely in the context of proofs by induction. In the works of Brotherston (2005), and Sprenger and Dam (2003), it is shown that cyclic proofs seem as powerful as systems dealing classically with induction. A particular advantage of cyclic proofs is that finding an invariant is not needed, making them particularly suited to automation. However, once again those studies are essentially theoretical and there are no completeness results for particular subclasses.

To summarize, known *decidable* logics (such as propositional $\mu$-calculus) or even *semi-decidable* ones such as first-order logic are not expressive enough to directly embed iterated schemata, whereas those that are sufficiently expressive (such as fixpoint or higher order logics) are not suitable for automation. Together with the potential applications mentioned in Section 2.4, this justifies to our opinion the interest of the considered language.

# 4. Reduction to Regular Schemata

In this section we reduce the satisfiability problem for bound-linear schemata (see Definition 2.9) to a much simpler class of schemata, called *regular*. This class is defined as follows:

**Definition 4.1**
A schema $\phi$ is:

- *flat* if for every iteration $\Pi_{i=a}^{b}\psi$ occurring in $\phi$, $\psi$ does not contain any iteration (i.e. iterations cannot be nested in $\phi$).

- *of bounded propagation* if every atom that occurs in an iteration $\Pi_{i=a}^{b}\psi$ in $\phi$ is of the form $p_{i+\gamma}$ for some $\gamma \in \mathbb{Z}$. Since the number of atoms is finite, there exist $\alpha, \beta \in \mathbb{Z}$ s.t. for every atom $p_{i+\gamma}$ occurring in an iteration we have $\gamma \in [\alpha, \beta]$. $\alpha, \beta$ are called the *propagation limits*.

- *aligned* on $[c, d]$ if all iterations occurring in $\phi$ are of the form $\Pi_{i=c}^{d}\psi$ (i.e. all iterations must have the same bounds).

- *regular* if it has a unique parameter $n$ and if it is flat, of bounded propagation and aligned on $[\alpha, n - \beta]$ for some $\alpha, \beta \in \mathbb{Z}$.

As an example, the schema *Adder* defined in the Introduction is regular, but the last example in Section 2.4 (three cells register with shift) is not. Obviously, every regular schema is also bound-linear (see Definition 2.9). We now define an algorithm that transforms every bound-linear schema into a sat-equivalent regular one. This result is somewhat surprising because the class of regular schemata seems much simpler than bound-linear schemata. In some sense, it points at regular schemata as a canonical decidable class of schemata.

## 4.1 Overview of the Transformation Algorithm

We first give an informal overview of the algorithm reducing every bound-linear schema into a regular one, together with examples illustrating each transformation steps. This very high level description is intended to help the reader to grasp the intuitive ideas behind the formal definitions and more technical explanations provided in the next section. The transformation is divided into several steps.

- The first step is the elimination of iterations occurring inside an iteration. Consider for instance the following schema $\phi : \bigvee_{i=1}^{n}(p_i \Rightarrow \bigwedge_{j=1}^{n} q_j)$. The reader can check that $\phi$ is bound-linear but non regular. It is easy to transform $\phi$ into a sat-equivalent regular schema: since $\bigwedge_{j=1}^{n} q_j$ does not depend on the counter $i$, one can simply replace this formula by a new propositional variable $r$ and add the equivalence $r \Leftrightarrow \bigwedge_{j=1}^{n} q_j$ outside the iteration. This yields the schema: $\bigvee_{i=1}^{n}(p_i \Rightarrow r) \wedge (r \Leftrightarrow \bigwedge_{j=1}^{n} q_j)$, which is clearly regular and sat-equivalent (but not equivalent) to $\phi$. This process can be generalized; however, replacing an iteration by a proposition is only possible if the iteration contains no variable that is bound in the original schema. Consider the schema: $\phi' : \bigvee_{i=1}^{n} \bigwedge_{j=1}^{n}(p_i \Rightarrow q_j)$. Here $\bigwedge_{j=1}^{n}(p_i \Rightarrow q_j)$ cannot be replaced by a variable $r$, since it depends on $i$. The solution is to get the variable $p_i$ containing $i$ out of the iteration $\bigwedge_{j=1}^{n}(p_i \Rightarrow q_j)$: as $p_i$ does not involve $j$, it is easily seen that we can turn $\bigwedge_{j=1}^{n}(p_i \Rightarrow q_j)$ into $p_i \Rightarrow \bigwedge_{j=1}^{n} q_j$. This transformation can be generalized by using case-splitting: indeed, it is well-known that every formula $\psi$ is equivalent to $(r \wedge \psi[\top/r]) \vee (\neg r \wedge \psi[\bot/r])$, for every propositional variable $r$. Applying this decomposition scheme to $\bigwedge_{j=1}^{n}(p_i \Rightarrow q_j)$ and $p_i$ we get: $\bigwedge_{j=1}^{n}(p_i \Rightarrow q_j) \equiv (p_i \wedge \bigwedge_{j=1}^{n}(\top \Rightarrow q_j)) \vee (\neg p_i \wedge \bigwedge_{j=1}^{n}(\bot \Rightarrow q_j))$, i.e. (by usual transformations): $\bigwedge_{j=1}^{n}(p_i \Rightarrow q_j) \equiv (p_i \wedge \bigwedge_{j=1}^{n} q_j) \vee \neg p_i$. Afterwards, the remaining iteration $\bigwedge_{j=1}^{n} q_j$ can be replaced by a new variable $r$.

The decomposition scheme just explained can be applied on every variable occurring in an iteration, but not containing the counter of this iteration. By definition of bound-linear schemata, the propositional symbols have only one index and this index contains at most one bound variable, thus this technique actually removes every atom containing a counter variable distinct from the one of the considered iteration. However, it does *not* remove the variables that occur in the bound of the iteration. Consider for instance the following formula: $\phi'' \stackrel{\text{def}}{=} \bigvee_{i=1}^{n} \bigwedge_{j=1}^{i} q_j$. Here $i$ occurs in the bound of the iteration and thus cannot be removed by the previous technique. The idea is then to encode the formula $\bigwedge_{j=1}^{i} q_j$ by a new variable $r_i$, that can be defined inductively as follows: $r_0$ is $\top$ and $r_{i+1}$ is $r_i \wedge q_{i+1}$. This is expressed by the schema: $r_0 \wedge \bigwedge_{i=0}^{n-1}(r_{i+1} \Leftrightarrow (r_i \wedge q_{i+1}))$.

Notice that $r_i$ needs only to be defined for $i = 0, \ldots, n$ because $i$ ranges over the interval $[1, n]$ in $\phi''$.

- In order to get a regular schema one has to guarantee that every iteration ranges over the *same* interval of the form $[\alpha, n - \beta]$ (where $\beta \in \mathbb{Z}$). This is actually simple to ensure by unfolding and shifting the iterations. For instance a schema $\bigvee_{i=1}^{2n} p_i$ can be transformed into $\bigvee_{i=1}^{n} p_i \vee \bigvee_{i=n+1}^{2n} p_i$ and then into $\bigvee_{i=1}^{n} p_i \vee \bigvee_{i=1}^{n} p_{i+n}$. Similarly $\bigvee_{i=2}^{n} p_i \vee \bigvee_{j=1}^{n-1} q_j$ can be reduced to $\bigvee_{i=2}^{n-1} p_i \vee p_n \vee q_1 \vee \bigvee_{j=2}^{n-1} q_j$ to get iterations defined on the same interval.

- A major difference between regular schemata and bound-linear ones is that, in a regular schema, the indexed variables occurring inside an iteration cannot contain parameters (e.g., an iteration such as $\bigvee_{i=1}^{n} p_{i+n}$ is forbidden). Therefore we have to replace every variable of the form $p_{\alpha.n+\beta\pm i}$ by a new variable $q_i$, depending only on $i$. The problem is that in order to preserve sat-equivalence, one also has to encode the *relation* between these variables. For instance, assume that $p_{n+i}$ is replaced by $q_i$ and that $p_{2n-j}$ is replaced by $r_j$. Then obviously, we must have $q_i \equiv r_j$ if $n+i = 2n-j$, i.e. $q_i \equiv r_{n-i}$. This step may be problematic because in general there are infinitely many such axioms. However, by defining the translation carefully, we will show that actually only finitely many equivalences are required. To this aim, we have to assume that the initial coefficient of the parameter is even in every index (see Definition 4.2), which is easy to ensure by case splitting. Then the maximal number of overlaps between the newly defined variable is actually bounded (this is shown by the crucial lemma 4.6).

For instance, a formula $\bigvee_{i=0}^{n}(\neg p_i \vee p_{2n-i})$ is replaced by $\bigvee_{i=0}^{n}(\neg p_i \vee q_i) \wedge (p_n \Leftrightarrow q_n)$. $q_i$ denotes the atom $p_{2n-i}$ and the equivalence encodes the fact that $q_n \equiv p_{2n-n} = p_n$.

Since $i$ ranges over the interval $[0..n]$ this is the only equation which is relevant w.r.t. $\phi$ (e.g. $p_0 \Leftrightarrow q_{2n}$ is useless).

The algorithm for transforming every bound-linear schema $\phi$ into a sat-equivalent regular schema $\psi$ is specified as a sequence of rewriting rules, operating on schemata and preserving sat-equivalence. The rules are depicted in Figure 1. *They must be applied in the order of their presentation.* As we shall see in Section 4.3, the rewrite system terminates (in exponential time). Moreover satisfiability is preserved and irreducible schemata are regular (see Section 4.4).

### 4.2 Formal Definition of the Algorithm

We now give a more detailed and precise description of the transformation algorithm (readers not interested in technical details can skip this section). We assume that the *initial* schema satisfies the following condition:

**Definition 4.2**
A bound-linear schema is *normalized* if the coefficient of the parameter $n$ is even in any expression occurring in the formula (either as the index of a symbol in $\mathcal{P}$ or as the bound of an iteration).

Considering exclusively normalized schemata is not restrictive because a schema $\phi$ not satisfying this property can be replaced by $\phi[2n/n] \vee \phi[2n+1/n]$ (e.g. $p_{3n}$ is turned into $p_{6n} \vee p_{6n+3}$). The obtained schema is obviously sat-equivalent to $\phi$ and normalized[4]. The use of normalized schemata will be explained later (see Remark 4.7).

**Remark 4.3**
The property of being normalized is only useful for the algorithm of Figure 1 to be well-defined. But the schema obtained after application of this algorithm is actually not normalized in general.

We now explain in more details the different steps of the transformation.

#### 4.2.1 Elimination of Nested Iterations

As explained in Section 4.1, the first step is to remove the iterations $\Pi_{i=a}^{b}\phi$ occurring inside another iteration $\Gamma_{j=c}^{d}\psi$. This is done by the rules $\tau_1, \tau_2, \tau_3, \tau_4$. $\tau_2$ moves $\Pi_{i=a}^{b}\phi$ out by introducing a new variable $p$ as explained before. This is possible only if $\phi$ does not contain any free variable except $i$ and the parameter $n$. Removing all other variables is precisely the role of $\tau_1$:

$\tau_1 \quad \Pi_{i=a}^{b}\phi \quad \rightarrow \quad (p_c \wedge \Pi_{i=a}^{b}\phi[\top/p_c]) \ \vee (\neg p_c \wedge \Pi_{i=a}^{b}\phi[\bot/p_c])$
    If the variables in $c$ are free in $\Pi_{i=a}^{b}\phi$, $p_c$ occurs in $\phi$
    and if for every iteration $\Gamma_{j=c}^{d}\phi'$ containing $\Pi_{i=a}^{b}\phi$, $p_c$ contains either $j$ or a
    variable bound in $\Gamma_{j=c}^{d}\phi'$.

---

4. But the two formulae are not equivalent in general. For instance, if $\phi = p_n$, then the interpretation defined by $\mathcal{I}(n) \overset{\text{def}}{=} 1$ and $\mathcal{I}(p_\kappa) \overset{\text{def}}{=} \mathbf{T}$ iff $\kappa = 1$ validates $p_n$ but obviously *not* $p_{2n} \vee p_{2n+1}$.

| | | | |
|---|---|---|---|
| $\tau_1$ | $\Pi_{i=a}^b \phi$ | $\rightarrow$ | $(p_c \wedge \Pi_{i=a}^b \phi[\top/p_c]) \vee (\neg p_c \wedge \Pi_{i=a}^b \phi[\bot/p_c])$ |

If the variables in $c$ are free in $\Pi_{i=a}^b \phi$, $p_c$ occurs in $\phi$
and if for every iteration $\Gamma_{j=c}^d \phi'$ containing $\Pi_{i=a}^b \phi$, $p_c$ contains either $j$ or a
variable bound in $\Gamma_{j=c}^d \phi'$.

| | | | |
|---|---|---|---|
| $\tau_2$ | $\psi$ | $\rightarrow$ | $(p \Leftrightarrow \Pi_{i=a}^b \phi) \wedge \psi[p/\Pi_{i=a}^b \phi]$ |

If $p$ is a fresh symbol, $\psi$ is the global schema, $\Pi_{i=a}^b \phi$ occurs in an iteration in $\psi$
and contains no free variable except $n$.

| | | | |
|---|---|---|---|
| $\tau_3$ | $\phi$ | $\rightarrow$ | $\bigwedge_{j=\min_\phi(j)}^{a-b-1} \neg p_j$ |
| | | | $\wedge \bigwedge_{j=a-b}^{\max_\phi(j)} (p_j \Leftrightarrow (p_{j-1} \vee \psi[b+j/i])) \wedge (\phi[p_j/\bigvee_{i=a}^{b+j} \psi])$ |

If $p$ is a fresh symbol, $\bigvee_{i=a}^{b+j} \psi$ occurs in an iteration of $\phi$, $j$ is bound in $\phi$,
$a$, $b$ and $\psi$ contain no free variable except $n$, $\phi$ is the global schema.

| | | | |
|---|---|---|---|
| $\tau_3'$ | $\phi$ | $\rightarrow$ | $\bigwedge_{j=\min_\phi(j)}^{a-b-1} \neg p_j$ |
| | | | $\wedge \bigwedge_{j=a-b}^{\max_\phi(j)} (p_j \Leftrightarrow (p_{j-1} \wedge \psi[b+j/i])) \wedge (\phi[p_j/\bigwedge_{i=a}^{b+j} \psi])$ |

If $p$ is a fresh symbol, $\bigwedge_{i=a}^{b+j} \psi$ occurs in an iteration of $\phi$, $j$ is bound in $\phi$,
$a$, $b$ and $\psi$ contain no free variable except $n$, $\phi$ is the global schema.

| | | | |
|---|---|---|---|
| $\tau_4$ | $\phi$ | $\rightarrow$ | $\bigwedge_{j=b-a}^{\max_\phi(j)} \neg p_j$ |
| | | | $\wedge \bigwedge_{j=\min_\phi(j)}^{b-a} (p_j \Leftrightarrow (p_{j+1} \vee \psi[b-j/i])) \wedge (\phi[p_j/\bigvee_{i=a}^{b-j} \psi])$ |

If $p$ is a fresh symbol, $\bigvee_{i=a}^{b-j} \psi$ occurs in an iteration of $\phi$, $j$ is bound in $\phi$,
$a$, $b$ and $\psi$ contain no free variable except $n$, $\phi$ is the global schema.

| | | | |
|---|---|---|---|
| $\tau_4'$ | $\phi$ | $\rightarrow$ | $\bigwedge_{j=b-a+1}^{\max_\phi(j)} p_j$ |
| | | | $\wedge \bigwedge_{j=\min_\phi(j)}^{b-a} (p_j \Leftrightarrow (p_{j+1} \wedge \psi[]/b-j]) \wedge (\phi[p_j/\bigwedge_{i=a}^{b-j} \psi])$ |

If $p$ is a fresh symbol, $\bigwedge_{i=a}^{b-j} \psi$ occurs in an iteration of $\phi$, $j$ is bound in $\phi$,
$a$, $b$ and $\psi$ contain no free variable except $n$, $\phi$ is the global schema.

| | | | |
|---|---|---|---|
| $\tau_5$ | $\Pi_{i=\alpha.n+\beta}^{\gamma.n-\delta} \phi$ | $\rightarrow$ | $\Pi_{i=\beta}^{(\gamma-\alpha).n-\delta} \phi[i+\alpha.n/i]$ |

If $\alpha \neq 0, \beta \in \mathbb{Z}$.

| | | | |
|---|---|---|---|
| $\tau_6$ | $\psi$ | $\rightarrow$ | $[\psi]_{n \mapsto 0} \vee \ldots \vee [\psi]_{n \mapsto \kappa} \vee (n > \kappa \wedge \psi[\diamond/\Pi_{i=\gamma}^{\alpha.n-\beta} \phi])$ |

If $\psi$ contains $\Pi_{i=\gamma}^{\alpha.n-\beta} \phi$, with $\alpha, \beta, \gamma \in \mathbb{Z}$, $\alpha < 0$ and $\Pi \in \{\bigwedge, \bigvee\}$,
where $\kappa = \lceil \frac{\gamma-\beta}{\alpha} \rceil$ and $\Pi$ is $\bigvee$ then $\diamond = \bot$ and if $\Pi = \bigwedge$ then $\diamond = \top$.

| | | | |
|---|---|---|---|
| $\tau_7$ | $\psi$ | $\rightarrow$ | $((\alpha-1).n - \beta \geq \gamma \wedge \psi[\psi'/\Pi_{i=\gamma}^{\alpha.n-\beta} \phi]) \vee ([\psi]_{n \mapsto 0} \vee \ldots \vee [\psi]_{n \mapsto \kappa})$ |

where $\psi$ contains an iteration $\Pi_{i=\gamma}^{\alpha.n-\beta} \phi$ with $\alpha > 1$,
$\psi'$ is $\Pi_{i=\gamma}^{(\alpha-1).n-\beta} \phi \star \Pi_{i=1}^n \phi[i+(\alpha-1).n-\beta/i]$, with $\Pi \in \{\bigwedge, \bigvee\}$,
where $\kappa = \lfloor \frac{\gamma-\beta}{\alpha-1} \rfloor$, $\Pi = \bigwedge$ then $\star = \wedge$, if $\Pi = \bigvee$ then $\star = \vee$.

| | | | |
|---|---|---|---|
| $\tau_8$ | $\Pi_{i=\gamma}^{n-\beta} \phi$ | $\rightarrow$ | $\Pi_{i=\beta}^{n-\gamma} \phi[n-i/i]$ |

If the indices of the variables in $\phi$ are of the form $(2\alpha+1).n+c$, where $c \in \mathcal{N}_i$.

| | | | |
|---|---|---|---|
| $\tau_9$ | $\phi$ | $\rightarrow$ | $\overleftrightarrow{\phi} \wedge \bigwedge_{\psi \in \Psi(\phi)} \psi$ |

If $\phi$ contains a variable $p$ not occurring in $V^- \cup V^+$,
and where $\Psi(\phi)$ is defined by Definitions 4.4, 4.5 and Lemma 4.6.

| | | | |
|---|---|---|---|
| $\tau_{10}$ | $\Pi_{i=\alpha}^{n-\beta} \phi$ | $\rightarrow$ | $(n < \alpha + \beta \wedge \diamond) \vee$ |
| | | | $(n \geq \alpha + \beta \wedge \Pi_{i=\alpha'}^{n-\beta-1+\alpha'-\alpha} \phi[i-\alpha'+\alpha/i] \star \phi[n-\beta/i])$ |

where $\alpha'$ is the maximal lower bound of an iteration occurring in the
whole formula and $\beta'$ is the minimal upper bound, $\alpha \neq \alpha'$ or $\beta \neq \beta'$,
and if $\Pi$ is $\bigvee$ then $\diamond = \bot, \star = \vee$ and if $\Pi = \bigwedge$ then $\diamond = \top, \star = \wedge$.

Figure 1: Transformation Into Regular Schemata

This rule aims at eliminating, in the body of an iteration $\Pi_{i=a}^b \phi$, every variable distinct from the iteration counter $i$ and from the (unique) parameter $n$. This is feasible because no index can contain two variables distinct from $n$ (by definition of bound-linear schemata). This implies that the indexed variables containing an arithmetic variable distinct from $i$ and $n$ cannot contain $i$ thus they can be taken out of the iteration $\Pi_{i=a}^b \phi$ by case splitting. Notice that the rule $\tau_1$ can increase exponentially the size of the formula.

Once $\phi$ contains no free variable except $n$ and $i$, $\Pi_{i=a}^b \phi$ may be taken out of the global iteration $\Gamma_{j=c}^d \psi$ by renaming. This is very easy if the bounds of the iteration only depend on $n$, because in this case $\Pi_{i=a}^b \phi$ contains no free variable except $n$, thus it may be replaced by a fresh variable $p$ and the equivalence $p \Leftrightarrow \Pi_{i=a}^b \phi$ may be added as an axiom. This is done by the rule $\tau_2$:

$\tau_2 \quad \psi \quad \rightarrow \quad (p \Leftrightarrow \Pi_{i=a}^b \phi) \wedge \psi[p / \Pi_{i=a}^b \phi]$
  If $p$ is a fresh symbol, $\psi$ is the global schema, $\Pi_{i=a}^b \phi$ occurs in an iteration in $\psi$ and contains no free variable except $n$.

Things get more complicated if the bounds of the iteration contain a bound variable $j$ (e.g., the schema $\bigvee_{j=1}^n (q_i \Rightarrow \bigvee_{i=1}^j r_i)$) because in this case the iteration cannot be taken out and $j$ cannot be eliminated by $\tau_1$. Notice that, in this case, the lower bound $a$ cannot contain $j$ and the coefficient of $j$ in the upper bound $b$ must be $\pm 1$. In this case, $\Pi_{i=a}^b \phi$ can be replaced by a new variable $p_j$ that can be defined inductively. For instance in the previous example, $\bigvee_{i=1}^j r_i$ is replaced by a variable $p_j$ defined as follows: $\neg p_0 \wedge \bigwedge_{i=1}^n [p_j \Leftrightarrow (r_j \vee p_{j-1})]$. The transformation is formally specified by the rules $\tau_3$ (if the coefficient of $j$ is 1) and $\tau_4$ (if the coefficient of $j$ is $-1$). Notice that if $\psi$ denotes the global schema, then $p_j$ must be defined for every $j \in [\min_\psi(j), \max_\psi(j)]$.

$\tau_3 \quad \phi \quad \rightarrow \quad \bigwedge_{j=\min_\phi(j)}^{a-b-1} \neg p_j$
  $\wedge \ \bigwedge_{j=a-b}^{\max_\phi(j)} (p_j \Leftrightarrow (p_{j-1} \vee \psi[b+j/i])) \wedge \ (\phi[p_j / \bigvee_{i=a}^{b+j} \psi])$
  If $p$ is a fresh symbol, $\bigvee_{i=a}^{b+j} \psi$ occurs in an iteration of $\phi$, $j$ is bound in $\phi$, $a$, $b$ and $\psi$ contain no free variable except $n$, $\phi$ is the global schema.

$\tau_4 \quad \phi \quad \rightarrow \quad \bigwedge_{j=b-a+1}^{\max_\phi(j)} \neg p_j$
  $\wedge \ \bigwedge_{j=\min_\phi(j)}^{b-a} (p_j \Leftrightarrow (p_{j+1} \vee \psi[b-j/i])) \wedge \ (\phi[p_j / \bigvee_{i=a}^{b-j} \psi])$
  If $p$ is a fresh symbol, $\bigvee_{i=a}^{b-j} \psi$ occurs in an iteration of $\phi$, $j$ is bound in $\phi$, $a$, $b$ and $\psi$ contain no free variable except $n$, $\phi$ is the global schema.

The rules $\tau_3'$ and $\tau_4'$ for $\bigwedge$ are defined in a similar way (see Figure 1).

### 4.2.2 TRANSFORMING EVERY ITERATION INTO ITERATIONS OVER INTERVALS OF THE FORM $[\alpha, n - \beta]$

The next step is to ensure that for every iteration $\Pi_{i=a}^b \phi$, $a$ is an integer $\alpha$ and that $b$ is of the form $n - \beta$, where $\beta$ is a constant (initially both $a$ and $b$ must be of the form $2.\delta.n + \gamma$ (since the initial schema is normalized and no iteration is contained inside another one so no bound variable occurs in the upper bound). The first point is easily performed by an appropriate translation of the iteration counter (rule $\tau_5$):

$$\tau_5 \quad \Pi_{i=\alpha.n+\beta}^{\gamma.n-\delta}\phi \quad \rightarrow \quad \Pi_{i=\beta}^{(\gamma-\alpha).n-\delta}\phi[i+\alpha.n/i]$$
$$\text{If } \alpha \neq 0, \beta \in \mathbb{Z}.$$

Then we ensure that the coefficient of $n$ in $b$ is positive. Fortunately, if this coefficient is negative then there is $\kappa \in \mathbb{N}$ s.t. for every interpretation $I$ s.t. $I(n) > \kappa$, the interval $[I(a), I(b)]$ is empty, in which case $\Pi_{i=a}^{b}\phi$ is either $\top$ or $\bot$ (depending on $\Pi$). Since the value of $n$ is positive, there exist finitely many values for $n$ s.t. the iteration is non empty. One can eliminate the iteration by considering these cases separately. This is done by the rule $\tau_6$:

$$\tau_6 \quad \psi \quad \rightarrow \quad [\psi]_{n\mapsto 0} \vee \ldots \vee [\psi]_{n\mapsto \kappa} \vee \ (n > \kappa \wedge \psi[\diamond/\Pi_{i=\gamma}^{\alpha.n-\beta}\phi])$$
$$\text{If } \psi \text{ contains } \Pi_{i=\gamma}^{\alpha.n-\beta}\phi, \text{ with } \alpha, \beta, \gamma \in \mathbb{Z}, \ \alpha < 0 \text{ and } \Pi \in \{\bigwedge, \bigvee\},$$
$$\text{where } \kappa = \lceil \tfrac{\gamma-\beta}{\alpha} \rceil \text{ and } \Pi \text{ is } \bigvee \text{ then } \diamond = \bot \text{ and if } \Pi = \bigwedge \text{ then } \diamond = \top.$$

Finally, we obtain the desired result by (recursively) decomposing an iteration interval of the form $[\gamma, \alpha.n + \beta]$ (where $\alpha > 1$) into two smaller intervals $[\gamma, (\alpha-1).n + \beta]$ and $[(\alpha-1).n + \beta + 1, \alpha.n + \beta]$. Obviously, this is possible only if $(\alpha-1).n + \beta \geq \gamma$, thus the case where $(\alpha-1).n + \beta < \gamma$ must be considered separately. This is easy to achieve, since in this case there are only finitely many possible values of $n$, namely $0, 1, \ldots, \lfloor \tfrac{\gamma-\beta}{\alpha-1} \rfloor$.

$$\tau_7 \quad \psi \quad \rightarrow \quad ((\alpha-1).n - \beta \geq \gamma \wedge \psi[\psi'/\Pi_{i=\gamma}^{\alpha.n-\beta}\phi]) \vee ([\psi]_{n\mapsto 0} \vee \ldots \vee [\psi]_{n\mapsto \kappa})$$
$$\text{where } \psi \text{ contains an iteration } \Pi_{i=\gamma}^{\alpha.n-\beta}\phi \text{ with } \alpha > 1,$$
$$\psi' \text{ is } \Pi_{i=\gamma}^{(\alpha-1).n-\beta}\phi \star \Pi_{i=1}^{n}\phi[i + (\alpha-1).n - \beta/i], \text{ with } \Pi \in \{\bigwedge, \bigvee\},$$
$$\text{where } \kappa = \lfloor \tfrac{\gamma-\beta}{\alpha-1} \rfloor, \ \Pi = \bigwedge \text{ then } \star = \wedge, \text{ if } \Pi = \bigvee \text{ then } \star = \vee.$$

### 4.2.3 Removing the Parameter from the Indices in the Iterations

The next phase consists in removing the indexed variables of the form $p_{\alpha.n+\epsilon.i+\beta}$ where $\beta \in \mathbb{Z}$ and either $\alpha \neq 0$ or $\epsilon = -1$ (to get variables indexed by expressions of the form $i + \beta$ only). We first ensure that $\alpha$ is even. Although initially the coefficient of every occurrence of $n$ is even, this property does not hold anymore at this point because of the rule $\tau_7$. Suppose a variable $p_{(2\gamma+1).n+c}$, where $c$ does not contain $n$, occurs in an iteration $\Pi_{i=a}^{b}\phi$. Then (since the schema is normalized) this variable must have been introduced by the rule $\tau_7$ and $i$ has been shifted by $(\alpha - \kappa).n$ for some $\kappa$ (by definition of $\tau_7$). This shift is applied to every index containing $i$ (by definition of $\tau_7$), i.e. to every index of a variable occurring in $\Pi_{i=a}^{b}\phi$ (otherwise the iteration would be reducible by $\tau_1$). As a consequence *every index* in this iteration has an odd coefficient for $n$. Hence if we add $n$ to each index we retrieve even coefficients in all the iteration. Fortunately by commutativity of $\vee$ and $\wedge$, any iteration $\Pi_{i=a}^{b}\phi$ is equivalent to $\Pi_{i=0}^{b-a}\phi[b - i/i]$. In our case $b$ is of the form $n - \beta$ for some $\beta \in \mathbb{Z}$ so applying this transformation precisely adds $n$ to each index (and substracts a $\beta$). For instance, the iteration $\bigvee_{i=1}^{n}(p_{n+i} \vee p_{n-i})$ can be replaced by $\bigvee_{i=0}^{n-1}(p_{2n-i} \vee p_i)$. This idea is formalized by the rule $\tau_8$:

$$\tau_8 \quad \Pi_{i=\gamma}^{n-\beta}\phi \quad \rightarrow \quad \Pi_{i=\beta}^{n-\gamma}\phi[n-i/i]$$
$$\text{If the indices of the variables in } \phi \text{ are of the form } (2\alpha+1).n + c, \text{ where } c \in \mathcal{N}_i.$$

Once the coefficient of $n$ in every indexed variable is even, we introduce, for every variable $p$ and for every integer $\kappa$, two new (fresh) variables $\overline{p}^{\kappa^+}$ and $\overline{p}^{\kappa^-}$ s.t. $\overline{p}_a^{\kappa^+}$ and $\overline{p}_a^{\kappa^-}$ denote respectively $p_{2.\kappa.n+a}$ and $p_{2.\kappa.n-a}$ where $a \in \mathcal{N}_i \cup \mathbb{Z}$ i.e. $a$ is of the form $\beta.i + \gamma$ where $\beta \in \{0, 1\}, \gamma \in \mathbb{Z}$ (rule $\tau_9$). Then the index of $\overline{p}_a^{\kappa^+}$ does not contain $n$ anymore. Furthermore, the index of $\overline{p}_a^{\kappa^-}$ now contains $+i$ instead of $-i$. Thus this transformation indeed achieves our goal however it does *not* preserve sat-equivalence because two variables $p_{2\alpha.n+a}$ and $p_{2\beta.n-b}$ (respectively $p_{2\alpha.n+a}$ and $p_{2\beta.n+b}$, $p_{2\alpha.n-a}$ and $p_{2\beta.n-b}$) s.t. $2\alpha.n + a = 2\beta.n - b$ (respectively $2\alpha.n + a = 2\beta.n + b$ and $2\alpha.n - a = 2\beta.n - b$) may be replaced by distinct variables $\overline{p}_a^{\alpha^+}$ and $\overline{p}_b^{\beta^-}$ (respectively $\overline{p}_a^{\alpha^+}$ and $\overline{p}_b^{\beta^+}$, $\overline{p}_a^{\alpha^-}$ and $\overline{p}_b^{\beta^-}$). Notice that it is important to distinguish the sign $+$ or $-$ in front of $a$ and $b$, as both are not integers but expressions of $\mathcal{N}_i \cup \mathbb{Z}$. In order to preserve sat-equivalence one would have to explicitly add the following axioms to the schema:

$$2\alpha.n + \gamma = 2\beta.n - \delta \Rightarrow (\overline{p}_\gamma^{\alpha^+} \Leftrightarrow \overline{p}_\delta^{\beta^-})$$

and

$$2\alpha.n + \gamma = 2\beta.n + \delta \Rightarrow (\overline{p}_\gamma^{\alpha^+} \Leftrightarrow \overline{p}_\delta^{\beta^+})$$

and

$$2\alpha.n - \gamma = 2\beta.n - \delta \Rightarrow (\overline{p}_\gamma^{\alpha^-} \Leftrightarrow \overline{p}_\delta^{\beta^-})$$

for every tuple $(\alpha, \beta, \gamma, \delta) \in \mathbb{Z}^4$.

This transformation is problematic, because there exist infinitely many such formulae. Fortunately, we do not have to add all these equivalences, but only those concerning propositional variables that occur in a propositional realization of the schema. As we shall see, this set (denoted by $\Psi(\phi)$) is finite, because each expression $\gamma, \delta$ ranges over a set of the form $[-\iota, \iota] \cup [n - \iota, n + \iota]$, where $\iota \in \mathbb{N}$.

More formally, let $V^+$ and $V^-$ be two disjoint subsets of $\mathcal{P}$, distinct from the symbols already occurring in the considered formula. We assume that every pair $(p, \alpha)$ where $p$ is a variable occurring in the formula and $\alpha$ an integer is mapped to two variables $\overline{p}^{\alpha^+} \in V^+$ and $\overline{p}^{\alpha^-} \in V^-$. $\overline{p}_i^{\alpha^+}$ and $\overline{p}_i^{\alpha^-}$ will denote the atoms $p_{2\alpha.n+i}$ and $p_{2\alpha.n-i}$ respectively. We denote by $\overline{\phi}$ the schema obtained from $\phi$ by replacing every variable of the form $p_{2\alpha.n+a}$ (where $a \in \mathcal{N}_i \cup \mathbb{N}$ for some bound variable $i$) by $\overline{p}_a^{\alpha^+}$ and each variable of the form $p_{2\alpha.n-a}$ by $\overline{p}_a^{\alpha^-}$ (in both cases we may have $\alpha = 0$, moreover, if $a = 0$ then the replacement may be done arbitrarily by $\overline{p}_0^{\alpha^+}$ or $\overline{p}_0^{\alpha^-}$). Notice that all atoms in $\overline{\phi}$ are of the form $\overline{p}_a^{\alpha^+}$ or $\overline{p}_a^{\alpha^-}$, where $a \in \mathcal{N}_i \cup \mathbb{N}$ for some bound variable $i$. $\tau_9$ is defined as follows:

$\tau_9 \quad \phi \quad \rightarrow \quad \overline{\phi} \wedge \bigwedge_{\psi \in \Psi(\phi)} \psi$
 If $\phi$ contains a variable $p$ not occurring in $V^- \cup V^+$,
 and where $\Psi(\phi)$ is defined by Definitions 4.4, 4.5 and Lemma 4.6.

### 4.2.4 ALIGNING ITERATIONS

Finally, it remains to ensure that all the iterations have the same bounds. At this point every iteration is of the form $\Pi_{i=\alpha}^{n-\beta}\phi$ where $\alpha, \beta \in \mathbb{Z}$. Let $\alpha', \beta'$ be the greatest integers $\alpha, \beta$. If we have $\alpha \neq \alpha'$ or $\beta \neq \beta'$, then we unfold the iteration once, yielding $\Pi_{i=\alpha}^{n-\beta-1}\phi \star \phi[n-\beta/i]$. By translation of the iteration counter, $\Pi_{i=\alpha}^{n-\beta-1}$ is equivalent to $\Pi_{i=\alpha'}^{n-\beta-1+\alpha'-\alpha}\phi[i-\alpha'+\alpha/i]$. The lower bound of the obtained iteration is now identical to $\alpha'$ and its length has been decreased. This is repeated until we obtain an iteration on the interval $[\alpha', \beta']$. The rule $\tau_{10}$ formalizes this transformation:

$$\tau_{10} \quad \Pi_{i=\alpha}^{n-\beta} \phi \quad \rightarrow \quad (n < \alpha + \beta \wedge \diamond) \vee$$
$$(n \geq \alpha + \beta \wedge \Pi_{i=\alpha'}^{n-\beta-1+\alpha'-\alpha} \phi[i - \alpha' + \alpha/i] \star \phi[n - \beta/i])$$

where $\alpha'$ is the maximal lower bound of an iteration occurring in the whole formula and $\beta'$ is the minimal upper bound, $\alpha \neq \alpha'$ or $\beta \neq \beta'$, and if $\Pi$ is $\bigvee$ then $\diamond = \bot, \star = \vee$ and if $\Pi = \bigwedge$ then $\diamond = \top, \star = \wedge$.

### 4.2.5 Definition of $\Psi(\phi)$

The most difficult part of the transformation is the removal of the variable $n$ in the index performed by the rule $\tau_9$, and more precisely the definition of $\Psi(\phi)$. We now establish the results ensuring the feasability of this transformation.

**Definition 4.4**
We denote by $\Psi$ the set of schemata of the form:

$$2\alpha.n + a = 2\beta.n - b \Rightarrow (\overline{p}_a^{\alpha^+} \Leftrightarrow \overline{p}_b^{\beta^-})$$

or

$$2\alpha.n + a = 2\beta.n + b \Rightarrow (\overline{p}_a^{\alpha^+} \Leftrightarrow \overline{p}_b^{\beta^+})$$

or

$$2\alpha.n - a = 2\beta.n - b \Rightarrow (\overline{p}_a^{\alpha^-} \Leftrightarrow \overline{p}_b^{\beta^-})$$

where $\alpha, \beta \in \mathbb{Z}$, $a, b \in \mathcal{N}_n \cup \mathbb{Z}$.

The set $\Psi$ is infinite. Thus we add a further restriction:

**Definition 4.5**
Let $\phi$ be a schema containing a unique parameter $n$. A schema $\psi \Rightarrow (p \Leftrightarrow q)$ occurring in $\Psi$ is said to be *relevant* w.r.t. $\phi$ iff the following conditions hold:

- $p$ and $q$ are not syntactically identical.

- There exists a natural number $\kappa$ s.t. $\psi[\kappa/n]$ is true and $\overline{\phi}[\kappa/n]$ contains both $p[\kappa/n]$ and $q[\kappa/n]$

Notice that $p$ and $q$ do not necessarily occur in $\overline{\phi}$ itself. For instance, take $\phi = \bigwedge_{i=1}^{n}(p_{2n-i} \vee \neg p_i)$. So $\overline{\phi} = \bigwedge_{i=1}^{n}(\overline{p}_i^{2^-} \vee \neg \overline{p}_i^{0^+})$. Then $2n - n = 4 \Rightarrow (\overline{p}_n^{2^-} \Leftrightarrow \overline{p}_4^{0^+})$ is easily seen to be relevant, however both $\overline{p}_n^{2^-}$ and $p_4^{0^+}$ do not occur in $\overline{\phi}$.

The next lemma provides a very simple necessary condition on relevant equivalences in $\Psi$. It also shows that for every schema $\phi$ the number of relevant equivalences in $\Psi$ is finite (up to equivalence).

**Lemma 4.6**
Let $\phi$ be a schema containing a unique parameter $n$. Assume that the coefficient of $n$ is even in every index in $\phi$ and that every iteration in $\phi$ is of the form $\Pi_{i=\epsilon}^{n+\zeta}\psi$, where $\epsilon, \zeta \in \mathbb{Z}$ ($\epsilon, \zeta$ may depend on the iteration). Let $\iota$ be the greatest natural number occurring in $\phi$ (possibly as a coefficient of $n$ or in an expression of the form $-\iota$).

For every relevant formula of the form $2\alpha.n + a = 2\beta.n - b \Rightarrow (\overline{p}_a^{\alpha^+} \Leftrightarrow \overline{p}_b^{\beta^-})$, $2\alpha.n + a = 2\beta.n + b \Rightarrow (\overline{p}_a^{\alpha^+} \Leftrightarrow \overline{p}_b^{\beta^+})$ or $2\alpha.n - a = 2\beta.n - b \Rightarrow (\overline{p}_a^{\alpha^-} \Leftrightarrow \overline{p}_b^{\beta^-})$ in $\Psi$, we have, for every $\kappa \in \mathbb{N}$: $\alpha, \beta \in [-\iota, \iota]$ and $a[\kappa/n], b[\kappa/n] \in [-2\iota, 6\iota] \cup [\kappa - 2\iota, \kappa + 2\iota]$.

PROOF

Let $\sigma$ stand for the substitution $[\kappa/n]$. By definition of a relevant formula, there must exist $\kappa \in \mathbb{N}$ such that $\overline{p}_a^{\alpha^+} \sigma$ and $\overline{p}_b^{\beta^-} \sigma$ (respectively $\overline{p}_b^{\beta^-} \sigma$) occur in $[\overline{\phi}]_\sigma$ (but notice that $\overline{p}_a^{\alpha^+}$, $\overline{p}_b^{\beta^-}$ and $\overline{p}_b^{\beta^+}$ do not necessarily occur in $\overline{\phi}$). Furthermore we must have $2\alpha.\kappa + a\sigma = 2\beta.\kappa - b\sigma$ (resp. $2\alpha.\kappa + a\sigma = 2\beta.\kappa + b\sigma$).

Since the coefficient of $n$ is even in every index in $\phi$ and since $a, b \in \mathcal{N}_n \cup \mathbb{Z}$, $2\alpha, 2\beta$ necessarily occur in $\phi$. Thus $\alpha, \beta \in [-\iota/2, \iota/2] \subseteq [-\iota, \iota]$.

Moreover, by Lemma 2.8, there exist two atoms $\overline{p}_{a'}^{\alpha^+}$ and $\overline{p}_{b'}^{\beta^-}$ (respectively $\overline{p}_{b'}^{\beta^+}$) which occur in $\overline{\phi}$ and two $\phi$-expansions $\sigma'$ and $\sigma''$ of $\sigma$ for $\overline{p}_{a'}^{\alpha^+}$ and $\overline{p}_{b'}^{\beta^-}$ (respectively $\overline{p}_{b'}^{\beta^+}$) s.t. we have $a\sigma = a'\sigma'$ and $b\sigma = b'\sigma''$. By definition, $a', b'$ come from the replacement of some proposition $p_{2\alpha.n + a'}$ (resp. $p_{2\beta.n - b'}$ and $p_{2\beta.n + b'}$) by $\overline{p}_{a'}^{\kappa^+}$ (resp. $\overline{p}_{b'}^{\kappa^-}$ and $\overline{p}_{b'}^{\kappa^+}$). Thus $a'$ and $b'$ do not contain $n$. Thus $a'$ and $b'$ are either in $\mathbb{Z}$ (and in this case we must have $a\sigma, b\sigma \in [-\iota, \iota] \subseteq [-2\iota, \kappa + 2\iota]$) or respectively of the form $i + \gamma$ and $i + \delta$ where $i$ is a bound variable and $\gamma, \delta \in \mathbb{Z}$. Then since $\sigma', \sigma''$ are $\phi$-expansions of $\sigma$ we have $i\sigma', i\sigma'' \in [\min_\phi(i)\sigma, \max_\phi(i)\sigma]$. We have $\min_\phi(i) = \epsilon \geq -\iota$ and $\max_\phi(i) = n + \zeta \leq n + \iota$. Thus $a\sigma, b\sigma \in [-2\iota, \kappa + 2\iota]$.

Assume that we have $2\alpha.\kappa + a\sigma = 2\beta.\kappa - b\sigma$. Then $a\sigma + b\sigma = 2.(\beta - \alpha).\kappa$.

- If $\beta \leq \alpha$ then $a\sigma + b\sigma \leq 0$. Since $a\sigma, b\sigma \geq -2\iota$, we deduce $a\sigma, b\sigma \leq 2\iota$. Thus $a\sigma, b\sigma \in [-2\iota, 6\iota]$.

- If $\beta > \alpha$ then $a\sigma + b\sigma \geq 2\kappa$. Since $a\sigma \leq \kappa + 2\iota$ and $b\sigma \leq \kappa + 2\iota$ we must have $a\sigma \geq \kappa - 2\iota$ and $b\sigma \geq \kappa - 2\iota$. Thus $a\sigma, b\sigma \in [\kappa - 2\iota, \kappa + 2\iota]$.

Now, assume that $2\alpha.\kappa + a\sigma = 2\beta.\kappa + b\sigma$. Then $a\sigma - b\sigma = 2.(\beta - \alpha).\kappa$.

- If $\alpha = \beta$ then we must have $a\sigma = b\sigma$. This contradicts the first condition in Definition 4.5 (the indexed variables cannot be syntactically identical).

- If $\alpha < \beta$ then $a\sigma - b\sigma > 2\kappa$. This is possible only if $a\sigma > 2\kappa + b\sigma > 2\kappa - 2\iota$, hence $\kappa + 2\iota > 2\kappa - 2\iota$, i.e. $4\iota > \kappa$. Then since we must have $a\sigma, b\sigma \in [-2\iota, \kappa + 2\iota]$ we deduce $a\sigma, b\sigma \in [-2\iota, 6\iota]$.

- The proof is symmetric if $\alpha > \beta$.

Finally if $2\alpha.\kappa - a\sigma = 2\beta.\kappa - b\sigma$ then $a\sigma - b\sigma = 2.(\alpha - \beta).\kappa$ and the proof follows exactly as in the previous case. $\qquad\square$

Lemma 4.6 implies that the set of relevant formulae is finite (up to equivalence). Indeed, it suffices to instantiate $\alpha, \beta$ by every integer in $[-\iota, \iota]$ and $a, b$ either by elements of $[-\iota, 6\iota]$ or by expressions of the form $n + \gamma$, where $\gamma$ is an integer in $[-2\iota, 2\iota]$. Thus we denote by $\Psi(\phi)$ a finite subset of $\Psi$ containing all relevant formulae (up to equivalence). Such a set can be easily computed by applying Lemma 4.6, but using refined criteria is possible, thus we opt for a generic definition.

**Remark 4.7**

The fact that the coefficient of $n$ is even (see Definition 4.2 of normalized schemata) is *essential* at this point. If arbitrary coefficients are allowed for $n$, then the coefficients $2\alpha$ and $2\beta$ must be replaced by $\alpha$ and $\beta$ respectively. Then in the second item in the proof of Lemma 4.6 we obtain $\alpha\sigma + b\sigma \geq \kappa$ (instead of $a\sigma + b\sigma \geq 2\kappa$). Thus we get eventually $\alpha\sigma, b\sigma > -2\iota$ (instead of $a\sigma \geq \kappa - 2\iota$). This means that $a\sigma, b\sigma$ range over the interval $[-2\iota, \kappa + 2\iota]$ instead of $[-\iota, 6\iota] \cup [\kappa - 2\iota, \kappa + 2\iota]$. But this interval is unbounded, thus $\Psi(\phi)$ is infinite (even up to equivalence).

For instance, suppose that we allow any coefficient for $n$ (i.e. odd or even) and that $p_{\alpha.n+\beta}$ is turned into $\overline{p}_\beta^{1^+}$. Consider then $\phi = \bigvee_{i=1}^n (p_i \vee p_{n-i})$. We get: $\overline{\phi} = \bigvee_{i=0}^n (\overline{p}_i^{0^+} \vee \overline{p}_i^{1^-})$. But the equivalence $\overline{p}_i^{0^+} \Leftrightarrow \overline{p}_{n-i}^{1^-}$ is obviously needed for every $i \in [1, n]$, which cannot be expressed by a finite number of equivalences.

On the other hand, if we only allow normalized schemata, i.e. even coefficients for $n$, then we first have to turn $\phi$ into $\psi = \bigvee_{i=1}^{2n} (p_i \vee p_{2n-i})$ hence (by $\tau_7$) $\psi = \bigvee_{i=1}^n (p_i \vee p_{2n-i}) \vee \bigvee_{i=1}^n (p_{n+i} \vee p_{n-i})$, and (by $\tau_8$) $\psi = \bigvee_{i=1}^n (p_i \vee p_{2n-i}) \vee \bigvee_{i=0}^{n-1} (p_{2n-i} \vee p_i)$. Then $\overline{\psi} = \bigvee_{i=1}^n (\overline{p}_i^{0^+} \vee \overline{p}_i^{1^-}) \vee \bigvee_{i=0}^{n-1} (\overline{p}_i^{1^-} \vee \overline{p}_i^{0^+})$. No equivalence is needed in this simple case.

**Lemma 4.8**

Let $\phi$ be a schema containing a unique parameter $n$ s.t. every iteration in $\phi$ is of the form $\Pi_{i=\alpha}^{n+\beta}\psi$, where $\alpha, \beta \in \mathbb{Z}$. $\phi$ is satisfiable iff $\overline{\phi} \cup \Psi(\phi)$ is satisfiable.

Proof

Let $\mathcal{I}$ be an interpretation satisfying $\phi$. Let $\kappa = \mathcal{I}(n)$. We define an interpretation $\mathcal{J}$ as follows: $\mathcal{J}(n) \overset{\text{def}}{=} \kappa$ and for every pair of integers $(\alpha, \beta)$: $\mathcal{J}(\overline{p}_\beta^{\alpha^+}) \overset{\text{def}}{=} \top$ iff $\mathcal{I}(p_{2\alpha.\kappa+\beta}) = \top$ and $\mathcal{J}(\overline{p}_\beta^{\alpha^-}) \overset{\text{def}}{=} \top$ iff $\mathcal{I}(p_{2\alpha.\kappa-\beta}) = \top$. By definition for all $\psi \in \Psi$, $\mathcal{J} \models \psi$. $\overline{\phi}$ is obtained from $\phi$ by replacing every atom of the form $p_{2\alpha.n+a}$ (respectively $p_{2\alpha.n-a}$) where $a \in \mathcal{N}_i \cup \mathbb{Z}$ (for some bound variable $i$) by $\overline{p}_a^{\alpha^+}$ (respectively $\overline{p}_a^{\alpha^-}$). By definition of $\mathcal{J}$, $\mathcal{J} \models \overline{p}_a^{\alpha^+}$ iff $\mathcal{I} \models p_{2\alpha.n+\beta}$ and $\mathcal{J} \models \overline{p}_\beta^{\alpha^-}$ iff $\mathcal{I} \models p_{2\alpha.n-\beta}$. Since $\mathcal{I} \models \phi$ it is clear that we have $\mathcal{J} \models \overline{\phi}$. Thus $\mathcal{J} \models \overline{\phi} \cup \Psi(\phi)$.

Conversely, let $\mathcal{I} \models \overline{\phi} \cup \Psi(\phi)$. Let $\kappa = \mathcal{I}(n)$. Let $\mathcal{J}$ be the interpretation defined as follows. $\mathcal{J}(n) \overset{\text{def}}{=} \kappa$, $\mathcal{J}(p_{2\alpha.\kappa+\beta}) = \mathcal{I}(\overline{p}_\beta^{\alpha^+})$ if $\overline{p}_\beta^{\alpha^+}$ occurs in $[\overline{\phi}]_{\mathcal{I}}$, and $\mathcal{J}(p_{2\alpha.\kappa-\beta}) = \mathcal{I}(\overline{p}_\beta^{\alpha^-})$ if $\overline{p}_\beta^{\alpha^-}$ occurs in $[\overline{\phi}]_{\mathcal{I}}$. It is easy to check that $\mathcal{J}$ is well-defined since $\mathcal{I} \models \Psi(\phi)$ and $\Psi(\phi)$ contains all the necessary equivalences. By definition, $\overline{p}_a^{\alpha^+}$ (respectively $\overline{p}_a^{\alpha^-}$) occurs in $\overline{\phi}$ iff $p_{2\alpha.n+a}$ (respectively $p_{2\alpha.n+a}$) occurs in $\phi$. Thus, since $\mathcal{I} \models \overline{\phi}$ we have $\mathcal{J} \models \phi$. □

### 4.3 Termination and Complexity

In this section, we investigate the complexity of the transformation algorithm and show that it is exponential. For every schema $\phi$, we denote by $|\phi|$ the size of $\phi$, i.e. the number of symbols occurring in $\phi$. $\tau$ denotes the system of rewrite rules of Figure 1.

**Theorem 4.9**
Let $\phi$ be a normalized bound-linear schema. A normal form $\psi$ of $\phi$ w.r.t. $\tau$ can be computed in $O(2^{|\phi|})$ rewriting steps. Moreover, $|\psi| = O(2^{|\phi|})$.

Proof
We first notice that the rules are always applied sequentially: it is easy to check that a rule cannot introduce a formula on which a previous rule applies. Thus we consider each rule in sequence.

First, we consider the rule $\tau_1$. We call $\tau_1$-*atoms* the atoms $p_c$ on which the rule possibly applies, i.e. the atom occurring in an iteration $\Pi_{i=a}^b \psi$ but not containing the iteration counter $i$. This rule removes an atom occurring in an iteration but not containing the iteration counter. Due to the control (i.e. the application conditions of the rules), no atom satisfying this condition can be introduced into the formula (indeed, if the atom $p_c$ occurs in an iteration then, because of the second application condition of the rule, it must contain the corresponding iteration counter of this iteration). Therefore, the number of applications of this rule on an iteration is bounded by the number of $\tau_1$-atoms it contains. Since the rule duplicates the considered iteration the total number of applications of the rule is bounded by $2^m$, where $m$ is the total number of $\tau_1$-atoms. Obviously $m \leq |\phi|$.

This is not sufficient to prove the second result, i.e. that the size of the formula is $O(2^{|\phi|})$, since each application of the rule can double the size of the formula (which would yield a double exponential blow-up since there are $2^m$ rule applications). Consider the set of leaf positions of the considered formula. For each position $p$ in this set, we denote by $|p|$ the length of $p$ and by $r_p$ the number of possible applications of the rule $\tau_1$ along $p$. Each application of the rule $\tau_1$ removes some positions $p$ from this set (those corresponding to the leaves of the subformula on which the rule is applied) and replaces them by new positions $p'_1, \ldots, p'_\kappa$. Both the *number* of these positions and their *length* possibly increase. However, we remark that the rule can only increase the length of these positions by 2 (by adding a disjunction of conjunctions), i.e. we have $\forall \iota \in [1, \kappa], |p'_\iota| \leq |p| + 2$. Furthermore, the number $r_p$ necessarily decreases: $\forall \iota \in [1, \kappa], r_{p'_\iota} < r_p$. Consequently, the value $|p| + 2 \times r_p$ cannot increase (i.e. we have $\forall \iota \in [1, \kappa], |p'_\iota| + 2 \times r_{p'_\iota} \leq |p| + 2 \times r_p$), which implies that the length of the final positions (when $r_{p'_\iota} = 0$) are lower than $|p_{\max}| + 2 \times r_{\max}$, where $r_{\max}$ denotes the maximal number of possible applications of the rule $\tau_1$ along some position in the initial formula (i.e. the max of the $r_p$ in the initial formula) and $p_{\max}$ is the position of maximal length in the initial formula. Both $|p_{\max}|$ and $r_{\max}$ are $O(|\phi|)$, thus the depth of the final formula is $O(|\phi|)$, which implies that its size is $O(2^{|\phi|})$.

We now consider the other rules. First we analyze the transformation due to a single application of each of those rules (then we will analyze the number of such applications). Since the proofs for the different cases are actual very similar, we do not consider each rule separately, but we rather factorize some part of the analysis.

- Each application of the rule $\tau_2$ only increases the size of the formula by a constant number of symbols, since a fixed number of new connectives is added and no part of the formula is duplicated.

- The application of the rules $\tau_3$, $\tau'_3$, $\tau_4$, $\tau'_4$, $\tau_5$, $\tau_8$ and $\tau_{10}$ adds a constant number of new connectives in the formula and replaces each occurrence of the counter $i$ in the formula $\phi$ by an expression of the form $b + j$, $b - j$, $i + \alpha . n$ or $n - i$. The size of these expressions is bounded by the size of the original formula, thus the size of the formula increases quadratically (since the number of occurrences of $i$ is also bound by the size of the formula).

- Now consider the rules $\tau_6$ and $\tau_7$. These rules introduce a constant number of new connectives and occurrences of atoms and duplicate $\kappa$ times a subformula $\psi$. The value of $\kappa$ is bounded by the natural number $\gamma$ that occurs in $\phi$, thus the size of the formula increases polynomially (since natural numbers are encoded as unary terms $s(\dots(s(0))\dots)$ is our setting, hence $\kappa$ is bounded by the size of the formula – notice that this would not be the case if the numbers were encoded as sequences of digits[5]).

Thus we only have to show that the number of applications of each of these rules is polynomially bounded by the size of the initial formula. Once again, we distinguish several cases:

- The rules $\tau_2, \tau_3, \tau_3', \tau_4, \tau_4'$ only apply on iterations occurring inside another iteration. During the application of the rule, this iteration is replaced by an atom, hence removed from the outermost iteration. The rule introduces new iterations, however they only occur at the root level, outside the scope of any iteration. Thus the total number of possible applications of these rules is bounded by the number of iterations initially occurring inside another iteration, hence by $|\phi|$.

- The rules $\tau_5$, $\tau_6$ and $\tau_8$ apply at most once on each iteration: $\tau_5$ applies on an iteration in which the lower bound contain $n$ and gets rid of any occurrence of $n$ in the lower bound. $\tau_6$ applies on iterations in which the upper bound contains $-n$ and replaces these iterations by purely propositional formulae. $\tau_8$ applies if the coefficient of $n$ in every index is odd. Since the rule adds $n$ to each index, after the application of the rule, the coefficient of $n$ must be even and the rule cannot apply again on the same iteration.

- The rule $\tau_7$ decreases the value of the coefficient $\alpha$ of $n$ in the upper bound by 1. Thus the number of applications of the rule $\tau_7$ on each iteration is lower than the initial value of $\alpha$ (which is bound by the size of the formulae since integers are encoded as terms). Similarly, since $\tau_{10}$ unfolds an iteration until an iteration of length $n - \beta' - \alpha'$ is obtained, the number of applications of the rule $\tau_{10}$ on each iteration is bound by the value of $-\beta + \alpha + \beta' - \alpha'$.

- Finally, the rule $\tau_9$ applies only once on the whole schema. The rule adds a conjunction of equivalence to the schema, but by Lemma 4.6, the size of the conjunction is polynomially bounded by the greatest natural number $\iota$ occurring in the schema, hence by the size of the formula. $\qquad\square$

For every schema $\phi$, we denote by $\phi \downarrow_\tau$ a normal form of $\phi$ w.r.t. the rules in $\tau$.

---

5. Actually the translation is doubly exponential in this case.

## 4.4 Soundness and Completeness

We prove that the rules in $\tau$ preserve sat-equivalence and that every irreducible formula is regular. We need the two propositions below:

**Lemma 4.10**

Let $\psi$, $\phi$ and $\phi'$ be schemata. Let $\mathcal{I}$ be an interpretation such that for every ground substitution $\sigma$ of the parameters of $\psi$ and for every $\psi$-expansion $\theta$ of $\sigma$ for $\phi, \phi'$, we have: $[\![\phi\theta]\!]_{\mathcal{I}} = [\![\phi'\theta]\!]_{\mathcal{I}}$. Then $[\![\psi]\!]_{\mathcal{I}} = [\![\psi[\phi'/\phi]]\!]_{\mathcal{I}}$.

PROOF

The proof is by induction on $\psi$. If $\psi$ does not contain $\phi$ the proof is trivial. If $\psi = \phi$ then $\psi[\phi'/\phi] = \phi'$. By definition $[\![\phi]\!]_{\mathcal{I}} = [\![\phi\sigma_{\mathcal{I}}]\!]_{\mathcal{I}}$ and $[\![\phi']\!]_{\mathcal{I}} = [\![\phi'\sigma_{\mathcal{I}}]\!]_{\mathcal{I}}$. But $\sigma_{\mathcal{I}}$ is a ground substitution of the parameters of $\psi = \phi$ and thus is of course a $\psi$-expansion of itself for $\phi$ and $\phi'$. Thus $[\![\phi\sigma_{\mathcal{I}}]\!]_{\mathcal{I}} = [\![\phi'\sigma_{\mathcal{I}}]\!]_{\mathcal{I}}$ hence $[\![\psi]\!]_{\mathcal{I}} = [\![\psi']\!]_{\mathcal{I}}$.

Assume that $\psi = \neg\psi'$. We have $[\![\psi[\phi'/\phi]]\!]_{\mathcal{I}} = \neg[\![\psi'[\phi'/\phi]]\!]_{\mathcal{I}} = \neg[\![\psi']\!]_{\mathcal{I}}$ (by induction). Thus $[\![\psi[\phi'/\phi]]\!]_{\mathcal{I}} = [\![\psi]\!]_{\mathcal{I}}$. The proof is similar if $\psi = (\psi_1 \vee \psi_2)$ or if $\psi = (\psi_1 \wedge \psi_2)$.

Now assume that $\psi = \bigwedge_{i=a}^{b}\psi'$. $\mathcal{I} \models \psi$ iff for every integer $\kappa \in [\![a]\!]_{\mathcal{I}}, [\![b]\!]_{\mathcal{I}}]$ we have $\mathcal{I}[\kappa/i] \models \psi'$. Let $\sigma'$ be the substitution such that $\sigma'(i) = \kappa$ and $\sigma'(x) \stackrel{\text{def}}{=} \sigma(x)$ if $x \neq i$. Let $\theta$ be a $\psi$-expansion of $\sigma'$ for $\psi'$. By definition $\kappa \in [[\![\min_{\psi}(i)]\!]_{\mathcal{I}}, [\![\max_{\psi}(i)]\!]_{\mathcal{I}}]$, thus $\theta$ is also a $\psi$-expansion of $\sigma$. Therefore we have $[\![\phi\theta]\!]_{\mathcal{I}} = [\![\phi'\theta]\!]_{\mathcal{I}}$, hence $[\![\phi\theta]\!]_{\mathcal{I}[\kappa/i]} = [\![\phi'\theta]\!]_{\mathcal{I}[\kappa/i]}$ (since $\phi\theta$ and $\phi'\theta$ do not contain $i$). Consequently, by the induction hypothesis, we have $[\![\psi']\!]_{\mathcal{I}[\kappa/i]} = [\![\psi'[\phi'/\phi]]\!]_{\mathcal{I}[\kappa/i]}$. Hence $\mathcal{I} \models \psi$ iff for every integer $\kappa \in [\![a]\!]_{\mathcal{I}}, [\![b]\!]_{\mathcal{I}}]$ we have $\mathcal{I}[\kappa/i] \models \psi'[\phi'/\phi]$ i.e. iff $\mathcal{I} \models \psi[\phi'/\phi]$. The proof is similar if $\psi = \bigvee_{i=a}^{b}\psi'$. □

**Lemma 4.11**

For every schema $\phi$ and for every indexed proposition $p$ that does not contain any variable bound in $\phi$:

$$\phi \equiv (p \wedge \phi[\top/p]) \vee (\neg p \wedge \phi[\bot/p])$$

PROOF

We have $p \vee \neg p \equiv \top$ hence by distributivity $\phi \equiv (p \wedge \phi) \vee (\neg p \wedge \phi)$. We now show that for every interpretation $\mathcal{I}$, $[\![p \wedge \phi]\!]_{\mathcal{I}} = [\![p \wedge \phi[\top/p]]\!]_{\mathcal{I}}$. If $[\![p]\!]_{\mathcal{I}} = \mathbf{F}$ then both $p \wedge \phi$ and $p \wedge \phi[\top/p]$ are false in $\mathcal{I}$. Otherwise, by Lemma 4.10, we have $[\![\phi]\!]_{\mathcal{I}} = [\![\phi[\top/p]]\!]_{\mathcal{I}}$. Similarly, we have $[\![\neg p \wedge \phi]\!]_{\mathcal{I}} = [\![\neg p \wedge \phi[\bot/p]]\!]_{\mathcal{I}}$. Hence $\phi \equiv (p \wedge \phi[\top/p]) \vee (\neg p \wedge \phi[\bot/p])$. □

**Theorem 4.12**

Let $\phi$ be a normalized bound-linear schema. $\phi$ is satisfiable iff $\phi \downarrow_{\tau}$ is satisfiable.

PROOF

The proof is by inspection of the different rules (see the definition of the rules for the notations):

- $\tau_1$. The proof is a direct application of Lemma 4.11.

- $\tau_2$. For every model $\mathcal{I}$ of $\psi$, one can construct an interpretation $\mathcal{J}$ of $(p \Leftrightarrow \Pi_{i=a}^{b}\phi) \wedge \psi[p/\Pi_{i=a}^{b}\phi]$ by interpreting $p$ as $[\![\Pi_{i=a}^{b}\phi]\!]_{\mathcal{I}}$. By definition we have $\mathcal{J} \models (p \Leftrightarrow \Pi_{i=a}^{b}\phi)$.

Since $\mathcal{I} \models \psi$ we have $\mathcal{J} \models \psi$. By Lemma 4.10 we deduce that $\mathcal{I} \models \psi[p/\Pi_{i=a}^b \phi]$. Hence $\mathcal{J} \models (p \Leftrightarrow \Pi_{i=a}^b \phi) \wedge \psi[p/\Pi_{i=a}^b \phi]$.

Conversely, if $\mathcal{I}$ is a model of $(p \Leftrightarrow \Pi_{i=a}^b \phi) \wedge \psi[p/\Pi_{i=a}^b \phi]$, then due to the first conjunct $\Pi_{i=a}^b \phi$ and $p$ have the same truth value in $\mathcal{I}$ hence since $\mathcal{I} \models \psi[p/\Pi_{i=a}^b \phi]$, we deduce $\mathcal{I} \models \psi$, by Lemma 4.10.

- $\tau_3$. Assume that $\mathcal{I} \models \phi$. Let $\mathcal{J}$ be the extension of $\mathcal{I}$ obtained by interpreting $p_\kappa$ as $[\![\bigvee_{i=a}^{b+\kappa} \psi]\!]_\mathcal{I}$. By Lemma 4.10 we have $\mathcal{J} \models (\phi[p_j/\bigvee_{i=a}^{b+j} \psi])$. Furthermore by definition of the semantics, we have $[\![\bigvee_{i=a}^{b+\kappa} \psi]\!]_\mathcal{I} = \mathbf{F}$ if $[\![b+\kappa-a]\!] < 0$ hence $\mathcal{J} \models \neg p_\kappa$ if $\kappa < a-b$. Thus $\mathcal{J} \models \neg p_{a-b-1} \wedge \bigwedge_{j=\min_\phi(j)}^{a-b-1}(p_j \Leftrightarrow p_{a-b-1})$. Furthermore, for every $\iota \geq [\![a-b]\!]_\mathcal{I}$, we have $[\![\bigvee_{i=a}^{b+\iota} \psi]\!]_\mathcal{I} = \mathbf{T}$ iff either $[\![\bigvee_{i=a}^{b+\iota-1} \psi]\!]_\mathcal{I} = \mathbf{T}$ or $[\![\psi[b+\iota/j]]\!]_\mathcal{I} = \mathbf{T}$. Hence $[\![p_\iota]\!]_\mathcal{I} = \mathbf{T}$ iff either $[\![p_{\iota-1}]\!]_\mathcal{I} = \mathbf{T}$ or $[\![\psi[b+\iota/j]]\!]_\mathcal{I} = \mathbf{T}$. Therefore $\mathcal{I} \models \bigwedge_{j=a-b}^{\max_\phi(j)}(p_j \Leftrightarrow (p_{j-1} \vee \psi))$.

  Conversely, let $\mathcal{I}$ be a model of $\neg p_{a-b-1} \wedge \bigwedge_{j=\min_\phi(j)}^{a-b-1}(p_j \Leftrightarrow p_{a-b-1}) \wedge \bigwedge_{j=a-b}^{\max_\phi(j)}(p_j \Leftrightarrow (p_{j-1} \vee \psi[b+j/i])) \wedge (\phi[p_j/\bigvee_{i=a}^{b+j} \psi])$. We show by induction on $\iota$ that $\mathcal{I} \models (p_\iota \Leftrightarrow \bigvee_{i=a}^{b+\iota} \psi)$ for every $\iota \in [[\min_\phi(j)]\!]_\mathcal{I}, [\![\max_\phi(j)]\!]_\mathcal{I}]$:

  - If $\iota < [\![a-b]\!]_\mathcal{I}$ then by definition $[\![\bigvee_{i=a}^{b+\iota} \psi]\!]_\mathcal{I} = \mathbf{F}$. Moreover by the first two conjuncts in the previous formula we must have $[\![p_\iota]\!]_\mathcal{I} = \mathbf{F}$.
  - Otherwise, we have $[\![\bigvee_{i=a}^{b+\iota} \psi]\!]_\mathcal{I} = [\![\bigvee_{i=a}^{b+\iota-1} \psi \vee \psi[b+\iota/i]]\!]_\mathcal{I}$. Hence by the induction hypothesis: $[\![\bigvee_{i=a}^{b+\iota} \psi]\!]_\mathcal{I} = [\![p_{\iota-1}]\!]_\mathcal{I} \vee \psi[b+\iota/i]$, and by the third conjunct in the formula above, we get: $[\![\bigvee_{i=a}^{b+\iota} \psi]\!]_\mathcal{I} = [\![p_\iota]\!]_\mathcal{I}$.

  Then by Lemma 4.10 we deduce that $\mathcal{I} \models \psi$. The proofs for the rules $\tau_3'$, $\tau_4$ and $\tau_4'$ are similar.

- $\tau_5$. Assume that $\Pi = \bigvee$ (the case $\Pi = \bigwedge$ is similar). By definition $\mathcal{I} \models \bigvee_{i=\alpha.n+\beta}^{\gamma.n+\epsilon} \phi$ iff there exists $\kappa \in [[\![\alpha.n+\beta]\!]_\mathcal{I}, [\![\gamma.n+\epsilon]\!]_\mathcal{I}]$ such that $\mathcal{I} \models \phi[\kappa/i]$, i.e. iff there exists $\kappa \in [[\![\beta]\!]_\mathcal{I}, [\![(\gamma-\alpha).n+\epsilon]\!]_\mathcal{I}]$ such that $\mathcal{I} \models \phi[\kappa+[\![\alpha.n]\!]_\mathcal{I}/i]$, i.e. iff $\mathcal{I} \models \bigvee_{i=\beta}^{(\gamma-\alpha).n+\epsilon} \phi[i+\alpha.n/i]$.

- $\tau_6$. We assume that $\Pi = \vee$ and $\diamond = \bot$ (the case $\Pi = \wedge, \diamond = \top$ is similar). Since we assume that $\mathcal{I}(n) \geq 0$ for every parameter $n$, we have $\mathcal{I} \models (n = 0 \vee \ldots \vee n = \kappa \vee n > \kappa)$ hence $\psi$ is equivalent to: $(n = 0 \vee \ldots \vee n = \kappa \vee n > \kappa) \wedge \psi$. By distributivity we get $\psi \equiv (n = 0 \wedge \psi) \vee \ldots \vee (n = \kappa \wedge \psi) \vee (n > \kappa \wedge \psi)$. But $\bigvee_{i=\gamma}^{\alpha.n+\beta} \phi$ is empty (thus equivalent to $\bot$) if $\mathcal{I}(n) > \kappa \geq \frac{\gamma-\beta}{\alpha}$, hence, by Lemma 4.10, we have $\psi \equiv (n = 0 \wedge \psi) \vee \ldots \vee (n = \kappa \wedge \psi) \vee (n > \kappa \wedge \psi[\bot/\bigvee_{i=\gamma}^{\alpha.n+\beta} \phi])$. For every $\iota \in [0, \kappa]$, we have $n = \iota \wedge \psi \models [\psi]_{n \mapsto \iota}$, hence $\psi \models [\psi]_{n \mapsto 0} \vee \ldots \vee [\psi]_{n \mapsto \kappa} \vee (n > \kappa \wedge \psi[\bot/\bigvee_{i=\gamma}^{\alpha.n+\beta} \phi])$.

  Conversely, if $\mathcal{I} \models [\psi]_{n \mapsto \iota}$ holds, then $\mathcal{I}$ can be straightforwardly extended into a model of $n = \iota \wedge \psi$ by interpreting $n$ as $\iota$. Thus for any model of $[\psi]_{n \mapsto 0} \vee \ldots \vee [\psi]_{n \mapsto \kappa} \vee (n > \kappa \wedge \psi[\bot/\bigvee_{i=\gamma}^{\alpha.n+\beta} \phi])$ there exists a model of $\psi$, and $\tau_6$ preserves satisfiability.

- $\tau_7$. Again, we assume that $\Pi = \bigvee$ and $\diamond = \bot$. We have $((\alpha-1).n+\beta < \gamma \vee (\alpha-1).n+\beta \geq \gamma) \equiv \top$ hence $\psi \equiv ((\alpha-1).n+\beta < \gamma \vee (\alpha-1).n+\beta \geq \gamma) \wedge \psi \equiv ((\alpha-1).n+\beta \geq \gamma \wedge \psi) \vee ((\alpha-1).n+\beta < \gamma \wedge \psi)$. Since the parameters are interpreted as natural numbers, we have $\mathcal{I} \models (\alpha - 1).n + \beta < \gamma$ iff $\mathcal{I}(n) \in [0, \lceil \frac{\gamma - \beta}{\alpha - 1} \rceil]$. Then by definition $[\![\psi]\!]_{\mathcal{I}} = [\![[\psi]_{n \mapsto \mathcal{I}(n)}]\!]_{\mathcal{I}}$. If $\mathcal{I} \models (\alpha - 1).n + \beta \geq \gamma$ then, by unfolding, $[\![\bigvee_{i=\gamma}^{\alpha.n+\beta} \phi]\!]_{\mathcal{I}} = [\![\bigvee_{i=\gamma}^{(\alpha-1).n+\beta} \phi \vee \bigvee_{i=(\alpha-1).n+\beta+1}^{\alpha.n+\beta} \phi]\!]_{\mathcal{I}} = [\![\bigvee_{i=\gamma}^{(\alpha-1).n+\beta} \phi \vee \bigvee_{i=1}^{n} \phi[i + (\alpha - 1).n + \beta/i]]\!]_{\mathcal{I}}$. Hence $\tau_7$ preserves satisfiability.

- $\tau_8$: the proof is similar to the one of $\tau_6$.

- The soundness of the rule $\tau_9$ is a direct consequence of Lemma 4.8.

- $\tau_{10}$. We assume that $\Pi = \bigvee$ and $\diamond = \bot$. We have $\bigvee_{i=\alpha}^{n-\beta} \phi \equiv (n < \alpha + \beta \wedge \bigvee_{i=\alpha}^{n-\beta} \phi) \vee (n \geq \alpha + \beta \wedge \bigvee_{i=\alpha}^{n-\beta} \phi)$. For every interpretation $\mathcal{I}$, if $\mathcal{I}(n) < \alpha + \beta$ then $[\![\bigvee_{i=\alpha}^{n-\beta} \phi]\!]_{\mathcal{I}} = \mathbf{F}$ thus $n < \alpha + \beta \wedge \bigvee_{i=\alpha}^{n-\beta} \phi \equiv (n < \alpha + \beta \wedge \diamond)$. If $\mathcal{I}(n) \geq \alpha + \beta$, then $[\![\bigvee_{i=\alpha}^{n-\beta} \phi]\!]_{\mathcal{I}} \equiv [\![\phi[\alpha/i] \vee \bigvee_{i=\alpha+1}^{n-\beta} \phi]\!]_{\mathcal{I}}$. Furthermore by translation of the iteration counter we have $\bigvee_{i=\alpha+1}^{n-\beta} \phi \equiv \bigvee_{i=\alpha+1-\beta'+\beta}^{n-\beta'} \phi[i + \beta' - \beta/i]$. Hence $\tau_{10}$ preserves equivalence. $\qquad\square$

**Theorem 4.13**

Let $\phi$ be a normalized bound-linear schema. $\phi \downarrow_\tau$ is regular.

PROOF

Firstly, we remark that the application of the rules in $\tau$ on a bound-linear schema generates a schema that is still bound-linear. Notice however that the obtained schema is not normalized in general.

Let $\phi$ be a bound-linear formula, irreducible by $\tau$. Assume that $\phi$ has been obtained from a normalized schema by application of the rules in $\tau$. We need to prove that $\phi$ is regular.

We first prove that $\phi$ contains no nested iteration. Let $\psi = \Pi_{i=a}^{b} \chi$ be an iteration occurring in $\phi$. Assume that $\chi$ contains an iteration $\Gamma_{j=c}^{d} \gamma$. W.l.o.g. we assume that $\gamma$ contains no iteration (otherwise we could simply take $\psi = \chi$). By irreducibility w.r.t. the rule $\tau_1$, all the indices in $\gamma$ must contain $j$. By definition of the class of bound-linear schemata, this implies that these indices cannot contain $i$. If $j$ occurs in $d$ then one of the rule $\tau_3, \tau_3', \tau_4$ or $\tau_4'$ applies. Consequently the only free variable in $\Gamma_{j=c}^{d} \gamma$ is $n$. Thus the rule $\tau_2$ applies which is impossible by irreducibility.

Then we remark that for all iterations $\Pi_{i=a}^{b} \psi$ in $\phi$, $a \in \mathbb{Z}$ and $b$ is of the form $n + \alpha$ where $\alpha \in \mathbb{Z}$. Indeed, if $a$ contains $n$ then the rule $\tau_5$ applies and if the coefficient of $n$ in $b$ is different from 1 then the rule $\tau_6$ or $\tau_7$ applies.

The rule $\tau_8$ eliminates all indexed propositions in which the coefficient of $n$ is odd (since the initial schema is normalized, these indexed variables have been necessarily introduced by the rule $\tau_7$, thus they must occur in an iteration and all the indices in the iteration must have an odd coefficient in front of $n$).

$\tau_9$ eliminates all the variables of the form $p_{2\alpha.n \pm a}$, where $\alpha \in \mathbb{Z}$ and $a \in \mathcal{N}_i \cup \mathbb{N}$, for some bound variable $i$, and replaces them by variables indexed only by $a$.

Finally $\tau_{10}$ ensures that all the iterations have the same bounds. $\qquad\square$

## 5. STAB: A Decision Procedure for Regular Schemata

Now that we have shown how to transform a bound linear schema into a regular one, we show that the satisfiability problem is decidable for regular schemata. This is done by providing a set of block tableaux rules (Smullyan, 1968) that are complete w.r.t. satisfiability. Those rules are concise and natural, and, compared to the naive procedure described in the proof of Proposition 2.7, they are much more efficient and terminate more often (see the end of Section 5.1). The procedure is called STAB (standing for **s**chemata **tab**leaux). Notice that it applies on any schema (not only on regular ones). We assume (w.l.o.g) that schemata are in negative normal form.

### 5.1 Inference Rules

**Definition 5.1 (Tableau)**
A *tableau* is a tree $\mathcal{T}$ s.t. each node $N$ occurring in $\mathcal{T}$ is labeled by a set of schemata written $\Phi_{\mathcal{T}}(N)$.

As usual a tableau is generated from another tableau by applying some extension rules. Let $r = \dfrac{P}{C_1 \mid \ldots \mid C_\kappa}$ be a rule where $P$ denotes a set of schemata (the *premises*), and $C_1, \ldots, C_\kappa$ denote the conclusions. Let $N$ be a leaf of a tree $\mathcal{T}$. If a subset $\mathcal{S}$ of $\Phi_{\mathcal{T}}(N)$ matches $P$ then we can *extend* the tableau by adding $\kappa$ children to $N$, each of them labeled with $C_\iota \sigma \cup (\Phi_{\mathcal{T}}(N) \setminus \mathcal{S})$ where $\iota = 1, \ldots, \kappa$ and $\sigma$ is the matching substitution. A leaf $N$ is *closed* iff the set of arithmetic formulae (i.e. schemata containing only atoms of the form $\ldots < \ldots$ and no iteration) in $\Phi_{\mathcal{T}}(N)$ is unsatisfiable. This can be detected using decision procedures for arithmetic without multiplication (Cooper, 1972).

**Definition 5.2 (Extension rules)**
The *extension rules* of STAB are defined as follows.

- The usual rules of propositional tableaux:

$$(\wedge)\colon \quad \frac{\phi \wedge \psi}{\phi \quad \psi} \qquad (\vee)\colon \quad \frac{\phi \vee \psi}{\phi \mid \psi}$$

- Rules proper to schemata ("iteration rules")[6]:

$$(\textit{Iterated } \wedge)\colon \quad \frac{\bigwedge_{i=a}^{b} \phi}{\bigwedge_{i=a}^{b-1} \phi \wedge \phi[b/i] \ \Big| \ b < a} \qquad (\textit{Iterated } \vee)\colon \quad \frac{\bigvee_{i=a}^{b} \phi}{\bigvee_{i=a}^{b-1} \phi \vee \phi[b/i]}$$

---

6. The right branch in the conclusion of the Iterated $\wedge$ rule is required, e.g., to detect that $\bigwedge_{i=1}^{n} \bot$ is satisfiable with $n = 0$.

- The closure rule adds the constraints needed for the branch *not to be closed*. The rule is applied only if $a \neq b$ does not already occur in the branch.

$$\textit{(Closure):} \qquad \frac{p_a \quad \neg p_b}{p_a, \neg p_b, a \neq b}$$

STAB without the loop detection rule described in the next section is already better than the straightforward procedure introduced in the proof of Proposition 2.7. First, it terminates in some cases where the schema is unsatisfiable (whereas the naive procedure never terminates in such a case, unless the schema is just an unsatisfiable propositional formula). This is trivially the case for any schema $\bigwedge_{i=1}^{n} \phi$ with $n \geq 1$, where $\phi$ is propositionally unsatisfiable. Second, it can find a model much faster than the naive procedure. Consider, e.g., $(\bigwedge_{i=n}^{10000} p) \wedge (\neg p \vee \phi)$ where $\phi$ is an unsatisfiable formula. In this case STAB immediately finds a model where $n > 10000$ and $p$ is interpreted as $\mathbf{F}$.

**Remark 5.3**
Using a tableaux-based system for deciding regular schemata may seem surprising, since DPLL procedures (Davis, Logemann, & Loveland, 1962) are usually more efficient in propositional logic. However, extending such procedures to schemata is not straightforward. The main problem is that evaluating an atom in a schema is not immediate, since this atom may well appear in some realization of the schema without appearing in the schema itself. Thus, in contrast to the propositional case, it is not sufficient to replace syntactically the atom by its truth value. For instance, the atom $p_2$ (implicitly) appears in the schema $\bigvee_{i=1}^{n} p_i$ if $n > 1$. Thus evaluating $p_2$ to, say, $\mathbf{F}$ would yield two distinct branches: $(\bigvee_{i=1}^{n} p_i) \wedge n \leq 1$ and $(p_1 \vee \bigvee_{i=3}^{n} p_i) \wedge n > 1$. Thus one would have to define rules operating at deep positions in the schema in order to unfold the iterations and instantiate the counter variables when needed. In contrast, the tableaux method operates only on formulae occurring at root level and compares literals only after they have been instantiated (using unfolding). This makes the procedure much easier to define and reason with (in particular the termination behavior is easier to control). Actually a DPLL procedure for schemata is presented in our previous work (Aravantinos, Caferra, & Peltier, 2009a, 2010), but it is much more complicated than the calculus presented here.

Of course, one could combine the iteration rules of the tableaux procedure with a SAT-solver used as a "black box" that could be in charge of the purely propositional part. However this is also not straightforward, mainly due to the fact that a *partial* evaluation is needed to propagate the values of the propositional variables into the iterations.

## 5.2 Discarding Infinite Derivations: the Looping Rule

STAB does not terminate in general. The reason is that an iteration is, in general, infinitely unfolded by the iteration rules. Assume for instance that $\phi$ is a propositional unsatisfiable formula. Then starting from $\bigvee_{i=1}^{n} \phi$ one could derive an infinite sequence of formulae of the form $\bigvee_{i=1}^{n-1} \phi, \ldots, \bigvee_{i=1}^{n-\kappa} \phi$, for every $\kappa \in \mathbb{N}$. We now introduce a loop detection rule that aims at improving the termination behavior of STAB. Detecting looping is the most natural way to avoid this divergence: if, while extending the tableau, we find a schema that has already been seen, possibly up to a shift of arithmetic variables, then there is no need to consider it again and we can stop the procedure. Such loopings can also be interpreted as well-foundedness arguments in an inductive proof.

**Definition 5.4 (Looping)**
A *shift* is a substitution mapping every variable $n$ to an expression of the form $n - \iota$, where $\iota \in \mathbb{N}$ s.t. there is at least one variable $n$ s.t. $n\sigma < n$ (which is not always the case since we may have $\iota = 0$).

Если $\mathcal{I}, \mathcal{J}$ are two interpretations, we write $\mathcal{I} < \mathcal{J}$ iff there exists a shift $\sigma$ s.t. $\mathcal{J} = \mathcal{I}\sigma$.

Let $\phi, \psi$ be two schemata (or sets of schemata). We write $\phi \models_s \psi$ iff for every model $\mathcal{I}$ of $\phi$, there exists $\mathcal{J} < \mathcal{I}$ s.t. $\mathcal{J} \models \psi$.

Let $N, N'$ be two nodes of a tableau $\mathcal{T}$. Then $N'$ *loops* on $N$ iff $\Phi_{\mathcal{T}}(N') \models_s \Phi_{\mathcal{T}}(N)$.

In existing work on cyclic proofs, $N'$ is sometimes called a *bud node* and $N$ is the *companion node of $N'$* (Brotherson, 2005). When a leaf loops, it is treated as a closed leaf (though it is not necessarily unsatisfiable). To distinguish this particular case of closed leaf from the usual one, we say that it is *blocked* (blocked leaves are closed). Notice that $N$ and $N'$ may be on different branches, thus looping may occur more often, allowing more simplifications.

**Example 5.5**
Let $\Phi = \{\bigvee_{i=1}^n p_i\}$ and $\Psi = \{\bigvee_{i=2}^n q_i\}$. Intuitively, $\Phi$ and $\Psi$ have the same "structure": STAB will behave similarly on both formulae. The relation $\models_s$ is supposed to formalize this notion. We show on this example that it is the case, as expected, i.e. that we have $\Psi \models_s \Phi$. Indeed, consider a model $\mathcal{I}$ of $\Psi$. We construct an interpretation $\mathcal{J}$ as follows: $\mathcal{J}(n) \stackrel{\text{def}}{=} \mathcal{I}(n) - 1$ and for every $\kappa \in [1, \mathcal{J}(n)]$, $\mathcal{J}(p_\kappa) \stackrel{\text{def}}{=} \mathcal{I}(q_{\kappa+1})$. Since $\mathcal{I} \models \Psi$ there exists $\kappa \in [2, \mathcal{I}(n)]$ such that $\mathcal{I}(q_\kappa) = \mathbf{T}$. Thus there exists $\kappa \in [1, \mathcal{I}(n)-1]$ such that $\mathcal{I}(q_{\kappa+1}) = \mathbf{T}$, i.e. there exists $\kappa \in [1, \mathcal{J}(n)]$ such that $\mathcal{J}(p_\kappa) = \mathbf{T}$. Therefore $\mathcal{J} \models \Phi$.

**Proposition 5.6**
Let $\phi$ be a schema. If $\phi$ is satisfiable then $\phi$ has a model $\mathcal{I}$ that is minimal w.r.t. $<$ (i.e. for every interpretation $\mathcal{J}$, if $\mathcal{J} < \mathcal{I}$ then $\mathcal{J} \not\models \phi$).

Proof
Let $V$ be the set of parameters of $\phi$. Notice that $V$ is finite. For every interpretation $\mathcal{I}$ we denote by $\mathcal{I}(V)$ the integer: $\mathcal{I}(V) \stackrel{\text{def}}{=} \Sigma_{n \in V} \mathcal{I}(n)$. Since we assumed that $\mathcal{I}(n) \in \mathbb{N}$ for every variable $n$, we deduce that $\mathcal{I}(V) \geq 0$.

Let $\mathcal{I}$ be a model of $\phi$ such that $\mathcal{I}(V)$ is minimal. Since the truth value of $\phi$ does not depend on the values of the variables that are not in $V$, we may assume that $\forall n \notin V, \mathcal{I}(n) = 0$. Let $\mathcal{J}$ be a model of $\phi$ such that $\mathcal{J} < \mathcal{I}$. By definition there exists a shift $\sigma$ such that $\mathcal{J} = \mathcal{I}\sigma$. For every arithmetic variable $n$, we have $n\sigma = n - \iota_n$, where $\iota_n \in \mathbb{N}$; furthermore, there exists at least one variable $m$ such that $\iota_m > 0$. Thus $\mathcal{J}(n) = \mathcal{I}(\sigma(n)) \leq \mathcal{I}(n)$ and $\mathcal{J}(m) < \mathcal{I}(m)$. Consequently we must have $\mathcal{J}(V) \leq \mathcal{I}(V)$, thus $\mathcal{J}(V) = \mathcal{I}(V)$ (since $\mathcal{I}(V)$ is minimal). By definition, this entails that $\iota_n = 0$ for every $n \in V$. Thus $m \notin V$, but in this case $\mathcal{I}(m) = 0$ hence $\mathcal{J}(m) < 0$ which is impossible (since we assume that parameters are interpreted by natural numbers). $\qquad\square$

To apply the looping rule in practice one has to find a shift and check that the implication holds. Unfortunately, the relation $\models_s$ is obviously undecidable (for instance if $\psi = \bot$, then it can be easily checked that $\phi \models_s \psi$ iff $\phi$ is unsatisfiable, and as we shall see in Section 6 the satisfiability problem is undecidable for propositional schemata). Thus, in the following, we shall use a much stronger criterion that is sufficient for our purpose. An obvious solution would be to use set inclusion: indeed, $\phi \models_s \psi$ if there exists a shift $\sigma$ s.t. $\phi \supseteq \psi\sigma$. However, this criterion is too strong, as the following example shows.

**Example 5.7**
The schema $\phi = p_n \wedge (p_n \Rightarrow q_n) \wedge \neg q_0 \wedge \bigwedge_{i=1}^{n}(q_i \Rightarrow q_{i-1})$ is obviously unsatisfiable. The reader can easily check that STAB generates an infinite sequence of sets of schemata of the form:

$$\{p_n, q_n, \neg q_0, q_{n-1}, \ldots, q_{n-\kappa}, \bigwedge_{i=1}^{n-\kappa}(q_i \Rightarrow q_{i-1})\}, \text{ where } \kappa \in \mathbb{N}$$

None of these sets contains a previous one up to a shift on $n$ because of the indexed proposition $p_n$ that must occur in every set.

Thus we introduce a refinement of set inclusion based on the purity principle. The pure literal rule is standard in propositional theorem proving. It consists in evaluating a literal $L$ to $\top$ in a formula $\phi$ (in NNF) if the complement of $L$ does not occur in $\phi$. Such a literal is called *pure*. It is well-known that this operation preserves satisfiability and may allow many simplifications.

We show how to extend the pure literal rule to schemata. The conditions on $L$ have to be strengthened in order to take iterations into account. For instance, if $L = p_n$ and $\phi$ contains $\bigvee_{i=1}^{2n} \neg p_i$ then $L$ is not pure in $\phi$, since $\neg p_i$ is the complement of $L$ for $i = n$ (and since $1 \leq n \leq 2n$). On the other hand $p_{2n+1}$ may be pure in $\phi$ (since $2n+1 \notin [1, 2n]$).

For every set of schemata $\Phi$ we denote by $\Phi_{\mathcal{N}}$ the conjunction of purely arithmetic formulae in $\Phi$: $\Phi_{\mathcal{N}} \stackrel{\text{def}}{=} \bigwedge_{\phi \in \Phi, \phi \text{ is arithmetic}} \phi$.[7]

**Definition 5.8 (Pure literal)**
A literal $p_a$ (respectively $\neg p_a$) is *pure* in a set of schemata $\Phi$ iff for every occurrence of a literal $\neg p_b$ (respectively $p_b$) in $\Phi$, the arithmetic formula $\Phi_{\mathcal{N}} \wedge IC(\Phi) \wedge a = b$ is unsatisfiable[8].

**Definition 5.9**
Let $\Phi, \Psi$ be two sets of schemata. We write $\Phi \supseteq_s \Psi$ iff there exists a shift $\sigma$ for the set of parameters in $\Phi$ and $\Psi$ s.t. for every $\psi \in \Psi$:

- Either $\psi$ is an arithmetic formula and $\Phi_{\mathcal{N}} \models \psi\sigma$.

- Or $\psi$ is a pure literal in $\Psi$.

- Or $\psi\sigma \in \Phi$.

The first and third items correspond roughly to set inclusion (up to arithmetic properties). The second item only deals with $\Psi$ and not with $\Phi$. It corresponds to the informal idea that a pure literal can be removed. Of course it is the most important one.

---

7. A possible improvement would be to add in $\Phi_{\mathcal{N}}$ formulae that are obvious logical consequences of $\Phi$. For instance, if $\Phi = \{p_n \wedge (n > 1), \neg p_1\}$ then $\Phi_{\mathcal{N}}$ would contain $n > 1$. This would make the notion of 'pure literal' slightly more general, e.g., $p_n$ would be pure in $\Phi$, which is not the case with our current definition.

8. See page 606 for the definition of $IC(\Phi)$.

**Example 5.10**
Let $\Psi = \{n \geq 0, p_{n+1}, p_n, \bigwedge_{i=1}^{n}(\neg p_i \vee p_{i-1}), \neg p_0\}$ and $\Phi = \{n \geq 1, p_{n-1}, \bigwedge_{i=1}^{n-1}(\neg p_i \vee p_{i-1}), \neg p_0\}$. We have $\Phi \supseteq_s \Psi$. Indeed, consider the shift $\sigma = \{n \mapsto n-1\}$. By definition $\Phi_{\mathcal{N}} = \{n \geq 1\}$. We have $(n \geq 0)\sigma = n - 1 \geq 0 \equiv n \geq 1$, thus $\Phi_{\mathcal{N}} \models (n \geq 0)\sigma$. Since $n \geq 0$ and $i \in [1, n]$, $p_i$ cannot be identical to $p_{n+1}$, thus $p_{n+1}$ is pure in $\Psi$. Finally, we have $p_n\sigma = p_{n-1} \in \Phi$ and $\bigwedge_{i=1}^{n}(\neg p_i \vee p_{i-1})\sigma = \bigwedge_{i=1}^{n-1}(\neg p_i \vee p_{i-1}) \in \Phi$.

We now show that $\supseteq_s$ is decidable. First of all, it is trivial that syntactic equality is decidable as shown by the following definition and proposition:

**Definition 5.11**
Let $\mathcal{U}(\phi, \psi)$ be the arithmetic formula defined as follows:

- If $\phi = p_a$ and $\psi = p_b$ then $\mathcal{U}(\phi, \psi) \stackrel{\text{def}}{=} (a = b)$.

- If $\phi = (a \triangleleft b)$ and $\psi = (c \triangleleft d)$ (with $\triangleleft \in \{\leq, <\}$) then $\mathcal{U}(\phi, \psi) \stackrel{\text{def}}{=} (a = c) \wedge (b = d)$.

- If $\phi = \neg\phi'$ and $\psi = \neg\psi'$ then $\mathcal{U}(\phi, \psi) = \mathcal{U}(\phi', \psi')$.

- If $\phi = (\phi_1 \pi \phi_2)$ (with $\pi \in \{\vee, \wedge\}$) and $\psi = (\psi_1 \pi \psi_2)$ then $\mathcal{U}(\phi, \psi) = \mathcal{U}(\phi_1, \psi_1) \wedge \mathcal{U}(\phi_2, \psi_2)$.

- If $\phi = \Pi_{i=a}^{b}\phi'$ and $\psi = \Pi_{i=c}^{d}\psi'$ then $\mathcal{U}(\phi, \psi) \stackrel{\text{def}}{=} (a = c) \wedge (b = d) \wedge \mathcal{U}(\phi', \psi')$.

- Otherwise $\mathcal{U}(\phi, \psi) \stackrel{\text{def}}{=} \bot$.

**Proposition 5.12**
Let $\phi, \psi$ be two schemata. For every substitution $\sigma$, $\mathcal{U}(\phi, \psi)\sigma$ is valid iff $\phi\sigma$ and $\psi\sigma$ are syntactically identical.

PROOF
By a straightforward induction on the formulae. □

We can prove the decidability of $\supseteq_s$:

**Proposition 5.13**
$\supseteq_s$ is decidable.

PROOF
Since linear arithmetic is decidable, it is possible to check whether a literal is pure or not in a set of formulae $\Psi$. Then these pure literals can be simply removed from $\Psi$ (since they satisfy the second condition in Definition 5.9). One now has to find a shift $\sigma$ such that every remaining formula in $\Psi$ satisfies the first or third condition. Let $n_1, \ldots, n_\kappa$ be the variables in $\Phi, \Psi$. Let $\sigma$ be a substitution mapping every parameter $n_\iota$ $(1 \leq \iota \leq \kappa)$ to $n_\iota - l_\iota$, where the $l_\iota$ are distinct *variables* not occurring in $\Phi, \Psi$. One has to check that there exists a substitution $\theta$ mapping every variable $l_\iota$ to an integer such that:

- $\forall \iota \in [1, \kappa], \theta(l_\iota) \geq 0$ and $\exists \iota \in [1, \kappa], \theta(l_\iota) > 0$. Since $\kappa$ is fixed, this condition can be stated as an arithmetic formula.

- For every formula $\psi \in \Psi$, one of the following conditions holds:

  – $\psi$ is an arithmetic formula and $\Phi_{\mathcal{N}} \models \psi\sigma\theta$, i.e. the formula $\forall n_1, \ldots, n_\kappa.\Phi_{\mathcal{N}} \Rightarrow \psi\sigma\theta$ is valid.

  – $\psi\sigma\theta$ occurs in $\Phi$. This holds iff $\Phi$ contains a formula $\phi$, such that $\psi\sigma\theta$ and $\phi$ are identical for every value of the parameters, i.e., by Proposition 5.12, iff $\forall n_1, \ldots, n_k.\mathcal{U}(\phi, \psi\sigma\theta)$ is valid.

Since every condition above is equivalent to an arithmetic formula, the whole condition can be expressed as an arithmetic formula (taking the conjunction of the formulae corresponding to each $\psi \in \Psi$ and $\phi \in \Phi$). This formula is satisfiable iff there exists a substitution $\theta$ satisfying the desired property. Then the proof follows straightforwardly from the decidability of linear arithmetic. $\qquad \square$

Now we prove that $\supseteq_s$ is stronger than the relation $\models_s$.

**Proposition 5.14**
Let $\Phi, \Psi$ be two sets of schemata. If $\Phi \supseteq_s \Psi$ then $\Phi \models_s \Psi$.

PROOF
Let $\sigma$ be the shift satisfying the conditions of Definition 5.9. Let $\mathcal{I}$ be an interpretation satisfying $\Phi$. Let $\theta = \sigma_{\mathcal{I}}$. We have to show that there exists $\mathcal{J} < \mathcal{I}$ s.t. $\mathcal{J} \models \psi$, i.e. that there exists a shift $\sigma'$ s.t. $\mathcal{J} = \mathcal{I}\sigma'$ and $\mathcal{J} \models \psi$. Equivalently, we can show that there exists a model $\mathcal{J}$ of $\psi\sigma$, i.e. that $\sigma' = \sigma$ is convenient. Let $\mathcal{J}$ be an interpretation s.t. $\mathcal{J}(L) = \mathbf{T}$ if $L$ is a literal that is pure in $\Psi\sigma$ and $\mathcal{J}(L) \overset{\text{def}}{=} \mathcal{I}(L)$ otherwise. Let $\psi \in \Psi$. We have to show that $\mathcal{J} \models \psi\sigma$. We distinguish three cases, according to the three items in Definition 5.9.

- If $\Phi_{\mathcal{N}} \models \psi\sigma$, then since $\mathcal{I} \models \Phi$ and since $\mathcal{J}$ and $\mathcal{I}$ coincide on every arithmetic variable we must have $\mathcal{J} \models \psi\sigma$.

- If $\psi$ is a literal pure in $\Psi$ then $\psi\sigma$ is pure in $\Psi\sigma$, thus we have $\mathcal{J} \models \psi\sigma$ by definition.

- If $\psi\sigma \in \Phi$, then $\mathcal{I} \models \psi\sigma$. Thus every literal that is pure in $\Phi$ must be pure in $\psi\sigma$. The complementary of these literals cannot occur in $[\psi\sigma]_\theta$. Since $\mathcal{I}$ and $\mathcal{J}$ coincide on all other literals and since $\psi$ is in negative normal form, we must have $\mathcal{J} \models \psi\sigma$.

Consequently $\mathcal{J} \models \psi\sigma$, hence $\mathcal{J}\sigma \models \psi$. $\qquad \square$

$\supseteq_s$ is strictly less general than $\models_s$ as evidenced by the following:

**Example 5.15**
Let $\Phi = \{\bigvee_{i=1}^n p_i\}$ and $\Psi = \{\bigvee_{i=2}^n p_i\}$. We have shown that $\Psi \models_s \Phi$ (see Example 5.5). However, we have $\Psi \not\supseteq_s \Phi$, since there is no shift $\sigma$ such that $(\bigvee_{i=1}^n p_i)\sigma = \bigvee_{i=2}^n p_i$ (this is obvious since $1\sigma$ cannot be equal to 2 whatever is $\sigma$).

### 5.3 Examples

Before proving the soundness, completeness and termination of STAB, we provide some examples of tableaux.

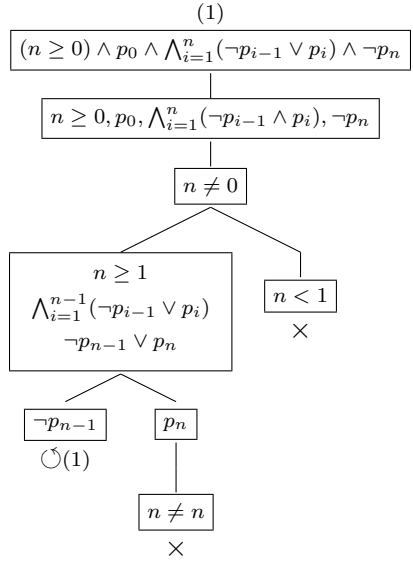

Figure 2: A Simple Example of Closed Tableau

### 5.3.1 A Simple Example

Let $\phi$ be the following formula: $(n \geq 0) \wedge p_0 \wedge \bigwedge_{i=1}^{n}(\neg p_{i-1} \vee p_i) \wedge \neg p_n$.

We construct a tableau for $\phi$. First the $\wedge$-rule applies to transform the conjunction into a set of schemata. The closure rule applies on $p_n$ and $p_0$, yielding the constraint $n \neq 0$. Then the iteration rule applies on the schema $\bigwedge_{i=1}^{n}(\neg p_{i-1} \vee p_i)$, yielding two branches. The first one corresponds to the case in which the iteration is non empty and can be unfolded, yielding $\bigwedge_{i=1}^{n-1}(\neg p_{i-1} \vee p_i)$ and $\neg p_{n-1} \vee p_n$ and the second one corresponds to the case where the iteration is empty (hence true), yielding the constraint $n < 1$. The latter branch can be closed immediately due to the constraints $n \geq 0$ and $n \neq 0$. In the former branch, the $\vee$-rule applies on the formula $\neg p_{n-1} \vee p_n$, yielding two branches with $\neg p_{n-1}$ and $p_n$ respectively. The closure rule applies on the latter one, yielding the unsatisfiable constraint $n \neq n$ hence the branch can be closed. The last remaining branch loops on the initial one, with the shift $n \mapsto n-1$. The obtained tableau is depicted in Figure 2. Closed leaves (resp. blocked leaves looping on $\alpha$) are marked by $\times$ (resp. $\circlearrowright(\alpha)$). Only new (w.r.t. the previous block) formulae are presented in the blocks.

### 5.3.2 $n$-Bit Adder

In this section we provide a slightly more complicated example. We use STAB to prove a simple property of the $n$-bit Adder defined in the Introduction. We aim at proving that $A + 0 = A$. A SAT-solver can easily refute this formula for a fixed $n$ (say $n = 10$). We prove it for all $n \in \mathbb{N}$. This simple example has been chosen for the sake of readability and conciseness, notice that commutativity or associativity of the $n$-bit adder could be proven too (see Section 5.7).

We express the fact that the second operand is null: $\bigwedge_{i=1}^{n} \neg q_i$, and the fact that the result equals the first operand: $\bigwedge_{i=1}^{n}(p_i \Leftrightarrow r_i)$, which gives $\bigvee_{i=1}^{n}(p_i \oplus r_i)$ by refutation. So we want to prove that $Adder \wedge \bigwedge_{i=1}^n \neg q_i \wedge \bigvee_{i=1}^n (p_i \oplus r_i)$ is unsatisfiable. Notice that this schema is regular.

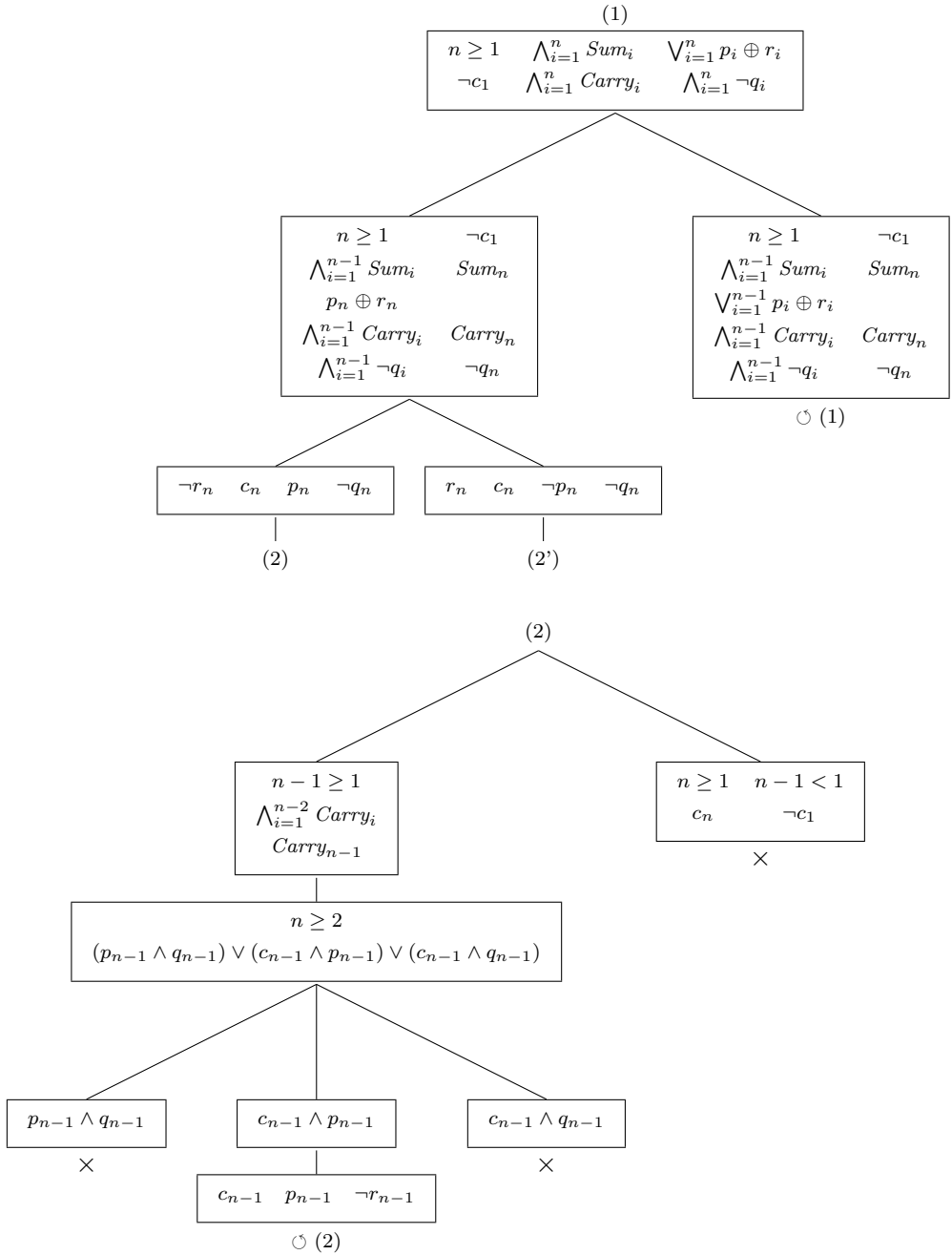

Figure 3: A Closed Tableau for $A + 0 = A$

The corresponding tableau is sketched in Figure 3. Sequences of propositional extension rules are not detailed.

**Explanations.** The first big step decomposes all the iterations. The branching is due to $\bigvee_{i=1}^{n} p_i \oplus r_i$: first we have $p_n \oplus r_n$, then $\bigvee_{i=1}^{n-1} p_i \oplus r_i$. The right branch loops after a few steps as all iterated conjunctions $\bigwedge_{i=1}^{n} \ldots$ contain $\bigwedge_{i=1}^{n-1} \ldots$ The left one is extended by propositional rules (the reader can easily check that $Sum_n$, $Carry_n$, $p_n \oplus r_n$ and $\neg q_n$ indeed lead to the presented branches, notice that $c_n$ must hold, otherwise we would have $p_n \Leftrightarrow r_n$).

In (2) we start by decomposing all iterations a second time. Iterations are aligned on $[1, n-1]$ so they all introduce the same constraints i.e. either $n - 1 \geq 1$ (first branch) or $n - 1 < 1$ (second branch). In the second case, the introduced constraint implies that $n = 1$, thus $c_n = c_1$ which closes the branch. In the first case we decompose $Carry_{n-1}$ and consider the various cases. Two of them are trivially discarded as they imply $q_{n-1}$, whereas we easily obtain $\neg q_{n-1}$ by an unfolding of $\bigwedge_{i=1}^{n} \neg q_i$. It only remains one case which is easily seen to loop on (2). The branch (2') is very similar to (2).

## 5.4 Soundness and Completeness

A leaf is *irreducible* if no extension rule applies to it. A *derivation* is a (possibly infinite) sequence of tableaux $(\mathcal{T}_\iota)_{\iota \in I}$ s.t. $I$ is either $[0, \kappa]$ for some $\kappa \geq 0$, or $\mathbb{N}$ and s.t. for all $\iota \in I \setminus \{0\}$, $\mathcal{T}_\iota$ is obtained from $\mathcal{T}_{\iota - 1}$ by applying one of the extension rules. A derivation is *fair* if either there is $\iota \in I$ s.t. $\mathcal{T}_\iota$ contains an irreducible not closed leaf or if for all $\iota \in I$ and every not closed and not blocked leaf $N$ in $\mathcal{T}_\iota$ there is $\lambda \geq \iota$ s.t. a rule is applied on $N$ in $\mathcal{T}_\lambda$ (i.e. no leaf can be "freezed").

### Definition 5.16 (Tableau Semantics)
For every node $N$ in a tableau $\mathcal{T}$, $\Phi_{\mathcal{T}}(N)$ is interpreted as the conjunction of its elements. $\mathcal{T}$ is *satisfied* in an interpretation $\mathcal{I}$ iff there exists a leaf $N$ in $\mathcal{T}$ s.t. $\mathcal{I} \models \Phi_{\mathcal{T}}(N)$.

### Lemma 5.17
If $\mathcal{T}'$ is a tableau obtained by applying one of the extension rules on a leaf $N$ of a tableau $\mathcal{T}$ then $\mathcal{I} \models \Phi_{\mathcal{T}}(N)$ *iff* there exists a leaf $N'$ of $\mathcal{T}'$ s.t. $N'$ is a child of $N$ in $\mathcal{T}'$ and $\mathcal{I} \models \Phi_{\mathcal{T}'}(N')$ (i.e. the rules are sound and *invertible*).

Proof
Obvious, by inspection of the extension rules. □

### Lemma 5.18
If a leaf $N$ in $\mathcal{T}$ is irreducible and not closed then $\mathcal{T}$ is satisfiable.

Proof
Let $\Psi$ be the set of arithmetic formulae in $\Phi_{\mathcal{T}}(N)$ and $\Phi \overset{\text{def}}{=} \Phi_{\mathcal{T}}(N) \setminus \Psi$. As $N$ is not closed $\Psi$ is satisfiable (by definition), so let $\sigma$ be a solution of $\Psi$. If $\Phi$ contains a formula $\phi$ that is not a literal, one of the extension rules applies and deletes $\phi$, which is impossible. Let $c_{\mathcal{T}}(N)$ be the number of pairs $p_a$, $\neg p_b \in \Phi_{\mathcal{T}}(N)$ s.t. there is an interpretation $\mathcal{I}$ validating $\Psi$ s.t. $[\![a]\!]_{\mathcal{I}} = [\![b]\!]_{\mathcal{I}}$. If $c_{\mathcal{T}}(N) \neq 0$, then the closure rule applies on $p_a$, $p_b$ which is impossible. Hence $c_{\mathcal{T}}(N) = 0$ and in particular this implies that $\Phi\sigma$ is propositionally satisfiable (i.e. contains no pair of complementary literals). Thus $\Phi_{\mathcal{T}}(N)\sigma$ is satisfiable and by definition $\mathcal{T}$ is satisfiable. □

**Theorem 5.19 (Soundness and Completeness w.r.t. Satisfiability)**
Let $(\mathcal{T}_\kappa)_{\kappa \in I}$ be a derivation.

- If there exists $\iota \in I$ s.t. $\mathcal{T}_\iota$ contains an irreducible, not closed leaf then $\mathcal{T}_0$ is satisfiable.

- If the derivation is fair and if $\mathcal{T}_0$ is satisfiable then there exist $\iota \in I$ and a leaf in $\mathcal{T}_\iota$ that is irreducible and neither closed nor blocked.

PROOF
The first item (i.e. soundness) follows from Lemmata 5.17 and 5.18.

We now prove that the procedure is complete w.r.t. satisfiability (the second item). Let $\mathcal{I}$ be an interpretation and $\phi$ a schema. We define $m_\mathcal{I}(\phi)$ as follows:

- $m_\mathcal{I}(\phi) \overset{\text{def}}{=} 0$ if $\phi$ is an arithmetic atom (i.e. an atom of the form $\ldots < \ldots$).

- $m_\mathcal{I}(\phi) \overset{\text{def}}{=} 1$ if $\phi$ is an indexed proposition or its negation, or $\phi$ is $\top$ or $\bot$.

- $m_\mathcal{I}(\phi_1 \star \phi_2) \overset{\text{def}}{=} m_\mathcal{I}(\phi_1) + m_\mathcal{I}(\phi_2)$ if $\star \in \{\vee, \wedge\}$.

- $m_\mathcal{I}(\Pi_{i=a}^b \phi) \overset{\text{def}}{=} 2$ if $[\![b]\!]_\mathcal{I} < [\![a]\!]_\mathcal{I}$

- $m_\mathcal{I}(\Pi_{i=a}^b \phi) \overset{\text{def}}{=} \beta - \alpha + 2 + \Sigma_{\iota=\alpha}^\beta m_{\mathcal{I}[\iota/i]}(\phi)$ where $\Pi \in \{\bigwedge, \bigvee\}$, $\alpha = [\![a]\!]_\mathcal{I}$, $\beta = [\![b]\!]_\mathcal{I}$, and $\beta \geq \alpha$.

If $\Phi$ is a set, then $m_\mathcal{I}(\Phi) \overset{\text{def}}{=} \{m_\mathcal{I}(\phi) \mid \phi \in \Phi\}$. If $\mathcal{T}$ is a tableau and $N$ is a leaf in $\mathcal{T}$ then $m_\mathcal{I}(N, \mathcal{T}) \overset{\text{def}}{=} (m_\mathcal{I}(\Phi_\mathcal{T}(N)), c_\mathcal{T}(N))$ where $c_\mathcal{T}(N)$ is defined in the proof of Lemma 5.18. This measure is ordered using the multiset and lexicographic extensions of the usual ordering on natural numbers. Thus, it is obviously well-founded. We need the following:

**Lemma 5.20**
Let $\mathcal{I}$ be an interpretation. Let $\mathcal{T}$ be a tableau. If $\mathcal{T}'$ is deduced from $\mathcal{T}$ by applying an extension rule on a leaf $N$ s.t. $\mathcal{I} \models \Phi_\mathcal{T}(N)$, then for every child $N'$ of $N$ in $\mathcal{T}'$ s.t. $\mathcal{I} \models \Phi_{\mathcal{T}'}(N')$, we have $m_\mathcal{I}(N', \mathcal{T}') < m_\mathcal{I}(N, \mathcal{T})$.

PROOF
All the rules except the iteration rule and the closure rule replace a formula by simpler ones, hence it is easy to see that $m_\mathcal{I}(\Phi_\mathcal{T}(N))$ decreases. The iteration rules replace an iteration of length $\iota$ either by $\top$ or by a disjunction/conjunction of an iterated disjunction/conjunction of length $\iota - 1$, and a smaller formula. Since $\iota > \iota - 1$, $m_\mathcal{I}(\Phi_\mathcal{T}(N))$ decreases. The closure rule does not affect $m_\mathcal{I}(\Phi_\mathcal{T}(N))$ but obviously decreases $c_\mathcal{T}(N)$. $\square$

Let $\mathcal{I}$ be a model of $\mathcal{T}_0$. By Proposition 5.6, we can assume that $\mathcal{I}$ is minimal w.r.t the ordering $<$ introduced in Definition 5.4.

By Lemma 5.17, for all $\iota \in I$, $\mathcal{T}_\iota$ contains a leaf $N_\iota$ s.t. $\mathcal{I} \models \Phi_{\mathcal{T}_\iota}(N_\iota)$. Let $\kappa \in I$ s.t. $m_\mathcal{I}(N_\kappa, \mathcal{T}_\kappa)$ is minimal ($\kappa$ exists since $m_\mathcal{I}(N_\iota, \mathcal{T}_\iota)$ is well-founded). Assume a rule is applied on $N_\kappa$ in the derivation, on some tableau $\mathcal{T}_\lambda$. By Lemma 5.17 there is a child $N'$ of $N_\kappa$ s.t. $\mathcal{I} \models \Phi_{\mathcal{T}_\lambda}(N')$. By Lemma 5.20 we have $m_\mathcal{I}(N', \mathcal{T}_\lambda) < m_\mathcal{I}(N_k, \mathcal{T}_\kappa)$ which is impossible. Thus no rule is applied on $N_\kappa$. Assume that $N_\kappa$ is blocked. Then there exists a node $N'$ s.t. $N_\kappa$ loops on $N'$. By Definition 5.4 there exists an interpretation $\mathcal{J}$ s.t. $\mathcal{J} \models N'$ and $\mathcal{J} <_V \mathcal{I}$. But then by Lemma 5.17 ("only if" implication), $\mathcal{J} \models \mathcal{T}_0$, which contradicts the minimality of $\mathcal{I}$. Since the derivation is fair, $N_\kappa$ is irreducible (or there is another leaf that is irreducible). Furthermore, $N_\kappa$ cannot be closed since it is satisfiable ($\mathcal{I} \models \Phi_{\mathcal{T}_\kappa}(N_\kappa)$).

It is worth emphasizing that STAB is sound and complete (w.r.t. satisfiability) for any schema, not only for bound-linear or regular ones. But the termination result in the next section only holds for regular schemata.

## 5.5 Termination on Regular Schemata

We consider the following strategy ST for applying the extension rules:

- The propositional extension rules, the looping and closure rules are applied as soon as possible on all leaves, with the highest priority. These rules obviously terminate on any schema.

- The iteration rules are applied only on iterations of maximal length (w.r.t. the natural partial ordering on arithmetic expressions). For instance if we have the schema $\bigwedge_{i=1}^n p_i \vee \bigvee_{j=1}^{n-1} q_j$ then the iteration rules will only apply on the first iteration $\bigwedge_{i=1}^n p_i$.

- The relation $\supseteq_s$ introduced in Section 5.2 is used to block looping nodes.

### Theorem 5.21
ST terminates on every regular schema.

Proof

Let $\alpha, \beta, \gamma, \delta \in \mathbb{Z}$ and $\phi$ be a regular schema aligned on $[\alpha, n - \beta]$, of propagation limits $\gamma, \delta$. Assume that an infinite branch is constructed. By definition of the strategy, after some time, the $\kappa$ last ranks of every iteration have been unfolded by the iteration rules. Thus all the remaining iterations are of the form $\Pi_{i=\alpha}^{n-\beta-\kappa} \phi'$ and we have the arithmetic constraint $n - \beta - \kappa - \alpha + 1 \geq 0$, i.e. $n \geq \beta + \kappa + \alpha - 1$.

From now on, we only consider nodes that are irreducible w.r.t. propositional rules. We show that a finite set of formulae are generated by STAB, up to a shift on $n$. As a consequence the looping rule must apply, at worst when all possible formulae have been generated.

The arithmetic formulae occurring in the initial formula must be of the form $\mu.n > \nu$ or $\mu.n < \nu$. After the last $\kappa$ ranks have been unfolded, the constraint $n \geq \beta + \kappa + \alpha - 1$ must have been added. Thus if $\kappa$ is sufficiently big, $\mu.n > \nu$ is equivalent to $\top$ and $\mu.n < \nu$ is equivalent to $\bot$. Thus every arithmetic formula occurring in the initial formula is either false or redundant w.r.t. $n \geq \beta + \kappa + \alpha - 1$. The remaining arithmetic formulae must have been introduced by the closure rule (since the iterations contain no occurrence of $<$). They are necessarily of the form $a \neq b$ where $a, b$ are arithmetic expressions (appearing as indices in some formula of the derivation). If $a, b$ both contain $n$, or if $a, b \in \mathbb{Z}$ then $a \neq b$ is equivalent either to $\bot$ or to $\top$. Thus we only consider the case in which $a$ contains $n$ and $b \in \mathbb{Z}$. If $a$ occurs in the initial formula then it must be of the form $\mu.n + \nu$ for $\mu, \nu \in \mathbb{Z}$. Since $n \geq \beta + \kappa + \alpha - 1$, if $\kappa$ is sufficiently big, the disequation $\mu.n + \nu \neq b$ must be false. If $a$ did not occur in the initial formula then it must come from the $(\kappa - \iota)^{th}$ unfolding of some iteration, for some $\iota \in [0, \kappa - 1]$. Since (by definition of a regular schema) the indices are of the form $i + \lambda$, where $\lambda \in [\gamma, \delta]$, the disequation is actually of the form $n - \beta - \kappa + \iota + \lambda \neq b$, where $\lambda \in [\gamma, \delta], \iota \in [0, \kappa - 1]$ (since the iteration counter $i$ may be replaced by $n - \beta, n - \beta - 1, \ldots, n - \beta - \kappa + 1$) and $b$ occurs in the initial formula. If the previous equation is not equivalent to $\top$, then, since we have the constraint $n \geq \beta + \kappa + \alpha - 1$, we must have $\iota \in [0, b - \alpha + 1 - \lambda]$. Hence there are finitely many such formulae, up to the translation $n \mapsto n - \kappa$.

Now, consider the non arithmetic formulae occurring in the branch. These schemata must be either iterations or literals (by irreducibility w.r.t. the propositional extension rules).

All the iterations are of the form $\Pi_{i=\alpha}^{n-\beta-\kappa} \phi'$, where $\Pi_{i=\alpha}^{n-\beta} \phi'$ is an iteration occurring in the initial formula. Obviously, the number of such iterations is finite up to the translation $n \mapsto n - \kappa$.

The literals occurring in the branch (but not in the scope of an iteration) are either literals of the initial schema or literals introduced by previous applications of the iteration rules. The former are indexed by expressions of the form $\mu \times n + \nu$ for some $\mu, \nu \in \mathbb{Z}$ and the latter by $n - \beta - \kappa + \epsilon$, where $\epsilon \in [\gamma + 1, \delta + \kappa]$.

If a literal is indexed by an expression $\mu \times n + \nu$ that is outside $[\alpha + \gamma, n - \beta - \kappa + \delta]$, then it must be pure in every iteration, hence (by irreducibility w.r.t. the closure rule) must be pure in the node. Actually, if $\kappa$ is large enough then, by the above arithmetic constraints, $\mu \times n + \nu$ cannot be in $[\alpha + \gamma, n - \beta - \kappa + \delta]$ if $\mu \neq 0$. Indeed, if $\mu$ is negative, then it suffices to take $\kappa > \frac{\alpha + \gamma - \nu}{\mu} - \beta - \alpha + 1$ to ensure $\mu \times n + \nu < \alpha + \gamma$, otherwise $\kappa \geq \delta - \beta - \nu$ is enough to have $\mu \times n + \nu > n - \beta - \kappa + \delta$ (as $\mu, n \geq 1$). Thus every literal indexed by integer terms of this form are pure, since by definition its index cannot be unifiable with an index occurring in an iteration (after unfolding).

Similarly literals indexed by expressions of the form $n - \beta - \kappa + \epsilon$ where $\epsilon > \delta$ are pure, thus we may assume that $\epsilon \in [\gamma, \delta]$. Consequently there are finitely many such literals up to the shift $n \mapsto n - \kappa$.

This implies that the number of possible schemata obtained after $\kappa$ unfolding steps is finite, up to a translation of $n$. By the pigeonhole principle, the looping rule necessarily applies at some point in the branch, which contradicts our initial assumption that an infinite branch is constructed. □

Termination of the strategy also ensures fairness:

**Lemma 5.22**
Any derivation constructed by ST (applied until irreducibility) is fair.

PROOF
Let $(\mathcal{T}_\iota)_{\iota \in I}$ be a derivation constructed by ST. Since ST terminates, there cannot be any infinite derivation, thus $I$ is necessarily of the form $[0, \kappa]$ for some $\kappa \in \mathbb{N}$. By definition, every node in $\mathcal{T}_\kappa$ is either blocked or closed or irreducible (the strategy is applied until irreducibility). If $\mathcal{T}_\kappa$ contains a not closed irreducible leaf then the proof is completed (by definition of the notion of fairness). Otherwise, consider $\mathcal{T}_\iota$ with $\iota \leq \kappa$. Let then $N$ be a not irreducible, not closed and not blocked leaf occurring in $\mathcal{T}_\iota$. Assume that there is no $\lambda \geq \iota$ s.t. a rule is applied on $N$ in $\mathcal{T}_\lambda$ (which would contradict our definition of fairness). This means that no extension can possibly affect $N$, thus $N$ must also occur in the final tableau $\mathcal{T}_\kappa$ (and is labeled by the same set of schemata than in $\mathcal{T}_\iota$). Thus $N$ must be not closed and not irreducible. Moreover it cannot be blocked in $\mathcal{T}_\kappa$, since no rule can affect the nodes on the branch behind $N$. But this is impossible since the nodes in $\mathcal{T}_\kappa$ must be either blocked or closed or irreducible. □

As an immediate corollary, we have the following:

**Theorem 5.23**
The satisfiability problem is decidable for bound-linear schemata.

Proof
By Theorems 4.12 and 4.13, every bound-linear schema can be transformed into a sat-equivalent regular one. Theorem 5.21 shows that STAB terminates on every regular schema, hence by Theorem 5.19 and Lemma 5.22, STAB can be used to decide the satisfiability problem for regular schemata. □

A fine analysis of the previous termination proof ensures that we can solve the satisfiability problem for regular schemata in exponential time (if natural numbers are written in unary notation). As we have seen furthermore (Theorem 4.9) that the translation of bound-linear schemata into regular ones was exponential, we can conclude that the satisfiability problem for bound-linear schemata can be solved in double exponential time.

## 5.6 Model Building

The existence of a non closed irreducible branch ensures that the root schema is satisfiable, as shown in Theorem 4.12. The arithmetic constraints in the branch specify the possible values of the parameter. The remaining formulae must be literals, since the extension rules apply on any complex formula (in particular, there can be no iteration schema). These literals specify the truth value of propositional variables exactly as in the usual case of propositional logic (the value of the propositional variables that do not appear in the branch may be chosen arbitrarily). Since the branch is not closed, it cannot contain any pair of complementary literals.

We illustrate this construction by a simple example. We consider the following tableau:

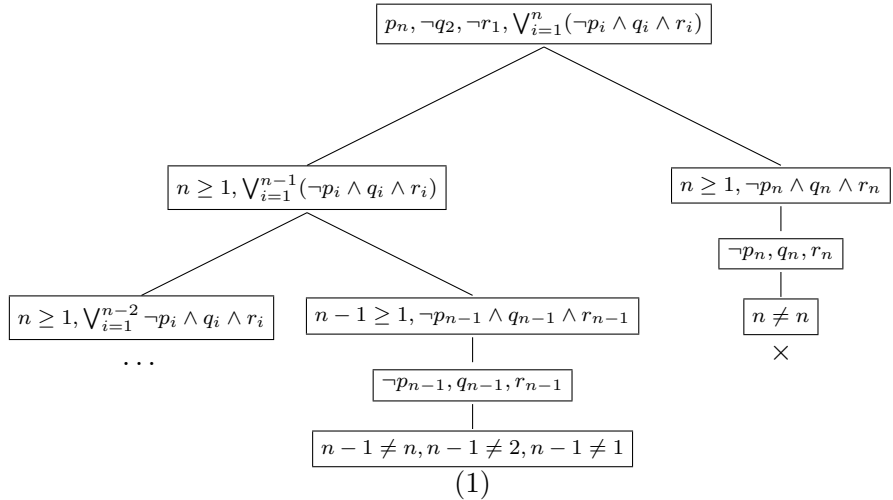

(1)

The branch (1) is irreducible. It contains the following formulae: $p_n$, $\neg q_2$, $\neg r_1$, $n-1 \geq 1$, $\neg p_{n-1}$, $q_{n-1}$, $r_{n-1}$, $n-1 \neq n$, $n-1 \neq 2$, $n-1 \neq 1$. The value of $n$ can be determined by finding a solution to the above arithmetic constraints. We choose for instance the solution $n = 4$. After instantiation we get the remaining formulae: $\{p_4, \neg q_2, \neg r_1, \neg p_3, q_3, r_3\}$, which gives for instance the following interpretation of $p, q$ and $r$: $p_\kappa$ is true iff $\kappa = 4$ and $q_\kappa$, $r_\kappa$ are true iff $\kappa = 3$. It is easy to check that the obtained interpretation satisfies the initial schema.

A possible extension of this simple algorithm would be, from a given tableau, to compute a symbolic representation of the whole set of models of the root schema. This set is infinite and must be defined by induction. The closed irreducible branches correspond to concrete models, or *base cases*, whereas the loops correspond to *inductive construction rules*. These rules take a model $\mathcal{I}$ and construct a new model $\mathcal{J}$ of a strictly greater cardinality (the values of the parameters increase strictly). This would require to define a formal language for denoting sets of interpretations (one could use, e.g., automata recognizing sequences of tuples of Boolean values).

## 5.7 The System

The decision procedure has been implemented and the program (called RegStab) is freely available on the web page `http://regstab.forge.ocamlcore.org/`. It is written in OCaml and was successfully tested on MacOSX (10.5), Win32 (Windows XP SP3) and GNU Linux (Ubuntu 9.04) x86 platforms. The system comes with a manual including installation and usage instructions and a description of the input syntax. Functions can be defined to make the input file more readable (see `Sum(i)` and `Carry(i)` below). Here is an input file for the adder example in Section 5.3.2.

```
// A+0=A
let Sum(i)      := S_i <-> (A_i (+) B_i (+) C_i) in
let Carry(i)    := C_i+1 <-> (A_i /\ B_i \/ C_i /\ A_i \/ C_i /\ B_i) in
let Adder       := /\i=1..n (Sum(i) /\ Carry(i)) /\ ~C_1 in
let NullB       := /\i=1..n ~B_i in
let Conclusion := \/i=1..n (A_i (+) S_i) in

Adder() /\ NullB() /\ Conclusion()
```

The software simply prints the status of the schema (satisfiable or unsatisfiable). Options are provided to get more information about the search space (number of inference rules, depth of unfolding etc.), see the manual for details. An additional tool is offered to expand the schema into a propositional formula in DIMACS format (by fixing the value of $n$).

Figure 4 gives some examples of problems that can be solved by RegStab and the corresponding running times (please refer to the distribution for input files and additional information).

Here is an example of output, proving that 0 is a neutral element for the carry-propagate adder. We ran the system in verbose mode, in which it prints some useful information about the search: number of application of extension rules, number of closed and looping leaves, unfolding depth and set of lemmata (companion nodes).

| Ripple-carry adder | |
|---|---|
| $x + 0 = x$ | 0.017s |
| commutativity | 0.267s |
| associativity | 28.902s |
| $3 + 4 = 7$ | 2.719s |
| $x + y = z_1 \wedge x + y = z_2 \Rightarrow z_1 = z_2$ | 0.490s |
| Carry-propagate adder | |
| $x + 0 = x$ | 0.016s |
| commutativity | 0.165s |
| associativity | 8.522s |
| equivalence between two different definitions of the same adder | 0.164s |
| equivalence with the ripple-carry adder | 0.194s |
| Comparisons between bit-vectors | |
| $x \geq 0$ | 0.004s |
| Symmetry of $\leq$ (i.e. $x \leq y \wedge x \geq y \Rightarrow x = y$) | 0.009s |
| Totality of $\leq$ (i.e. $x > y \vee x \leq y$) | 0.006s |
| Transitivity of $\leq$ | 0.011s |
| $1 \leq 2$ | 0.010s |
| Presburger arithmetic with bit vectors | |
| $x + y \geq x$ | 0.026s |
| $x_1 \leq x_2 \leq x_3 \Rightarrow x_1 + y \leq x_2 + y \leq x_3 + y$ | 1m42s |
| $x_1 \leq x_2 \wedge y_1 \leq y_2 \Rightarrow x_1 + y_1 \leq x_2 + y_2$ | 2.949s |
| $x_1 \leq x_2 \leq x_3 \wedge y_1 \leq y_2 \leq y_3 \Rightarrow x_1 + y_1 \leq x_2 + y_2 \leq x_3 + y_3$ | 46m57s |
| $1 \leq x + y \leq 5 \wedge x \geq 3 \wedge y \geq 4$ | 7m9s |
| same but with iterations factorized | 2m14s |
| Other | |
| automata inclusion | 2.324s |
| $\bigvee_{i=1}^{n} P_i \wedge \bigwedge_{i=1}^{n} \neg P_i$ | 0.001s |
| $P_1 \wedge \bigwedge_{i=1}^{n} (P_i \Rightarrow P_i + 1) \wedge \neg P_{n+1} \mid n \geq 0$ | 0.001s |
| model checking of some safety property | 5.251s |

Figure 4: Some Experimental Results

```
Conjecture:
  ((((/\i=1..n ((S_i <-> ((A_i (+) B_i) (+) C_i)) /\
  (C_i+1 <-> (((A_i /\ B_i) \/ (C_i /\ A_i))
  \/ (C_i /\ B_i))))) /\ ~C_1) /\ (/\i=1..n ~B_i)) /\
  (\/i=1..n (A_i (+) S_i))

Applications of tableau rules:
/\:  67
\/:  84
(+): 38
<->: 32
->:  0
Iterated /\: 12
Iterated \/:  3
-------
Total propositional rules: 221
Total iterated rules:        15

Number of closed leaves:  137
Number of looping leaves: 30
Number of lemmas:          4

Maximum number of unfoldings: 3
(if this number is surprising, notice that the tableau is
constructed depth-first)

Lemmas:
[\/i=1..n (A_i (+) S_i) ; /\i=1..n ((S_i <-> ((A_i (+) B_i) (+) C_i))
/\ (C_i+1 <-> (((A_i /\ B_i) \/ (C_i /\ A_i)) \/ (C_i /\ B_i)))) ;
/\i=1..n ~B_i ; ~C_1]
[\/i=1..n-1 (A_i (+) S_i) ; /\i=1..n-1 ((S_i <-> ((A_i (+) B_i) (+) C_i))
/\ (C_i+1 <-> (((A_i /\ B_i) \/ (C_i /\ A_i)) \/ (C_i /\ B_i)))) ;
/\i=1..n-1 ~B_i ; ~C_n ; ~C_1]  (n > 0)
[/\i=1..n-2 ((S_i <-> ((A_i (+) B_i) (+) C_i)) /\ (C_i+1 <-> (((A_i /\ B_i)
\/ (C_i /\ A_i)) \/ (C_i /\ B_i)))) ;
/\i=1..n-2 ~B_i ; C_n-1 ; ~C_1]  (n > 1)
[\/i=1..n-2 (A_i (+) S_i) ; /\i=1..n-2 ((S_i <-> ((A_i (+) B_i) (+) C_i))
/\ (C_i+1 <-> (((A_i /\ B_i) \/ (C_i /\ A_i)) \/ (C_i /\ B_i)))) ;
/\i=1..n-2 ~B_i ; C_n-1 ; ~C_1]  (n > 1)

UNSATISFIABLE
```

## 6. Undecidability Results

We provide some undecidability results for two natural extensions of the class of regular schemata.

### 6.1 Homothetic Transformations on the Iteration Counters

We consider the class of schemata $\mathfrak{C}_h$ defined as follows.

**Definition 6.1**

$\mathfrak{C}_h$ ($h$ stands for "homothetic") is the set of schemata $\phi$ satisfying the following properties:

- $\phi$ contains at most one parameter $n$.

- Every iteration in $\phi$ is of the form $\bigwedge_{i=1}^n \phi$ or $\bigvee_{i=1}^n \phi$, where:

  - $\phi$ contains no iteration.
  - Every atomic formula in $\phi$ belongs to $\{p_i, p_{2i}, p_{i\pm 1}, p_{2i\pm 1}\}$ where $p$ is a variable.

- The atomic formulae occurring in $\phi$ but not in the scope of an iteration are of the form $p_0$ or $p_n$ where $p$ is a variable[9].

$\mathfrak{C}_h$ is rather simple and very close to the class of regular schemata. There is only one parameter $n$, all the iterations have the same bounds 1 and $n$, there is no nested iteration and the indices of the symbol in $\mathcal{P}$ must be affine images of the iteration counter. The only difference with the regular class is that, in $\mathfrak{C}_h$ the coefficient of the iteration counter in the indexed variables may be equal to 2 whereas it must be equal to 0 or 1 in regular schemata. Thus regular schemata only contain translations of the iteration counter, whereas $\mathfrak{C}_h$ may involve (very simple) homothetic transformations.

Due to this closeness, one could expect that the satisfiability problem is decidable for $\mathfrak{C}_h$, but the next theorem shows that this is not the case.

**Theorem 6.2**

The set of unsatisfiable formulae in $\mathfrak{C}_h$ is not recursively enumerable.

The proof of Theorem 6.2 is difficult and the remaining part of this section is devoted to it. More precisely, we shall prove that the Post correspondence problem can be encoded into $\mathfrak{C}_h$. Notice that this problem is easily encoded with general schemata (Aravantinos et al., 2009b), whereas, here, the whole difficulty of the proof lies in the strong restrictions imposed by $\mathfrak{C}_h$. Observe that the difficult proof is really worth it as one would easily believe that just allowing multiplication by a constant is an unsignificant change.

### 6.1.1 NOTATIONS

We first recall some basic definitions and introduce some useful notations. Let $\mathcal{A}$ be an alphabet. Let $\kappa$ be a natural number. Let $a = (a_1, \ldots, a_\kappa)$ and $b = (b_1, \ldots, b_\kappa)$ be two sequences of words in $\mathcal{A}^*$. If $w \in \{a, b\}$ and $\iota \in [1, \kappa]$, $|w_\iota|$ denotes the length of $w_\iota$ and $w_\iota^\lambda$ denotes the $\lambda$-th character of the word $w_\iota$ ($1 \leq \lambda \leq |w_\iota|$).

---

9. Notice that $p_0$ and $p_n$ can occur in the scope of a negation.

If $\Delta = (\Delta_1, \ldots, \Delta_\iota)$ is a sequence of indices in $[1, \kappa]$ and if $w = (w_1, \ldots, w_\kappa)$ is a $\kappa$-tuple of words in $\mathcal{A}^*$ (where $w \in \{a, b\}$) we denote by $w_\Delta$ the word $w_{\Delta_1} . \cdots . w_{\Delta_\iota}$ (where "." denotes the concatenation operator). A *solution* of the Post correspondence problem is a sequence $\Delta$ s.t. $a_\Delta = b_\Delta$. The *witness* of this solution is the word $a_\Delta$.

For technical convenience, we assume (this is obviously not restrictive) that $\kappa > 1$, $\Delta_\iota = \kappa$, $\Delta_\lambda \neq \kappa$ if $\lambda < \iota$ and that $a_\kappa = b_\kappa = \perp$ where $\perp$ is a special character (not occurring in $a_1, \ldots, a_{\iota-1}, b_1, \ldots, b_{\iota-1}$) denoting the end of the sequence.

### 6.1.2 OVERVIEW OF THE ENCODING

The intuition behind the encoding is the following. We show how to encode any instance of the problem into a schema $\phi$ so that $\phi$ is satisfiable iff this instance has a solution. More precisely, we construct $\phi$ of parameter $n$ s.t. for all $\kappa \in \mathcal{N}$, $\phi[\kappa/n]$ is satisfiable iff there is a solution of length $\kappa$.

We first present the encoding used to represent the potential solutions $a_\Delta$ and $b_\Delta$; then we will see how to check that those are really solutions. We represent the potential solution $w_\Delta$ (where $w = a, b$) by a one-dimensional array of length $n$. More precisely, we do not store the characters themselves but rather, for each character, a pair containing the index $\Delta_\nu$ of the word $w_{\Delta_\nu}$ in which it occurs and its position in this word (as we shall see this is useful to find the next character in $w_\Delta$). For instance the first index should contain the pair $(\Delta_1, 1)$ (first word, first character). Then the next index contains either $(\Delta_1, 2)$ (if $|w_{\Delta_1}| > 1$, first word, second character) or $(\Delta_2, 1)$ (if $|w_{\Delta_1}| = 1$, second word, first character).

For example, if $\mathcal{A} = \{*, \circ, \star\}$, $a = (*\circ, \star)$ and $\Delta = (1, 2)$, then the obtained array would be the following one:

| Values | $(1, 1)$ | $(1, 2)$ | $(2, 1)$ |
|---|---|---|---|
| Indices | 1 | 2 | 3 |

However, the word $w_\Delta$ is not stored into consecutive indices in the array. Indeed, as we shall see, we also need to store, for each character $w_{\Delta_\lambda}$ of the witness, the indices $\Delta_{\lambda+1}, \ldots, \Delta_\iota$ of the remaining words, occurring after $w_{\Delta_\lambda}$ in $w_\Delta$. This sequence is called the *tail* of the potential solution. Since the length of this sequence is unbounded, it cannot be encoded simply by indexed propositions: it must be stored into the array and the simplest solution is to store these indices just after the character itself. Notice that only the indices of *words* are stored in the tail i.e. there is no character position. Thus we get:

| Values | $(1, 1)$ | 2 | $(1, 2)$ | 2 | $(2, 1)$ |
|---|---|---|---|---|---|
| Indices | 1 | 2 | 3 | 4 | 5 |

The easiest way to proceed would be to store the first character of the witness at position $0$, the indices of the remaining words at position $1, 2, \ldots, \iota$, then the second character of the witness at position $\iota + 1$ etc. That way, the $\lambda$-th character of the witness would be stored at position $(\lambda - 1) \times (\iota + 1)$ and the following characters in the sequence at positions $(\lambda - 1) \times (\iota + 1) + 1, \ldots, (\lambda - 1) \times (\iota + 1) + \iota$. For any character stored in an index $\lambda$, the next character would be stored at the index $\lambda + \iota + 1$. But this simple solution is not suitable because it is outside the considered class. Indeed, it requires the use of another parameter $\iota$ (the first parameter being $n$: the length of the array) and also the use of this parameter in the indices (to relate the character stored in index $\lambda$ to the one at index $\lambda + \iota$), which is forbidden in the class $\mathfrak{C}_h$.

Thus we need to find another encoding of the previous array. The idea is to store the first character at some index $\mu$ (where $\mu$ is assumed to be greater than $\iota$), the second character at the index $2 \times \mu$, ... and more generally the $\lambda$-th character at the index $\mu \times 2^{\lambda-1}$. The tail of the sequence is then stored at the indices $(\mu + 1) \times 2^{\lambda-1}, \ldots, (\mu + \iota) \times 2^{\lambda-1}$. This encoding ensures that the index of the next character after the one at index $i$ is simply $2.i$, and such homethetic transformations are precisely those allowed for the indices in $\mathfrak{C}_h$.

Finally, the array corresponding to our recurrent example is the following one (with $\mu = 2$):

| Values  |   | $(1,1)$ | $2$ | $(1,2)$ | $2$ |   |   | $(2,1)$ |
|---------|---|---------|-----|---------|-----|---|---|---------|
| Indices | 1 | 2       | 3   | 4       | 5   | 6 | 7 | 8       |

The witness is obtained by considering the characters stored at the indices 2,4 $(= 2 \times 2)$ and 8 $(= 2 \times 2^2)$, namely $*$ (first character of the first word), $\circ$ (first word, second character), and $\star$ (second word, first character). Obviously there are "holes" in the array, they are simply ignored.

### 6.1.3 The Signature

The array is encoded by two indexed propositions: $\mathrm{car}(w, \nu, \lambda)$ and $\mathrm{t}(w, \nu)$ ($t$ stands for "tail") where $w \in \{a, b\}$, $1 \leq \nu \leq \kappa$, $1 \leq \lambda \leq |w_\nu|$. The intuition behind $\mathrm{car}(w, \nu, \lambda)_l$ is that it holds iff the index $l$ in the array corresponding to $w_\Delta$ contains the pair $(\nu, \lambda)$ (representing the character $w_\nu^\lambda$). $\mathrm{t}(w, \nu)_l$ states that the index $l$ of the array corresponding to $w_\Delta$ contains $\nu$.

### 6.1.4 Formal Definition of the Encoding

Let $n$ be a variable (intended to denote the unique parameter of the schema).

As explained in the previous section, we store the characters in an array, at the indices $\mu, 2\mu, 4\mu$, etc. Intuitively, $\mu$ should be encoded as another parameter, but only one parameter $n$ is allowed. However, we can encode $\mu$ with a new proposition symbol in $\mathcal{P}$. We first define two symbols $p, q$ s.t. $p_\nu$ holds iff $\nu = \mu$ and s.t. $q_\nu$ holds iff $\nu \in [0, \mu - 1]$. The first schema defines $q$ in such a way that it holds exactly on an interval of the form $[0, \mu - 1]$:

$$q_0 \wedge \neg q_n \wedge \bigwedge_{i=1}^{n} (q_{i+1} \Rightarrow q_i)$$

The last formula obviously implies that if $q_\nu$ holds for some $\nu \in [1..n]$ then it must also hold for every $\lambda \in [1..\nu]$. Then $\mu$ is simply the first index $\nu$ such that $q_\nu$ does not hold (this element necessarily exists, since $q_n$ does not hold).

The second schema defines $p$ such that it holds exactly on the successor of the maximal element of the interval (i.e. $\mu$). Notice that due to the previous formula we must have $\mu \neq 0$ and $\mu \leq n$:

$$\bigwedge_{i=1}^{n} [p_i \Leftrightarrow (q_{i-1} \wedge \neg q_i)]$$

For the sake of clarity, we shall denote by $(\lambda = \mu)$ the atom $p_\lambda$ and by $(\lambda < \mu)$ the atom $q_\lambda$ (this makes the formulae much more readable).

We then define a variable $wt$ s.t. $wt_\nu$ holds iff there exists $\lambda \in \mathbb{N}$ s.t. $\nu = \mu.2^\lambda$: $wt$ stands for "witness", because $wt_\nu$ holds iff $\nu$ is the index of a character in the witness of a solution, as explained before:

$$\bigwedge_{i=1}^{n}[((i = \mu) \Rightarrow wt_i) \wedge ((i < \mu) \Rightarrow \neg wt_i) \tag{1}$$
$$\wedge (\neg(2i + 1 = \mu) \Rightarrow \neg wt_{2i+1}) \wedge (\neg(2i < \mu) \wedge \neg(2i = \mu)) \Rightarrow (wt_i \Leftrightarrow wt_{2i})]$$

The first line states that $wt_\mu$ holds and that $wt_i$ is false if $i < \mu$. The second line defines that value of $wt_i$ for $i > \mu$: $wt_{2i+1}$ is always false (except if $2i + 1 = \mu$) and $wt_{2i}$ is equivalent to $wt_i$ if $2i > \mu$. By an easy induction on the set of natural numbers, these properties imply that $wt_\nu$ holds iff $\exists \lambda.\nu = \mu.2^\lambda$. Notice the crucial use of the homothetic transformation here.

The following formula states that an index cannot represent two distinct characters (pairs) in the same sequence:

$$\bigwedge_{i=1}^{n}(\neg\mathrm{car}(w, \nu, \lambda)_i \vee \neg\mathrm{car}(w, \nu', \lambda')_i)$$

for every $w \in \{a, b\}$, $(\nu, \nu') \in [1, \kappa]^2, \lambda \in [1, |w_\nu|], \lambda' \in [1, |w_{\nu'}|]$ s.t. $(\nu, \lambda) \neq (\nu', \lambda')$

Similarly, we state that every index contains at most one word in each sequence:

$$\bigwedge_{i=1}^{n}(\neg\mathrm{t}(w, \nu)_i \vee \neg\mathrm{t}(w, \nu')_i) \text{ for every } w \in \{a, b\}, \nu, \nu' \in [1, \kappa]^2, \nu \neq \nu'$$

Both initial elements of the sequences corresponding to $a$ and $b$ must be of the form $(\nu, 1)$ ($\nu$ is the same in both sequences and is distinct from $\kappa$, since the word $|w_\kappa|$ marks the end of the sequence):

$$\bigwedge_{i=1}^{n}((i = \mu) \Rightarrow \exists \nu \in [1, \kappa - 1](\mathrm{car}(a, \nu, 1)_i \wedge \mathrm{car}(b, \nu, 1)_i))$$

We use existential quantification over intervals of natural numbers for the sake of clarity, but these quantifiers can be easily eliminated and transformed into *finite* (*not iterated*) disjunctions.

The next formula defines $e(w)$ to mark the end of the sequence corresponding to $w$. $e(w)_l$ should hold iff $l$ is of the form $\mu.2^\lambda$ for some $\lambda > 0$ and if the character stored at the index $l$ is the first character of the word $\kappa$ (remember that by convention $a_\kappa = b_\kappa = \top$ where $\top$ marks the end of the witness). Besides, we must ensure that the end of the sequence is eventually reached i.e. that there exists an index $l$ such that $e(a)_l$ and $e(b)_l$ both hold:

$$\bigvee_{i=1}^{n}(e(a)_i \wedge e(b)_i) \wedge \bigwedge_{i=1}^{n}((wt_i \wedge \mathrm{car}(w, \kappa, 1)_i) \Leftrightarrow e(w)_i) \tag{$\star$}$$

$$\text{for every } w \in \{a, b\}$$

We also have to ensure that the two sequences (i.e. the words $a_\Delta$ and $b_\Delta$) are identical. It suffices to check that for every index $l$ s.t. $wt_l$ holds (i.e. for every index $l$ of the form $\mu \times 2^\lambda$), the character stored in $l$ is the same in the sequences of $a$ and $b$:

$$\bigwedge_{i=1}^n (wt_i \Rightarrow (\neg\text{car}(a, \nu, \lambda)_i \vee \neg\text{car}(b, \nu', \lambda')_i)) \qquad (\star)$$

for every $\nu, \nu' \in [1, \kappa]^2$, $\lambda \in [1, |a_\nu|]$, $\lambda' \in [1, |b_{\nu'}|]$ s.t. $a_\nu^\lambda \neq b_{\nu'}^{\lambda'}$

So far, we have ensured that at most one character and word index can be stored in every index. We have defined the starting point and the end of the two sequences and ensured that the two represented words are identical. The next (and most difficult) step is to ensure that these sequences really encode two words of the form $a_\Delta$ and $b_\Delta$ respectively. To this aim, we shall relate the value of the character stored in every index $\mu.2^{\lambda+1}$ to the one stored in $\mu.2^\lambda$, to ensure that the former is really the successor of the latter in the witness. Since each character $c$ is represented by a pair $(\nu, \iota)$ where $\nu$ denotes the index of a word in $w$ and $\iota$ is the position of $c$ in $w_\nu$, it is easy to find the next character: if $\iota < |w_\nu|$ (i.e. if $c$ is not the last character in $w_\nu$) then the next character is simply $(\nu, \iota + 1)$ (same word $w_\nu$, next position $\iota + 1$). If $\iota = |w_\nu|$ (i.e. if $c$ is the last character in $w_\nu$) then the next character is $(\nu', 1)$ where $\nu'$ denotes the next word index in the solution sequence (word $w_{\nu'}$, first position).

In order to determine the index word $\nu'$ we use the fact that (as explained in the informal overview above) the remaining indices in the solution are stored in the index $\mu.2^\lambda + 1, \mu.2^\lambda + 2, \ldots$. Thus, we simply need to pick up the first element of this sequence.

After checking that the character stored at $\mu.2^{\lambda+1}$ is the successor of the one in $\mu.2^\lambda$ it remains to ensure that the indices stored at $\mu.2^{\lambda+1} + 1$, $\mu.2^{\lambda+1} + 2, \ldots$ correspond to the remaining part of the solution. If $\iota < |w_\nu|$ then the sequence must actually be identical to the one stored at $\mu.2^\lambda + 1$, $\mu.2^\lambda + 2, \ldots$ If $\iota = |w_\nu|$ then the first element of the sequence must be deleted (since we have entered into a new word).

The next formula states that if an index $l$ of the form $\mu.2^\lambda$ (i.e. an index s.t. $wt_l$ holds) contains a pair $(\nu, \iota)$ and if $w_\nu$ contains more that $\iota$ characters then $\mu.2^{\lambda+1}$ should encode the next character in the word $w_\nu$, namely $(\nu, \iota + 1)$. Moreover the tail of the sequence does not change, which is expressed using the variable $\text{c}(w)_l$ ($c$ stands for "copy") that will be specified thereafter:

$$\bigwedge_{i=1}^n [(wt_i \wedge \text{car}(w, \nu, \lambda)_i) \Rightarrow (\text{car}(w, \nu, \lambda + 1)_{2i} \wedge \text{c}(w)_{i+1})] \qquad (2)$$

for every $w \in \{a, b\}$, $\nu \in [1, \kappa]$, $\lambda \in [1, |w_\nu| - 1]$.

Now we define the formula encoding the copy of the tail. The most simple way to proceed would be to copy the values stored into the indices $l, l+1, \ldots, l+\mu-1$ into $2l+1, \ldots, 2l+\mu-1$. Unfortunately this cannot be done in this simple way because expressions of the form $l + j$ would be required in the indices, which is forbidden in our class (only $\pm 1$ can be added). As explained before, we overcome this problem by copying the indices $l+1, \ldots, l+\mu-1$ into $2l+2, 2l+4, \ldots, 2l+2\mu-2$, which can be done by doubling the iteration counter. The indices $2l+1, 2l+3, \ldots, 2l+2\mu-1$ are left empty (holes). This is not disturbing since such empty indices will simply be ignored. An important consequence is that the length of the sequence is doubled each time it is copied (we assume that the value of the parameter $n$ and the natural number $\mu$ are sufficiently large to ensure that there is enough "space" in the array).

This is expressed by the following formula:

$$\bigwedge_{i=1}^{n} (c(w)_i \Rightarrow [\neg t(w,\nu)_{2i-1} \wedge (t(w,\nu)_i \Leftrightarrow t(w,\nu)_{2i}) \wedge (\neg wt_{i+1} \Rightarrow c(w)_{i+1})]) \qquad (3)$$

$$\text{for every } \nu \in [1,\kappa], \ w \in \{a,b\}$$

We illustrate this construction by an example. Let $\mathcal{A} = \{*, \circ, \star, \diamond\}$, $a = (*\circ, \star, \diamond\circ)$ and $\Delta = (1,2,3)$. In the second line, we provide for every index $l$ the pair $(\nu, \iota)$ such that $car(a, \nu, \iota)_l$ holds (if any). The third line gives the represented character ($*, \circ, \star$ or $\diamond$). In the fourth line we provide the integer $\nu$ such that $t(a, \nu)_l$ holds. The fifth line gives the value of $c(a)$. The indices between $\mu+2$ and $2\mu$ are empty (we assume that $\mu=3$).

| $i$ | $\mu$ | $\mu+1$ | $\mu+2$ | $2\mu$ | $2\mu+1$ | $2\mu+2$ | $2\mu+3$ | $2\mu+4$ |
|---|---|---|---|---|---|---|---|---|
| car | $(1,1)$ | | | $(1,2)$ | | | | |
| character | $*$ | | | $\circ$ | | | | |
| t | | 2 | 3 | | | 2 | | 3 |
| c(a) | | **T** | **T** | | | | | |

By formula (2) we must have $c_{\mu+1}$. By formula (3), the value of $c(a)_{\mu+1}$ is propagated to $c(a)_{\mu+2}, \ldots, c(a)_{2\mu-1}$ (it is not propagated to $c(a)_{2\mu}$ since $wt_{2\mu}$ holds). Still by (2), if $c(a)_l$ holds then we have $t(a,\nu)_l \Leftrightarrow t(a,\nu)_{2l}$, and the cells corresponding to odd indices are left empty. Thus we get the array above.

If an index $\mu.2^{\lambda}$ contains a pair $(\nu, \iota)$ where $|w_\nu| = \iota$ (such as $2\mu$ in the previous example), then one must proceed to the next word. To this aim, we need to know what is the first character of the next word (after the current one). Because of the holes introduced by the special copying mechanism, the next word is not necessarily at index $l+1$. A simple solution is to change the contents of the tail so that each element contains not only the index of a *word* but also its first *character*. This is stated by the following formula:

$$\bigwedge_{i=1}^{n} [\neg wt_i \Rightarrow (t(w,\nu)_i \Rightarrow car(w,\nu,1)_i)] \qquad (4)$$

*Furthermore, we copy this character into all the holes preceding the element.* As a particular case we get what we wanted for the first non-empty word.[10] This is stated by

---

10. Notice that we could have as well copied the word's index instead of its first character, since the index contains all the information we need to retrieve the corresponding character. However it will be useful in the following to know that there is no word index stored in a particular cell, so we store only the information that is useful for the problem we want to solve at this point, i.e. the first character of the word.

the following formula:

$$\bigwedge_{i=1}^{n} [(\neg wt_{i-1} \wedge \neg wt_i \wedge \forall \lambda \in [1, \kappa] \ \neg t(w, \lambda)_{i-1}) \Rightarrow (\text{car}(w, \nu, 1)_i \Leftrightarrow \text{car}(w, \nu, 1)_{i-1})] \quad (5)$$

for every $\nu \in [1, \kappa]$, $w \in \{a, b\}$

Now, if the pair stored in $\iota$ is $(\nu, |w_\nu|)$ and if this word is not the final word in the sequence (i.e. $e(w)_\iota$ does not hold) then one has to store into $2\iota$ the first character of the next word, which is, due to the two previous formulae, the character represented by $\iota + 1$. The previous picture must thus be completed as follows:

| $i$ | | $\mu$ | $\mu + 1$ | $\mu + 2$ | $2\mu$ | $2\mu + 1$ | $2\mu + 2$ | $2\mu + 3$ | $2\mu + 4$ |
|-----------|---|--------|-----------|-----------|---------|------------|------------|------------|------------|
| car | | $(1,1)$ | $(2,1)$ | $(3,1)$ | $(1,2)$ | $(2,1)$ | $(2,1)$ | $(3,1)$ | $(3,1)$ |
| character | | $*$ | $\star$ | $\diamond$ | $\circ$ | $\star$ | $\star$ | $\diamond$ | $\diamond$ |
| t | | | 2 | 3 | | | 2 | | 3 |

By the formula (4) above, if $t(a, \nu)_i$ holds then $\text{car}(a, \nu, 1)_i$ also holds. Then by the formula (5), the value of $\text{car}(a, \nu, 1)_l$ is recursively propagated to $\text{car}(a, \nu, 1)_{l-1}$ until we have $l - 1 = \mu.2^\lambda$ or $t(a, \nu)_{l-1}$ holds for some $\nu$. Notice that a character is now stored in every index $l$ but only the characters in the indices $\mu.2^\lambda$ form the witness.

Thanks to this trick, finding the next character after the one stored in $\mu.2^\lambda$ is now trivial: this is simply the one stored in $\mu.2^\lambda + 1$, which, by the previous formula, actually corresponds to the first position of the word stored in $(\mu + 1).2^\lambda$ (of course, we also need to check that the character is not final). This is expressed by the following formula:

$$\bigwedge_{l=1}^{n} [(wt_l \wedge \neg e(w)_l \wedge \text{car}(w, \nu, |w_\nu|)_l) \Rightarrow (\text{car}(w, \lambda, 1)_{2l} \Leftrightarrow \text{car}(w, \lambda, 1)_{l+1}) \wedge s(w)_{l+1}]$$

for every $\nu, \lambda \in [1, \kappa]$, $w \in \{a, b\}$

The propositional variable $s(w)_{l+1}$ ($s$ stands for "shift") indicates that the tail at $2l$ is obtained by removing the first word in the tail at $l$. This is done as follows: the indices $2l + 2, \ldots, 2l + 2\mu - 1$ are obtained by copying the indices $l + 1, \ldots, l + \mu - 1$, except the first one, that is left empty. As for $c(w)$, the indices $2l - 1, \ldots, 2l + 2\mu - 3$ are empty. $s(w)$ is defined by the three following formulae.

$s(w)$ actually erases everything until it finds a non-empty index, which is expressed by the first formula: if $s(w)_l$ holds then the indices stored at $2l$ and $2l - 1$ must be empty (furthermore, we also check that the end of the tail has not been reached):

$$\bigwedge_{l=1}^{n} (s(w)_l \Rightarrow \neg wt_l \wedge \neg t(w, \nu)_{2l} \wedge \neg t(w, \nu)_{2l-1}) \quad \text{for every } \nu \in [1, \kappa], w \in \{a, b\} \quad (6)$$

The second one propagates the erasure if the current index is empty:

$$\bigwedge_{l=1}^{n} [(s(w)_l \wedge \neg wt_{l+1} \wedge \forall \nu \in [1, \kappa] \ \neg t(w, \nu)_l) \Rightarrow s(w)_{l+1} \quad \text{for every } w \in \{a, b\} \tag{7}$$

The third one states that once we have reached a non-empty index then we go on by copying everything (which is done by using the previous variable $c(w)$):

$$\bigwedge_{l=1}^{n} (s(w)_l \wedge \exists \lambda \in [1, \kappa] \ t(w, \lambda)_l \Rightarrow c(w)_{l+1} \text{ for every } w \in \{a, b\} \tag{8}$$

We illustrate this construction by showing how the erasure works on the previous example:

| $i$ | $2\mu$ | $2\mu+1$ | $2\mu+2$ | $2\mu+3$ | $2\mu+4$ |
|---|---|---|---|---|---|
| car | $(1,2)$ | $(2,1)$ | $(2,1)$ | $(3,1)$ | $(3,1)$ |
| character | $\circ$ | $\star$ | $\star$ | $\diamond$ | $\diamond$ |
| t | | | 2 | | 3 |
| c(a) | | | | **T** | **T** |
| s(a) | | **T** | **T** | | |

| $i$ | $4\mu$ | $4\mu+1$ | $4\mu+2$ | $4\mu+3$ | $4\mu+4$ | $4\mu+5$ | $4\mu+6$ | $4\mu+7$ | $4\mu+8$ |
|---|---|---|---|---|---|---|---|---|---|
| car | $(2,1)$ | $(3,1)$ | $(3,1)$ | $(3,1)$ | $(3,1)$ | $(3,1)$ | $(3,1)$ | $(3,1)$ | $(3,1)$ |
| character | $\star$ | $\diamond$ | $\diamond$ | $\diamond$ | $\diamond$ | $\diamond$ | $\diamond$ | $\diamond$ | $\diamond$ |
| t | | | | | | | | | 3 |

The character stored in $2\mu$ is the last one of the first word thus we have to remove the first word in the tail of the solution. As explained before, the character stored in $4\mu$ is the same as the one stored in $2\mu+1$, namely $(2,1)$, i.e. $\star$ (since we have $car(a, \nu, 1)_{4\mu} \Leftrightarrow car(a, \nu, 1)_{2\mu+1}$). Furthermore, $s(a)_{2\mu+1}$ holds. This implies by (6) that the indices $4\mu+2$ and $4\mu+1$ of $t$ must be empty. Since t is empty for $\iota = 2\mu+1$ (i.e. there is no $\nu$ such that $t(a, \nu)_{2\mu+1}$ holds), the value of $s(a)_{2\mu+1}$ is propagated to $s(a)_{2\mu+1}$, by (7). Thus by (6), the indices $4\mu+4$ and $4\mu+3$ of t must also be empty. This time, however, $t(a, 2)_{2\mu+2}$ holds. Thus the value of $s(a)$ is not propagated and $c(a)_{2\mu+3}$ must hold (by (8)). As before, this implies that the remaining part of the sequence (i.e. the cells $2\mu+3$, $2\mu+4$ of t) is copied (in the cells $4\mu+6$, $4\mu+8$, leaving the cells $4\mu+5$, $4\mu+7$ empty) until $4\mu$ is reached. This implies in particular that $t(a, 3)_{4\mu+8}$ holds (since $t(a, 3)_{2\mu+4}$ holds). Hence we have $car(a, 3, 1)_{4\mu+8}$. Since t is empty for $l \in [4\mu+1, \ldots, 4\mu+7]$, this value of $car(a, 3, 1)$ is propagated to the indices $4\mu+7, \ldots, 4\mu+1$ as explained before. We obtain the desired result, i.e. the first word in the sequence (namely 2) have been erased and the first character of the next word is stored into $4\mu+1$.

Finally, in order to ensure that the obtained sequence is really a solution to the Post correspondence problem, it only remains to check that the two sequences are identical, i.e. that the words contained in $\mu+1, \cdots, 2\mu-1$ are the same for both sequences $a$ and $b$. To this purpose we define a variable $r_l$ that is true iff $l < 2\mu$.

$$r_0 \wedge \neg r_n \wedge \bigwedge_{l=1}^{n} [(r_l \Rightarrow r_{l-1}) \wedge (l = \mu) \Rightarrow (r_{2l-1} \wedge \neg r_{2l})] \qquad (\star)$$

$$\bigwedge_{l=1}^{n} [(r_l \wedge \neg(l < \mu)) \Rightarrow (\mathrm{t}(a, \nu)_l \Leftrightarrow \mathrm{t}(b, \nu)_l)] \qquad (\star)$$

for every $\nu \in [1, \kappa]$

It is straightforward to check that the obtained formula is in $\mathfrak{C}_h$. The reader acquainted with Post's correspondence problem shall now be convinced that the obtained formula is satisfiable iff there exists a solution to the above Post problem, and can thus skip the end of this section. Otherwise we give in the following a sketch of the formal steps to this proof.

We denote by $\phi$ the conjunction of the above formulae, except the formulae marked $(\star)$. We first notice that $\phi$ is satisfiable (for every value of $n$). Indeed, as explained before, the formulae above impose that:

- There exists a unique natural number $\mu$ such that $p_\nu$ holds iff $\nu = \mu$ and $q_\nu$ holds iff $\nu \in [0, \mu - 1]$.

- $\mathrm{car}(w, \nu, \lambda)$ and $\mathrm{t}(w, \nu)$ encode (partial) functions $f_w, g_w$ mapping every index in $[1, n]$ to a pair $(\nu, \lambda)$ (where $\nu \in [1, \kappa], \lambda \in [1, |w_\nu|]$) and to a word index in $[1, \kappa]$ respectively. Moreover we must have $f_a(\mu) = (\nu, 1)$ and $f_b(\mu) = (\nu, 1)$ for some $\nu \in [1..\kappa - 1]$.

- $wt_\nu$ holds iff there exists $\lambda \in \mathbb{N}$ s.t. $\nu = \mu.2^\lambda$.

This obviously defines a partial interpretation. Then the remaining formulae in $\phi$ simply give the values of $\mathrm{car}(w, \nu, \lambda)_\iota$, $\mathrm{t}(w, \nu)_\iota$, $\mathrm{c}(w)_\iota$, $\mathrm{s}(w)_\iota$ for $\iota \geq 2\mu$. It is easy to check that distinct formulae cannot give distinct values to the same propositional variable, hence satisfiability is guaranteed.

Let $\mathcal{I}$ be an interpretation of $\phi$. Let $\iota \in [1..n]$. We define the following sequences.

- $h_w(\iota)$ is a sequence of word indices defined as follows: If $wt_{\iota+1}$ holds then $h_w(\iota)$ is empty. Otherwise, if $g_w(\iota) = \nu$ then $h_w(\iota) \overset{\mathrm{def}}{=} \nu.h_w(\iota + 1)$ and if $g_w(\iota)$ is undefined then $h_w(\iota) \overset{\mathrm{def}}{=} h_w(\iota + 1)$. Intuitively, $h_w(\iota)$ is the sequence of word indices stored juste after $\iota$ (i.e. the tail) ignoring empty cells.

- $j_w(\iota)$ is a word defined as follows: if $\iota > n$ then $j_w(\iota)$ is empty. Otherwise, $j_w(\iota) \overset{\mathrm{def}}{=} w_\nu^\lambda.j_w(2\iota)$ if $f_w(\iota)$ is a pair $(\nu, \lambda)$ distinct from $(\kappa, 1)$, $j_w(\iota) \overset{\mathrm{def}}{=} \top$ if $f_w(\iota) = (\kappa, 1)$ and $j_w(\iota) \overset{\mathrm{def}}{=} j_w(2\iota)$ if $f_w(\iota)$ is undefined. $j_w(\iota)$ denotes the word stored at the cells $\iota, 2\iota \ldots$ in the array corresponding to $w$ ($(\kappa, 1)$ marks the end of the word).

- If $f_w(\iota) = (\nu, \nu')$ then $k_w(\iota)$ denotes the suffix of length $|w_\nu| - \nu' + 1$ of the word $w_\nu$ (notice that by construction we must have $\nu' \leq |\nu|$).

By definition of the copying/erasing mechanism above, if $f_w(\mu.2^\lambda)$ is of the form $(\nu, |w_\nu|)$ (i.e. we are at the end of the word $\nu$) then $h_w(\mu.2^\lambda) = \nu'.h_w(\mu.2^{\lambda+1})$, where $f_w(\mu.2^{\lambda+1}) =$

651

$(\nu', 1)$ (i.e. the tail is equal to the next word followed by the next tail). Otherwise (i.e. if we are in the middle of a word) we have $h_w(\mu.2^\lambda) = h_w(\mu.2^{\lambda+1})$ and $f_w(\mu.2^{\lambda+1}) = (\nu, \nu' + 1)$ where $f_w(\mu.2^\lambda) = (\nu, \nu')$ ($\nu' \neq |w_\nu|$)). By an easy induction on the length of $j_w(\mu.2^\lambda)$, we deduce that $j_w(\mu.2^\lambda)$ is a prefix of $k_w(\mu.2^\lambda).w_{h_w(\mu.2^\lambda)}$: $k_w(\mu.2^\lambda)$ represents the end of the word considered at the character $\lambda$, and $w_{h_w(\mu.2^\lambda)}$ is the concatenation of all words in the tail.

For $\lambda = 0$ we get in particular that $j_w(\mu)$ is a prefix of $k_w.w_{h_w(\mu)}$. But by definition $k_w(\mu) = w_\nu$ for some $\nu$ (not depending on $w$). Thus $j_w(\mu)$ is a prefix of $w_{\nu.h_w(\mu)}$.

The formulae occurring in the conjunction but not in $\phi$ check that $h_a(\mu) = h_b(\mu)$ (same sequence of word indices for $a$ and $b$), that $j_a(\mu) = j_b(\mu)$ and that $j_a(\mu)$ ends with a character $\top$ (marking the end of the witness).

If $\mathcal{I}$ is a model of the whole formula, then $j_w(\mu)$ is a prefix of $w_{\nu.h_w(\mu)}$, ending with $\top$, thus must be of the form $w_{\nu.\Delta}$ where $\Delta$ is a prefix of $h_w(\mu)$. Hence $\nu.\Delta$ is a solution to the Post's correspondence problem.

Conversely, if such a solution $\nu.\Delta$ exists, then we simply consider a model $\mathcal{I}$ of $\phi$ such that $h_a(\mu) = h_b(\mu) = \Delta$ (this implies that $\mu > |\Delta|$, notice that the values of $f_w(l)$ and $g_w(l)$ can be fixed arbitrarily for $l < 2\mu$) and $\mathcal{I}(n) > \mu.2^\lambda$, where $\lambda = |a_{\nu.\Delta}|$. $j_w(\mu)$ is a prefix of $w_{\nu.h_w(\mu)}$. Since the length $j_w(\mu)$ cannot be greater than the one of $w_{\nu.\Delta}$, $j_w(\mu)$ must end with $\top$. Thus we must have $j_w(\mu) = w_{\nu.\Delta}$ (since $\top$ is the last character in $w_{\nu.\Delta}$). Moreover since $\nu.\Delta$ is a solution we have $j_a(\mu) = j_b(\mu)$. Thus $\mathcal{I}$ validates all the formulae above.

## 6.2 Unbounded Translation

One can wonder whether the decidability of the class of regular schemata still holds when unbounded translations are allowed in the indices, i.e. translations of the form $i + m$ where $i$ denotes the iteration counter and $m$ a *parameter* (the case $m \in \mathbb{Z}$ is covered by the regular class). The following definition and theorem show that the answer is negative.

### Definition 6.3
$\mathfrak{C}_t$ ($t$ stands for "translation") is the set of schemata $S$ satisfying the following properties.

- $S$ contains at most two parameters $n, m$.

- Every iteration in $S$ is of the form $\bigwedge_{i=1}^n \phi$ or $\bigvee_{i=1}^n \phi$, where:

  - $\phi$ contains no iteration.
  - Every atomic formula in $\phi$ is of the form $p_{\alpha.i+\beta+\gamma.m}$, where $p$ is a variable, $\alpha, \gamma \in \{0, 1\}$ and $\beta \in \{-1, 0, 1\}$.

- The atomic propositions occurring in $\phi$ but not in the scope of an iteration are of the form $p_0$ or $p_n$ where $p$ is a variable.

### Theorem 6.4
The set of unsatisfiable formulae in $\mathfrak{C}_t$ is not recursively enumerable.

### PROOF
(Sketch) We do not detail the proof since it is very similar to the previous one. We reuse the same encoding as in the proof of Theorem 6.2, except that the pairs $(\nu, \lambda)$ in the array are stored in indices of the form $\mu + m \times \iota$ instead of $\mu.2^\iota$. Formally, the formulae (1), (2), (3) and (6) are replaced by the following ones, respectively:

$$\bigwedge_{l=1}^{n} [((l = \mu) \Rightarrow wt_l) \wedge ((l < \mu) \Rightarrow \neg wt_l) \wedge (\neg(l < \mu) \wedge \neg(l = \mu)) \Rightarrow (wt_l \Leftrightarrow wt_{l+m})]$$

(i.e. $wt_l$ holds now iff there exists $\iota$ s.t. $l = \mu + m\iota$).

$$\bigwedge_{l=1}^{n} [wt_l \wedge \text{car}(w, \nu, \lambda)_l \Rightarrow (\text{car}(w, \nu, \lambda + 1)_{l+m} \wedge \text{c}(w)_{l+1})]$$

(i.e. the index $2l$ is now replaced by $l + m$).

$$\bigwedge_{l=1}^{n} (\text{c}(w)_l \Rightarrow [(\text{t}(w, \nu)_l \Leftrightarrow \text{t}(w, \nu)_{l+m}) \wedge (\neg wt_{i+1} \Rightarrow \text{c}(w)_{l+1})])$$

$$\bigwedge_{l=1}^{n} (\text{s}(w)_l \Rightarrow \neg wt_l \wedge \neg \text{t}(w, \nu)_{l+m}) \qquad \qquad \square$$

## 7. Conclusion

We introduced the first (to the best of our knowledge) logic for reasoning with iterated propositional schemata. We defined a class of schemata called *bound-linear* for which the satisfiability problem is decidable. The decidability proof is constructive and divided into two parts: first we show how to transform every bound-linear schema into a sat-equivalent schema of a simpler form, called *regular*. Then a proof procedure is defined to decide the satisfiability of regular schemata. This proof procedure is sound and complete w.r.t. satisfiability for *every* schema (even if it is not regular or not bound-linear) and terminates on *every regular schema*. Termination relies on a special looping detection rule. This procedure has been implemented in the software RegStab.

The class of bound-linear schemata is expressive enough to capture specifications of many important problems in AI, especially in automated (or interactive) theorem proving (e.g., parameterized circuit verification problems). We proved that even a very slight relaxation of the conditions on bound-linear schemata makes the satisfiability problem undecidable (this is shown by a tricky reduction to the Post correspondence problem). As a consequence, bound-linear schemata can be considered as a "canonical" decidable class, providing a good compromise between expressivity and tractability.

As for future work, two ways are the most promising. Firstly, the extension of the previous results to particular classes of non-monadic schemata (i.e. schemata containing symbols with several indices, e.g., $\bigvee_{i=1}^{n} \bigwedge_{j=1}^{n} p_{i,j}$) would enlarge considerably applications of propositional schemata. Secondly, extending our approach to more expressive logics, such as first-order logic, description logics or modal logics, also deserves to be considered. The presented results should extend straightforwardly to many-valued propositional logic (provided the number of truth values is fixed and finite). This would allow to capture infinite constraint satisfaction languages.

## Acknowledgments

This work has been partly funded by the project ASAP of the French *Agence Nationale de la Recherche* (ANR-09-BLAN-0407-01). The authors wish to thank the anonymous referees for their insightful comments which helped improve an earlier version of this paper.